\documentclass[twocolumn,tighten]{aastex61}
\usepackage{CJKutf8}
\usepackage{apjfonts}
\usepackage{booktabs,pbox}
\usepackage{amsmath}

\graphicspath{{./maps/}{./masshist/}{./sig-Sig/}{./summary/}{./sig-Tpeak/}{./mcmc/}}

\newcommand{\citeinprep}[1]{#1}

\newcommand{\halfpanel}[3]{
\hfill
\vbox{
\parskip=0pt\hsize=0.49\textwidth\centering
\includegraphics[height=#2]{#1}
\vskip2pt
\vtop{\centering\footnotesize\hsize=0.49\textwidth
#3\vskip1pt}
}
\hfill}

\newcommand{\kms}{\mbox{km\,s$^{-1}$}}
\newcommand{\Kkms}{\mbox{K\,km\,s$^{-1}$}}
\newcommand{\Msun}{\mbox{$\mathrm{M}_{\odot}$}}

\newcommand{\Tpeak}{\mbox{$T_\mathrm{peak}$}}
\newcommand{\Ico}{\mbox{$I_\mathrm{CO}$}}
\newcommand{\alphaco}{\mbox{$\alpha_\mathrm{CO}$}}
\newcommand{\Sig}{\mbox{$\Sigma$}}
\newcommand{\sig}{\mbox{$\sigma$}}
\newcommand{\alphavir}{\mbox{$\alpha_\mathrm{vir}$}}
\newcommand{\Pturb}{\mbox{$P_\mathrm{turb}$}}
\newcommand{\Pext}{\mbox{$P_\mathrm{ext}$}}
\newcommand{\PDE}{\mbox{$P_\mathrm{DE}$}}
\newcommand{\rgal}{\mbox{$r_\mathrm{gal}$}}
\newcommand{\rbeam}{\mbox{$r_\mathrm{beam}$}}

\accepted{2018 Apr 26}
\shorttitle{Cloud-Scale Molecular Gas Properties in 15 Nearby Galaxies}
\shortauthors{Sun et al.}

\begin{document}

\title{Cloud-Scale Molecular Gas Properties in 15 Nearby Galaxies}

\author{\begin{CJK*}{UTF8}{} Jiayi~Sun ({\CJKfamily{gbsn}孙嘉懿}) \end{CJK*}}
\affil{Department of Astronomy, The Ohio State University, 140 West 18th Avenue, Columbus, OH 43210, USA}

\author{Adam~K.~Leroy}
\affil{Department of Astronomy, The Ohio State University, 140 West 18th Avenue, Columbus, OH 43210, USA}

\author{Andreas~Schruba}
\affil{Max-Planck-Institut f\"ur Extraterrestrische Physik, Giessenbachstra{\ss}e 1, D-85748 Garching, Germany}

\author{Erik~Rosolowsky}
\affil{Department of Physics, University of Alberta, Edmonton, AB T6G 2E1, Canada}

\author{Annie~Hughes}
\affil{CNRS, IRAP, 9 av. du Colonel Roche, BP 44346, F-31028 Toulouse cedex 4, France}
\affil{Universit\'{e} de Toulouse, UPS-OMP, IRAP, F-31028 Toulouse cedex 4, France}

\author{J.~M.~Diederik~Kruijssen}
\affil{Astronomisches Rechen-Institut, Zentrum f\"{u}r Astronomie der Universit\"{a}t Heidelberg, M\"{o}nchhofstra\ss e 12-14, D-69120 Heidelberg, Germany}

\author{Sharon~Meidt}
\affil{Max-Planck-Institut f\"{u}r Astronomie, K\"{o}nigstuhl 17, D-69117, Heidelberg, Germany}

\author{Eva~Schinnerer}
\affil{Max-Planck-Institut f\"{u}r Astronomie, K\"{o}nigstuhl 17, D-69117, Heidelberg, Germany}

\author{Guillermo~A.~Blanc}
\affil{Observatories of the Carnegie Institution for Science, 813 Santa Barbara Street, Pasadena, CA 91101, USA}
\affil{Departamento de Astronom\'{i}a, Universidad de Chile, Camino del Observatorio 1515, Las Condes, Santiago, Chile}

\author{Frank~Bigiel}
\affil{Institute für theoretische Astrophysik, Zentrum f\"{u}r Astronomie der Universit\"{a}t Heidelberg, Albert-Ueberle Str. 2, D-69120 Heidelberg, Germany}

\author{Alberto D. Bolatto}
\affil{Department of Astronomy, University of Maryland, College Park, MD 20742, USA}

\author{M\'elanie~Chevance}
\affil{Astronomisches Rechen-Institut, Zentrum f\"{u}r Astronomie der Universit\"{a}t Heidelberg, M\"{o}nchhofstra\ss e 12-14, D-69120 Heidelberg, Germany}

\author{Brent~Groves}
\affil{Research School for Astronomy \& Astrophysics Australian National University Canberra, ACT 2611, Australia}

\author{Cinthya~N.~Herrera}
\affil{Institut de Radioastronomie Millim\'{e}trique (IRAM), 300 Rue de la Piscine, F-38406 Saint Martin d'H\`{e}res, France}

\author{Alexander~P.~S.~Hygate}
\affil{Max-Planck-Institut f\"{u}r Astronomie, K\"{o}nigstuhl 17, D-69117, Heidelberg, Germany}
\affil{Astronomisches Rechen-Institut, Zentrum f\"{u}r Astronomie der Universit\"{a}t Heidelberg, M\"{o}nchhofstra\ss e 12-14, D-69120 Heidelberg, Germany}

\author{J\'er\^ome~Pety}
\affil{Institut de Radioastronomie Millim\'{e}trique (IRAM), 300 Rue de la Piscine, F-38406 Saint Martin d'H\`{e}res, France}
\affil{Observatoire de Paris, 61 Avenue de l'Observatoire, F-75014 Paris, France}

\author{Miguel~Querejeta}
\affil{European Southern Observatory, Karl-Schwarzschild-Stra{\ss}e 2, D-85748 Garching, Germany}
\affil{Observatorio Astron\'{o}mico Nacional (IGN),C/Alfonso XII, 3, E-28014 Madrid, Spain}

\author{Antonio~Usero}
\affil{Observatorio Astron\'{o}mico Nacional (IGN),C/Alfonso XII, 3, E-28014 Madrid, Spain}

\author{Dyas~Utomo}
\affil{Department of Astronomy, The Ohio State University, 140 West 18th Avenue, Columbus, OH 43210, USA}

\begin{abstract}
We measure the velocity dispersion, $\sigma$, and surface density, $\Sigma$, of the molecular gas in nearby galaxies from CO spectral line cubes with spatial resolution $45$-$120$~pc, matched to the size of individual giant molecular clouds. Combining $11$ galaxies from the PHANGS-ALMA survey with $4$ targets from the literature, we characterize $\sim 30,000$ independent sightlines where CO is detected at good significance. $\Sigma$ and $\sigma$ show a strong positive correlation, with the best-fit power law slope close to the expected value for resolved, self-gravitating clouds. This indicates only weak variation in the virial parameter $\alpha_\mathrm{vir}\propto\sigma^2/\Sigma$, which is ${\sim}1.5$-$3.0$ for most galaxies. We do, however, observe enormous variation in the internal turbulent pressure $P_\mathrm{turb}\propto\Sigma\,\sigma^2$, which spans ${\sim}5$~dex across our sample. We find $\Sigma$, $\sigma$, and $P_\mathrm{turb}$ to be systematically larger in more massive galaxies. The same quantities appear enhanced in the central kpc of strongly barred galaxies relative to their disks. Based on sensitive maps of M31 and M33, the slope of the $\sigma$-$\Sigma$ relation flattens at $\Sigma\lesssim10\rm\;M_\odot\,pc^{-2}$, leading to high $\sigma$ for a given $\Sigma$ and high apparent $\alpha_\mathrm{vir}$. This echoes results found in the Milky Way, and likely originates from a combination of lower beam filling factors and a stronger influence of local environment on the dynamical state of molecular gas in the low density regime.
\end{abstract}

\keywords{}

\section{Introduction}\label{sec:intro}

Observational evidence, including the close association of {\sc Hii} regions with molecular clouds \citep[see references in][]{Morris_Rickard_1982} and the correlation between star formation rate (SFR) tracers and molecular gas content \citep[e.g.,][]{Rownd_Young_1999,Wong_Blitz_2002,Bigiel_etal_2008,Schruba_etal_2011}, suggests that cold molecular gas is the direct gas reservoir for star formation. On the other hand, the molecular interstellar medium (ISM) in galaxies is observed to have diverse physical properties and dynamical states \citep[e.g.,][]{Elmegreen_1989}. Within a galaxy and even within an individual cloud, molecular gas can show a range of surface and volume densities, turbulent velocities, and bulk motions. For a thorough understanding of how star formation happens in galaxies, it is critical to know how galactic environments (e.g., large-scale gas dynamics and stellar feedback) influence the physical properties of the molecular gas, and how the gas properties in turn determine its ability to form stars.

The evolution of a molecular cloud depends primarily on the balance between its kinetic and gravitational potential energy. Other mechanisms, e.g., magnetic fields and external pressure, might also act to support or confine the cloud. In both observational and theoretical work, this balance is commonly described by the virial parameter, $\alphavir \equiv 2K/U_g$, which captures the ratio between kinetic energy, $K$, and self-gravitational potential energy, $U_g$. Theoretical work by \citet{McKee_Zweibel_1992,Krumholz_McKee_2005,Padoan_Nordlund_2011,Federrath_Klessen_2012,Kruijssen_2012, Krumholz_etal_2012,Hennebelle_Chabrier_2013,Padoan_etal_2017} predict that \alphavir\ plays an important role in determining the ability of a cloud to form stars and stellar clusters. In these theories, clouds with high \alphavir\ (i.e., a relative excess of kinetic energy) form fewer stars per unit time for a given gas mass and density. As a result, the dynamical state of molecular gas at the scale of individual molecular clouds represents an important consideration for galactic-scale theories and simulations of star formation.

On the observational side, this topic has been investigated via studies of individual clouds both in the Galaxy and several nearby galaxies. Most commonly, analysis of the CO line emission from the molecular medium provides estimates of the cloud velocity dispersion, $\sig$, surface density, $\Sig$, size, $R$, and mass, $M$. The relationship between these quantities then gives information about the physical state of the cloud \citep[following][]{Larson_1981}. 

Recent work on this topic tends to emphasize the relationship between $\sig^2/R$ and $\Sig$ \citep[e.g.,][]{Heyer_etal_2009,Leroy_etal_2015}, as the position of a cloud in  $\sig^2/R$-$\Sig$ space probes its dynamical state and internal gas pressure \citep{Keto_Myers_1986,Field_etal_2011}. A nearly linear scaling relation between $\sig^2/R$ and $\Sig$, suggestive of bound clouds with velocity dispersion balancing their self-gravity, has been observed for clouds in the Milky Way disk \citep{Heyer_etal_2009}, the Large Magellanic Cloud \citep[LMC;][]{Wong_etal_2011}, nearby dwarf galaxies \citep{Bolatto_etal_2008}, spiral galaxies  \citep{DonovanMeyer_etal_2013,Colombo_etal_2014a}, starbursts \citep{Rosolowsky_Blitz_2005,Leroy_etal_2015}, and one lenticular galaxy \citep{Utomo_etal_2015}. In contrast, clouds situated in regions with low gas density \citep[e.g., the outer Galaxy;][]{Heyer_etal_2001} or high ambient pressure \citep[e.g., the Galactic Center;][]{Oka_etal_2001}, tend to show much higher velocity dispersions than are expected for self-virialized clouds. This suggests that in these environments external gravitational potential is at least as important as cloud self-gravity in regulating cloud dynamics \citep[][]{Heyer_etal_2009, Field_etal_2011, Kruijssen_etal_2014}.

Despite the insight obtained by these Galactic and early extragalactic studies, our knowledge of the physical state of molecular gas in other galaxies remains limited. Many extragalactic molecular cloud studies target Local Group galaxies and the nearest dwarf galaxies \citep[][]{Rosolowsky_2007,Bolatto_etal_2008, Wong_etal_2011,Druard_etal_2014}. Although nearby, these systems are not representative of where most stars in the present-day Universe form. Low mass galaxies tend to show faint, isolated CO emission \citep[][]{Fukui_etal_1999, Engargiola_etal_2003, Schruba_etal_2017}, distinct from the bright, spatially contiguous emission distributions observed in more massive galaxies \citep[][]{Hughes_etal_2013a}. As interferometers force trade-offs between surface brightness sensitivity and resolution, it remains challenging to access the entire cloud population of normal star-forming disk galaxies, (which are usually $\gtrsim$ 10 times more distant than Local Group targets), with most contemporary observing facilities. Most studies to date have either focused on a single galaxy \citep[][]{Colombo_etal_2014a,Egusa_etal_2018,Faesi_etal_2018} or on the most massive clouds in the inner regions of a small sample of galaxies \citep{DonovanMeyer_etal_2013}. 

In this paper, we take the logical next step, exploring the surface density, line width, and dynamical state of molecular gas across a significant sample of star-forming disk galaxies. This is made possible by the ongoing PHANGS-ALMA survey (PHANGS-ALMA: Physics at High Angular-resolution in Nearby GalaxieS with ALMA, \citeinprep{A. K. Leroy et al. 2018, in prep.}; ALMA: Atacama Large Millimeter-submillimeter Array). PHANGS-ALMA is mapping the \mbox{CO~(2-1)} emission from a large sample of nearby star-forming galaxies ($74$ in total) with sufficient sensitivity and resolution to detect individual giant molecular clouds (GMCs) across most of the galaxies' star-forming disks. In this paper, we combine the first 11 galaxies from PHANGS-ALMA with four targets (M31, M33, M51, and the Antennae Galaxies) previously observed at high spatial resolution. These $15$ galaxies span dwarf spirals to starburst galaxies, and yield tens of thousands independent measurements at spatial scales of $20-130$~pc, comparable to the characteristic size of a GMC. 

We present more details about our dataset in Section~\ref{sec:data}, and explain the measurements that we perform in Section~\ref{sec:method}. In Section~\ref{sec:expectation} we discuss expectations for the $\sig$-$\Sig$ scaling relation based on simple theoretical arguments. In Section~\ref{sec:result}, we present the empirical scaling relation that best describes the relationship between line width and surface density for our sample of nearby galaxies. Then, in Section~\ref{sec:discussion} we discuss their physical interpretation. We present a summary of our main results in Section~\ref{sec:summary}, along with prospects for future work.

\section{Data}\label{sec:data}

\begin{deluxetable*}{lccccccccccc}
\tabletypesize{\footnotesize}
\tablecaption{Sample of Galaxies\label{tab:galaxies}}
\tablewidth{0pt}
\tablehead{
\colhead{Galaxy} &
\colhead{Morphology} &
\colhead{Distance\tablenotemark{a}} &
\colhead{Inclination\tablenotemark{b}} &
\colhead{Stellar Mass\tablenotemark{c}} &
\colhead{Telescope} &
\colhead{Line} &
\colhead{Resolution} &
\colhead{Channel Width} &
\colhead{Sensitivity} \\
\colhead{} &
\colhead{} &
\colhead{[Mpc]} &
\colhead{[$\arcdeg$]} &
\colhead{[$10^{10}\;$\Msun]} &
\colhead{} &
\colhead{} &
\colhead{[$\arcsec$~/~pc]} &
\colhead{[\kms]} &
\colhead{[K]}
}
\startdata
NGC~628    & Sc-A   & 9.0   & 6.5  & 2.1     & ALMA     & CO(2-1) & 1.0~/~44  & 2.5 & 0.13 \\
NGC~1672   & Sb-B   & 11.9  & 40.0 & 3.0     & ALMA     & CO(2-1) & 1.7~/~98  & 2.5 & 0.09 \\
NGC~2835   & Sc-B   & 10.1  & 56.4 & 0.76    & ALMA     & CO(2-1) & 0.7~/~34  & 2.5 & 0.27 \\
NGC~3351   & Sb-B   & 10.0  & 41.0 & 3.2     & ALMA     & CO(2-1) & 1.3~/~63  & 2.5 & 0.12 \\
NGC~3627   & Sb-AB  & 8.28  & 62.0 & 3.6     & ALMA     & CO(2-1) & 1.3~/~52  & 2.5 & 0.09 \\
NGC~4254   & Sc-A   & 16.8  & 27.0 & 6.5     & ALMA     & CO(2-1) & 1.6~/~130 & 2.5 & 0.06 \\
NGC~4303   & Sbc-AB & 17.6  & 25.0 & 7.4     & ALMA     & CO(2-1) & 1.5~/~128 & 2.5 & 0.10 \\
NGC~4321   & Sbc-AB & 15.2  & 27.0 & 7.9     & ALMA     & CO(2-1) & 1.4~/~103 & 2.5 & 0.09 \\
NGC~4535   & Sc-AB  & 15.8  & 40.0 & 3.9     & ALMA     & CO(2-1) & 1.5~/~115 & 2.5 & 0.08 \\
NGC~5068   & Scd-AB & 9.0   & 26.9 & 1.1     & ALMA     & CO(2-1) & 0.9~/~39  & 2.5 & 0.24 \\
NGC~6744   & Sbc-AB & 11.6  & 40.0 & 8.1     & ALMA     & CO(2-1) & 1.0~/~56  & 2.5 & 0.18 \\
M51        & Sbc-A  & 8.39  & 21.0 & 7.7     & PdBI     & CO(1-0) & 1.2~/~49  & 5.0 & 0.31 \\
M31        & Sb-A   & 0.79  & 77.7 & 16      & CARMA    & CO(1-0) & 5.5~/~21  & 2.5 & 0.19 \\
M33        & Scd-A  & 0.92  & 58.0 & 0.3-0.6 & IRAM-30m & CO(2-1) & 12~/~54   & 2.6 & 0.04 \\
Antennae   & Merger & 22.0  & --   & --      & ALMA     & CO(3-2) & 0.6~/~64  & 5.0 & 0.13 \\
\enddata
\tablenotetext{a}{Distance values are taken from the Extragalactic Distance Database \citep{Tully_etal_2009}.}
\tablenotetext{b}{References for the inclination angle values: {\it NGC~628} - \citet{Fathi_etal_2007}; {\it NGC~1672} - \citet{Diaz_etal_1999}; {\it NGC~2835 \& 5068} - values from the HyperLeda database \citep{Makarov_etal_2014}; {\it NGC~3351} - \citet{Dicaire_etal_2008}; {\it NGC~3627} - \citet{DeBlok_etal_2008}; {\it NGC~4254, 4321, 4535} - \citet{Guhathakurta_etal_1988}; {\it NGC~4303} - \citet{Schinnerer_etal_2002}; {\it NGC~6744} - \citet{Ryder_etal_1999}; {\it M51} - \citet{Colombo_etal_2014b}; {\it M31} - \citet{Corbelli_etal_2010}; {\it M33} - \citeinprep{E. Koch et al. (2018, in prep.)}.}
\tablenotetext{c}{References for the stellar mass values: {\it NGC~2835 \& 6744} - \citeinprep{A. K. Leroy et al. (2018, in prep.)}; {\it M31} - \citet{Barmby_etal_2006}; {\it M33} - \citet{Corbelli_2003}; {\it all other galaxies} - S$^4$G global stellar masses \citep[Spitzer Survey of Stellar Structure in Galaxies;][]{Sheth_etal_2010,Querejeta_etal_2015}.}
\end{deluxetable*}

Our sample consists of 15 nearby galaxies with high resolution maps of low-$J$ carbon monoxide (CO) rotational line emission. Table \ref{tab:galaxies} lists their name, morphology, orientation, adopted distance and stellar mass for each galaxy and the basic parameters of the CO data. Our sample includes 11 targets from the PHANGS-ALMA survey (\citeinprep{A. K. Leroy et al. 2018, in prep.}). These targets were observed by ALMA in CO~(2-1) using the 12-meter and 7-meter interferometric arrays as well as the total-power antennas. Thus, the maps capture information from all spatial scales. The whole PHANGS-ALMA sample is designed to cover the star-forming main sequence of galaxies across the local volume. When finished, it will provide $\sim 1$-$1.5''$ resolution CO~(2-1) maps of $74$ nearby ($d \lesssim 17$~Mpc), ALMA-visible, actively star-forming galaxies down to a stellar mass of $\sim 5\times10^9 M_\odot$. A detailed description of the PHANGS-ALMA sample, observing strategy and data reduction is presented in \citeinprep{A. K. Leroy et al. (2018, in prep.)}.

Observations of the full PHANGS-ALMA sample are currently underway, but the first $11$ targets analyzed here were already observed in 2016 during ALMA's Cycle 3. These targets sparsely sample the star-forming main sequence with an emphasis on higher mass systems.

To supplement these first PHANGS-ALMA targets, we include the PdBI Arcsecond Whirlpool Survey (PAWS) CO~(1-0) map of M51 \citep{Pety_etal_2013, Schinnerer_etal_2013}, which incorporates short-spacing data from the IRAM 30-m telescope. M51 is a massive grand-design spiral on the star-forming main sequence. 

We also analyze the Local Group galaxies M31 and M33. For M31, we use the CARMA CO~(1-0) survey by \citeinprep{A. Schruba et al.\ (in prep.)}, which includes short- and zero-spacing data from the IRAM 30-m telescope \citep{Nieten_etal_2006}. For M33, we use the IRAM 30-m CO~(2-1) survey by \citet{Gratier_etal_2010} and \citet{Druard_etal_2014}. M31 and M33 extend the parameter space probed by our galaxy sample down to low gas surface density regimes (see Section~\ref{sec:scaling:low-end}).

Finally, we include the ALMA CO~(3-2) map of the interacting region of the Antennae galaxies presented by \citet{Whitmore_etal_2014}, and analyzed by \citet{Johnson_etal_2015} and \citet{Leroy_etal_2016}. The physical state of the gas in the Antennae may be strongly affected by the galaxy merger; we include this CO map here as a point of contrast to the normal, undisturbed disk galaxies targeted by PHANGS-ALMA.

This combination of PHANGS-ALMA and literature data gives us (by far) the largest sample of star-forming galaxies with cloud-scale-resolution CO maps, and the prospect to expand this analysis to ${\sim}80$ galaxies in the near future is clearly exciting. The angular resolution of these maps (see Table~\ref{tab:galaxies}) corresponds to linear scales of $20$-$130$~pc. These resolutions are comparable to the typical size of Galactic GMCs \citep{Solomon_etal_1987, Heyer_etal_2009, MivilleDeschenes_etal_2017}. As a result, we expect individual molecular clouds to be at least marginally resolved in these maps.

The channel width for most of these observations is $2.5$-$2.6$~\kms . The two exceptions are the Antennae and M51, which have $5.0$~\kms\ channel width. This velocity resolution should be sufficient to measure the velocity dispersion for larger GMCs, but may bias the measurement to higher values for smaller GMCs. We account for the effect of finite channel width in our analysis and discuss its implications in Section \ref{sec:method:sig}. We note that both M51 and the Antennae are gas-rich, and typically have high line widths \citep[][]{Colombo_etal_2014a, Whitmore_etal_2014}.

In Table~\ref{tab:galaxies}, we also quote the sensitivity (1$\sigma$ channel-wise rms noise) of each data cube in units of K at their native angular resolution before any convolution. For objects in the PHANGS-ALMA survey, the sensitivity of velocity-integrated intensity maps are typically ${\sim}0.5$~\Kkms\ , corresponding to a gas surface density of $3~\Msun~{\rm pc}^{-2}$ for our assumed CO-to-H$_2$ conversion factor and CO line ratios (see Section~\ref{sec:method:Sig}). This surface brightness sensitivity improves as we convolve our data cubes to coarser angular resolutions to achieve uniform linear resolution across our targets.

We provide additional notes on a few galaxies in our sample:
\begin{itemize}
\item {\it NGC~2835}: a low mass star-forming disk galaxy. The CO map has relatively low sensitivity due to the extended interferometer configuration used during the observations. Furthermore, the CO surface brightness is low in accordance with the galaxy's low stellar mass.
\item {\it NGC~3351}: a strongly barred galaxy with a prominent central molecular gas disk \citep[][]{Jogee_etal_2005}, a gas-poor bulge, and a ring of molecular gas at larger galactocentric radius. Star formation takes place both in the central disk and in the molecular gas ring outside the bulge.
\item {\it NGC~5068}: a low mass star-forming disk galaxy. Similar to NGC~2835, the sensitivity of the CO map for this target is lower than for our other targets.
\item {\it NGC~6744}: a weakly barred star-forming spiral galaxy. The PHANGS-ALMA CO map covers the north and south part of the disk, but has no coverage of the (gas-poor) center.
\item {\it M51}: a normal star-forming disk galaxy with grand-design spiral structure. The PAWS CO map covers the central $9 \times 6$ kpc$^2$ region \citep[see][]{Schinnerer_etal_2013}.
\item {\it M31}: this high stellar mass Local Group spiral is a ``green valley'' galaxy \citep[i.e., it is located between the ``blue cloud'' and the ``red sequence'' in a color-magnitude diagram, see e.g., Figure~4 of][for an illustration]{Mutch_etal_2011}, with relatively quiescent star formation \citep[e.g.,][]{Lewis_etal_2015}. The CARMA CO survey covers the north-eastern part of the star-forming ring and a part of the inner disk \citep[\citeinprep{A. Schruba et al. in prep.}; and see visualizations in][]{CalduPrimo_etal_2016, Leroy_etal_2016}. Because of its proximity \citep[$d\sim0.79$\,Mpc,][]{Tully_etal_2009}, the sensitivity of the CO data for this target is much better than for most other galaxies in our sample. The gas reservoir in M31 is dominated by the large {\sc Hi} disk \citep{Braun_etal_2009}, and most of its molecular gas content sits at relatively large galactic radius.
\item {\it M33}: this low stellar mass Local Group dwarf spiral is also {\sc Hi}-dominated \citep[see][]{Druard_etal_2014}. Again, due to its proximity \citep[$d\sim$0.92\,Mpc,][]{Tully_etal_2009}, the sensitivity and spatial resolution of the CO data for M33 are better than for most of our sample.
\item {\it The Antennae Galaxies}: the nearest major merger. The CO map presented in \citet{Whitmore_etal_2014} covers only the interacting (``overlap'') region. This is the only target in our sample that lacks short- and zero-spacing data.

\end{itemize}

\section{Measurements}\label{sec:method}

\subsection{Fixed-Spatial-Scale Measurement Approach}\label{sec:method:fixed-scale}

From the high resolution CO imaging described in Section \ref{sec:data}, we estimate the molecular gas surface density, \Sig , and velocity dispersion, \sig , {\it along each sightline at a range of fixed spatial scales}. This approach has been advocated by \citet{Leroy_etal_2016} based on earlier work by \citet{Ossenkopf_MacLow_2002}, \citet{Sawada_etal_2012}, \citet{Hughes_etal_2013b}, and \citet{Leroy_etal_2013}. Recent work analyzing high resolution and high sensitivity ALMA data has also adopted similar approaches \citep[e.g.,][]{Egusa_etal_2018}.

This approach gives us access to all essential physical properties that we would like to measure (e.g., gas surface density, velocity dispersion, dynamical state, and turbulent energy content). Such ``fixed-spatial-scale'' (or simply ``fixed-scale'') approach is non-parametric, minimal in assumptions, and easy to apply to many data sets in a uniform way. Our measurements are easy to replicate in synthetic observations, and thus offer a straightforward path for direct comparison between observations and simulations. Finally, this apporach characterizes all detected CO emission, and allows us to rigorously treat the selection function.

Our fixed-scale approach differs from the cloud identification approaches commonly used in previous studies (e.g., {\tt Clumpfind}, see \citealt{Williams_etal_1994}; {\tt cprops}, see \citealt{Rosolowsky_Leroy_2006}). These methods segment CO emission by associating CO emission with local maxima. While this is a useful strategy for identifying isolated structures, there are three overlaping drawbacks that makes us prefer the fixed-scale approach. First, the criteria for segmentation are usually not physically motivated. Instead, several recent studies \citep{Pineda_etal_2009,Hughes_etal_2013a,Leroy_etal_2016} have shown that cloud identification methods tend to find beam-sized objects when applied to data sets with moderate resolution and sensitivity. Second, at $45$--$120$~pc resolution, we observe many crowded regions since we target the molecular gas rich parts of star-forming galaxies (e.g., galaxy centers, spiral arms, and bar ends). The commonly used segmentation algorithms are not optimized for identifying marginally resolved clouds in this regime. Third, some segmentation algorithms do not characterize all emissions, which then makes the selection function complex. Based on all these, we believe that the fixed-scale approach is at least as appropriate in this context as cloud identification.

Our fixed-scale approach does not provide information on the spatial extent of the CO emitting structures. For the rest of this study, we adopt the spatial resolution of the data to be the relevant size scale when, e.g., estimating the mean volume density and virial parameter. We expect this to be a reasonable assumption as long as two conditions hold: (1) the beam size is within the scale-free range of the hierarchical molecular ISM \citep[often taken to be either the disk scale height or the turbulent driving scale, and typically about a hundred or a few hundreds of parsec, see e.g.,][]{Brunt_2003} and (2) bright CO emission fills a large fraction of the beam. When condition (2) is violated, the beam-averaged surface density is diluted by the ``dark area'' and we no longer expect the beam size to represent the relevant size scale (i.e., beam dilution, see Section~\ref{sec:expectation}). Nevertheless, this concern is not specific to the fixed-scale approach, but generally applicable to most analysis using data sets with moderate spatial resolution.

The sampling scale is an adjustable variable in the fixed-scale approach. We explore the impact of using different sampling scales by changing the linear resolution of the data (see Section~\ref{sec:method:procedure}), and repeat our measurements at each resolution. This lets us investigate molecular gas properties as a function of averaging scale.

\subsection{Measurement Procedure}\label{sec:method:procedure}

Before performing any measurements, we pre-process all data sets by convolving them to a set of round Gaussian-shaped beams with fixed linear sizes: $45$, $60$, $80$, $100$, and $120$~pc (at FWHM). When the native angular resolution of a data set is coarser than the target beam size, we exclude that galaxy from the analysis at that spatial scale\footnote{In practice, due to the uncertainty in the distances to our targets, we allow a 10\% tolerance so that a target with its native resolution between $72$ and $88$ pc will be labeled as $80$ pc.}. Then, at each resolution, we resample the data sets onto regular square grids so that the pixels Nyquist-sample the new beam, resulting in an (areal) over-sampling factor of $\pi/\ln{2} \approx 4.53$.

In each data set and at each resolution, we identify all sightlines\footnote{We use the word ``sightline'' to denote the Nyquist-sampled pixels throughout this paper.} with significant CO emission. To do this, we first identify all 3D regions in the data cube where $\geq 2$ consecutive channels show emission with signal-to-noise ratio ${\rm (S/N)} \geq 5$. Then, we expand this mask in both spatial and spectral directions to include all adjacent pixels that contain CO emission in $\geq 2$ consecutive channels with ${\rm S/N} \geq 2$. This signal identification scheme is similar to those adopted in other works \citep[e.g., {\tt cprops}, see][]{Rosolowsky_Leroy_2006}.

Note that for most of our targets, we detect CO emission at relatively high S/N and much of the emission is spatially connected. In these cases, our sensitivity limit mostly reduces to the 2-consecutive-2$\sigma$-channels criterion. However, our two low stellar mass PHANGS-ALMA galaxies, NGC~2835 and NGC~5068, have lower overall S/N. For these targets, the 2-consecutive-5$\sigma$-channel criterion becomes more relevant. We project the selection effects due to these limits into the key parameter space that we study (Section \ref{sec:scaling}).

Faint line wings extending to velocities far from the centroid can have an important effect on the line width. With this in mind, we expand the mask along the velocity axis to cover the entire probable velocity range of the CO line. The amount by which we grow the mask depends on the ratio between the intensity at the line peak and at the edge of the mask, and hence varies between sightlines. Based on the peak-to-edge ratio, we expand the mask so that it would extend to $\pm 3\sigma$ if the line profile has a Gaussian shape with our measured peak intensity centered at the peak velocity.

For each sightline in the mask, we calculate the line-integrated intensity $\Ico$ and peak intensity $\Tpeak$ of the emission within the expanded mask. We further convert these CO line measurements into the molecular gas surface density \Sig\ and velocity dispersion \sig\ following the procedures described in the next two sections (Section~\ref{sec:method:Sig}~and~\ref{sec:method:sig}).

For the line-integrated intensity, peak intensity, and all other measurements, we estimate their statistical errors by performing a Monte Carlo simulation using the data cube itself as model. We generate $1,000$ realizations of the data cube by artificially adding random noise to the original cube, and repeat all our measurements for each realization. Note that the mask is only generated once for each galaxy using the original cube, and then it is applied to all the $1,000$ mock cubes. We record the rms scatter of the repeated measurements and quote these numbers as the statistical uncertainties. 

\subsection{Surface Density}\label{sec:method:Sig}

We estimate a surface density, \Sig , from the line-integrated CO intensity, \Ico , along each detected sightline. Throughout the paper, we assume that the low-$J$ CO lines trace mass in a simple way, so that we can estimate \Sig\ from \Ico\ and an adopted CO-to-H$_2$ conversion factor \citep[][]{Bolatto_etal_2013}. We adopt the following conversion factors $\alphaco \equiv \Sig/\Ico$:

\begin{eqnarray}
\alpha_{\rm CO(1-0)} &= 4.35\;\Msun\,{\rm pc}^{-2}\,(\Kkms)^{-1}, \label{eq:co10}\\
\alpha_{\rm CO(2-1)} &= 6.25\;\Msun\,{\rm pc}^{-2}\,(\Kkms)^{-1}, \label{eq:co21}\\
\alpha_{\rm CO(3-2)} &= 17.4\;\Msun\,{\rm pc}^{-2}\,(\Kkms)^{-1}. \label{eq:co32}
\end{eqnarray}

\noindent These correspond to the Galactic value for \mbox{CO\,(1-0)} recommended by \citet{Bolatto_etal_2013} along with transition line ratios of \mbox{CO\,(2-1)}$/$\mbox{CO\,(1-0)}$=$0.7 \citep{Leroy_etal_2013,Saintonge_etal_2017} and \mbox{CO\,(3-2)}$/$\mbox{CO\,(1-0)}$=$0.25 for the Antennae \citep{Ueda_etal_2012, Bigiel_etal_2015}. These take the contribution from helium and other heavy elements into account.

A single conversion factor has the merit of showing our measurements directly, but we have good reason to believe that $\alpha_{\rm CO}$ varies across our sample \citep{Blanc_etal_2013,Sandstrom_etal_2013}. Despite a general understanding of the likely variations in $\alpha_{\rm CO}$ \citep[][]{Bolatto_etal_2013}, we still lack a quantitative, observationally verified prescription for $\alpha_{\rm CO}$ that applies at cloud scales. Moreover, cloud scale models of the CO-to-H$_2$ conversion factor often identify the line width of a cloud and/or its density as an important quantity \citep[][]{Maloney_Black_1988, Narayanan_etal_2012}. Thus our measurements may have a complicated, non-linear interaction with $\alpha_{\rm CO}$. We adopt a single $\alpha_{\rm CO}$ and then discuss possible variations when they become relevant. A more general exploration of $\alpha_{\rm CO}$ for the CO~(2-1) line is a broad goal of the PHANGS-ALMA science program.

As most of the galaxies in our sample have relatively high metallicity and PHANGS-ALMA targets their molecular-gas-rich inner regions, we expect most of the molecular gas to produce bright CO emission. That is, ``CO-dark'' molecular gas should not represent a dominant fraction of the molecular gas mass, and the CO line width should be representative of the true H$_2$ velocity dispersion. We do include some low mass spiral galaxies in our analysis (M33, NGC~2835, NGC~5068) and here the contributions of a CO-dark phase may be more significant \citep[though $\alpha_{\rm CO}$ likely varies by less than a factor of 2, see][]{Gratier_etal_2017}.

We assume that our spatial resolution is sufficient to reach the scale of individual GMCs or small collections of GMCs, and we take such structures to be spherically symmetric. Therefore, we do not apply any correction to \Sig\ to account for beam filling factors or the effect of galaxy disk inclination.

\subsection{Line Width}\label{sec:method:sig}

We use the width of the CO emission line to trace the velocity dispersion, \sig , of molecular gas along each sightline. For the most part, we expect that the CO line width is driven by turbulent broadening, and thus it directly traces the turbulent velocity dispersion at the spatial scale set by the data resolution.

Several methods exist to estimate the width of an emission line. Common approaches include fitting the line profiles with a Gaussian function or Hermite polynomials, calculating the second moment (i.e., the rms dispersion of the spectrum), or measuring an ``effective width'' (see below). These methods have varying levels of robustness against noise and make different assumptions about the shape of the line.

For our main results, we use the ``effective width''\footnote{\citet{Heyer_etal_2001} and following works refer to this quantity as ``equivalent width''. We notice that this name has different meaning in other contexts (e.g., it is also defined as the ratio between the total flux of an emission/absorption line and the underlying continuum flux density). We instead adopt the name ``effective width'' here to avoid confusions in terminology.} as a proxy for the line width. Following \citet{Heyer_etal_2001}, we define the effective width as

\begin{equation} \label{eq:ew}
\sig_{\rm measured} = \frac{\Ico}{\sqrt{2\pi}\,\Tpeak}~,
\end{equation}

\noindent where \Tpeak\ is the specific intensity at the line peak (in K). This proxy is less sensitive to noise in line wings than the second moment and, unlike direct profile fitting, it does not {\it a priori} assume any particular line shape. However, the effective width can be sensitive to the channel width.

To correct for the broadening caused by finite channel width and spectral response curve width, we subtract the effective width of the spectral response from the measured effective width \citep{Rosolowsky_Leroy_2006}:

\begin{equation} \label{eq:resolution}
\sig\;=\;\sqrt[]{\sig^2_{\rm measured} - \sig^2_{\rm response}},
\end{equation}

\noindent where $\sig_{\rm response}$ is estimated from the channel width and the channel-to-channel correlation coefficient, following \citet{Leroy_etal_2016}. Appendix \ref{apdx:response} presents the detailed procedures for estimating $\sig_{\rm response}$, as well as discussions on the applicability of this ``deconvolution-in-quadrature'' approach. This broadening correction should be accurate enough in most cases, except when the measured line width becomes close to the channel width. In the following sections, we indicate this regime in our plots and discuss resolution effects, if relevant, when presenting our results.

One important caveat related to the line width near the centers of galaxies is that we do not correct for unresolved rotational motions or contributions from multiple clouds along the line of sight. This ``beam smearing'' effect may be important where the rotation curve rises quickly, even at our high resolution. The ongoing efforts of measuring rotation curves from PHANGS-ALMA data (P. Lang et al. in preparation) will allow a more careful treatment of this effect in the future.

\subsection{Completeness}\label{sec:method:completeness}

\begin{deluxetable*}{lccccccccc}
\tabletypesize{\footnotesize}
\tablecaption{CO Detection Statistics\label{tab:stats}}
\tablewidth{0pt}
\tablehead{
\colhead{Galaxy} & 
\multicolumn{3}{c}{at 45 pc resolution} &
\multicolumn{3}{c}{at 80 pc resolution} &
\multicolumn{3}{c}{at 120 pc resolution} \\
\cmidrule(lr){2-4} \cmidrule(lr){5-7} \cmidrule(lr){8-10}
\colhead{} &
\colhead{$N_{\rm sightlines}$} &
\colhead{$M_{\rm mol}$} &
\colhead{$f_M$} &
\colhead{$N_{\rm sightlines}$} &
\colhead{$M_{\rm mol}$} &
\colhead{$f_M$} &
\colhead{$N_{\rm sightlines}$} &
\colhead{$M_{\rm mol}$} &
\colhead{$f_M$} \\
\colhead{} &
\colhead{} &
\colhead{[$10^8$\,\Msun]} &
\colhead{} &
\colhead{} &
\colhead{[$10^8$\,\Msun]} &
\colhead{} &
\colhead{} &
\colhead{[$10^8$\,\Msun]} &
\colhead{} 
}
\startdata
NGC~2835 & 408 & 0.45 & 26\% & 308 & 0.59 & 33\% & 183 & 0.58 & 32\% \\
NGC~5068 & 914 & 1.1 & 32\% & 637 & 1.4 & 40\% & 370 & 1.4 & 40\% \\
NGC~628 & 6,156 & 6.7 & 45\% & 4,307 & 8.9 & 66\% & 2,711 & 10 & 74\% \\
NGC~1672 & -- & -- & -- & -- & -- & -- & 1,239 & 21 & 78\% \\
NGC~3351 & -- & -- & -- & 1,180 & 5.7 & 63\% & 990 & 6.4 & 71\% \\
NGC~3627 & -- & -- & -- & 4,030 & 28 & 84\% & 2,185 & 28 & 86\% \\
NGC~4535 & -- & -- & -- & -- & -- & -- & 2,031 & 18 & 63\% \\
NGC~4254 & -- & -- & -- & -- & -- & -- & 5,485 & 62 & 79\% \\
NGC~4303 & -- & -- & -- & -- & -- & -- & 3,522 & 44 & 74\% \\
NGC~4321 & -- & -- & -- & -- & -- & -- & 4,432 & 47 & 68\% \\
M51 & 6,087 & 28 & 80\% & 2,656 & 29 & 87\% & 1,484 & 31 & 91\% \\
NGC~6744 & -- & -- & -- & 3,353 & 8.7 & 52\% & 2,351 & 10 & 61\% \\
M31 & 2,266 & 0.68 & 66\% & 1,384 & 0.88 & 86\% & 780 & 0.96 & 94\% \\
M33 & -- & -- & -- & 1,798 & 1.0 & 63\% & 1,147 & 1.1 & 70\% \\
Antennae & -- & -- & -- & 1,097 & 62 & 111\%\tablenotemark{$\dagger$} & 603 & 62 & 111\%\tablenotemark{$\dagger$} \\
\enddata
\tablecomments{For each galaxy at each resolution level, we report:
(1) $N_\mathrm{sightlines}$ - number of {\it independent} sightlines with confident CO detection (i.e., the number of all CO-detected sightlines divided by the areal over-sampling factor $4.53$);
(2) $M_\mathrm{mol}$ - total recovered molecular gas mass (in units of $10^8\,\Msun$); and
(3) $f_M$ - fraction of the recovered gas mass comparing to the total gas mass inside the same field of view.\\
Galaxies are ordered following such scheme: the 12 galaxies in our ``main sample'' (PHANGS-ALMA targets plus M51) appear first, among which the ordering is determined by increasing total stellar mass (see Table \ref{tab:galaxies}); then the Local Group objects and the Antennae galaxies follow.}
\tablenotetext{\dagger}{The derived fraction of recovered gas mass exceeds 100\% in the Antennae galaxies because the data cubes lack short-spacing information and thus exhibit ``clean bowls''. These features show up as large regions with unphysical negative signals adjacent to bright emission structures. Summing over these negative signals reduces the estimated total gas mass \citep[see also][]{Leroy_etal_2016}.}
\end{deluxetable*}

Table~\ref{tab:stats} reports the number of independent measurements and total emission within the mask, expressed as molecular gas mass, in each target at $45$, $80$, and $120$~pc resolution. In massive disk galaxies, thousands of independent sightlines show CO emission above our sensitivity limit. This number drops to hundreds in our lower mass targets. For our adopted CO-to-H$_2$ conversion factor, the molecular mass implied by the CO flux along each sightline is $\sim 10^4-10^6\;\Msun$, and the implied surface density is $\sim 10-10^3\;\Msun~{\rm pc}^{-2}$. 

In total, our detected sightlines include CO emission equivalent to $10^7-10^9\;\Msun$ of molecular gas per galaxy. Some emission remains outside the mask, however, as it is too faint to be detected at good significance using our CO emission identification method. We estimate the fraction of the CO flux included in our analysis, $f_M$, by dividing the total flux inside the mask by the sum of the unmasked cube. Because most of our data cubes (all but the Antennae galaxies) incorporate total power data, and most targets have strong enough CO emission, we expect a direct sum of the cube to yield a robust estimate of the total flux.

We report $f_M$ for each data set at each resolution in Table~\ref{tab:stats}. At $80$~pc resolution, our completeness is $50$-$100\%$ in most galaxies, and slightly less than this in the molecule-poor low mass galaxies NGC~2835 and NGC~5068. The completeness improves with increasing beam size, reflecting improved surface brightness sensitivity at coarser resolution. It also varies from source to source, depending on the typical brightness of CO in the galaxy (correlating with molecular gas surface density) and the distance to the galaxy (relating to the surface brightness sensitivity at fixed resolution).

\section{Expectations}\label{sec:expectation}

\subsection{Expectations about Cloud Properties}

In early studies of the Milky Way cloud population \citep[e.g.,][]{Solomon_etal_1987}, GMCs are usually assumed to be long-lived, roughly virialized structures. Departures from virial equilibrium can be expressed via the virial parameter, $\alphavir \equiv 2K/U_g$, where $K$ and $U_g$ denote the kinetic energy and self-gravitational potential energy. Virialized clouds without surface pressure or magnetic support have $\alphavir = 1$, while marginally bound clouds have $\alphavir \approx 2$, as do molecular clouds in free-fall collapse \citep[e.g.,][]{BallesterosParedes_etal_2011,IbanezMejia_etal_2016, Camacho_etal_2016}.

For nearly spherical clouds, \alphavir\ can be expressed as \citep{Bertoldi_McKee_1992}\footnote{Note that our definition of \alphavir\ is different from the original one in \citet{Bertoldi_McKee_1992}. Here we add the geometrical factor $f$ in the denominator so that \alphavir\ is simply twice the ratio of the kinetic and potential energy.}

\begin{equation} \label{eq:alphavir}
\alphavir
\equiv \frac{2K}{U_g}
= \frac{5\,\sig^2R}{fGM}~.
\end{equation}

\noindent Here $M$, $R$ and \sig\ refer to the cloud mass, radius, and the one-dimensional velocity dispersion. $f$ is a geometrical factor that quantifies the density structure inside the cloud. For spherical clouds with a radial density profile of $\rho(r) \propto r^{-\gamma}$, $f=(1-\gamma/3)/(1-2\gamma/5)$ \citep{Bertoldi_McKee_1992}.

Equation \ref{eq:alphavir} implies a relationship between the line width \sig, size $R$, virial parameter \alphavir\ and surface density \Sig\ of a cloud:

\begin{equation} \label{eq:scaling-virial}
\begin{split}
\sig &= \left( \frac{f\,\alphavir G}{5} \right)^{0.5} R^{-0.5} M^{0.5} \\
     &= \left( \frac{f\,\alphavir G\,\pi}{5} \right)^{0.5} R^{0.5} \Sig^{0.5}.
\end{split}
\end{equation}

\noindent Here \Sig\ is the surface density averaged over the projected area of the cloud on the sky. Equation~\ref{eq:scaling-virial} can be restated as $\sig^2 / R \propto \alphavir\,\Sig$, so that for a fiducial size--line width relation $\sig = v_0\,R^{0.5}$, the coefficient $v_0$ depends on the cloud surface density and virial parameter \citep[][]{Solomon_etal_1987, Heyer_etal_2009}. Following \citet{Heyer_etal_2009} and the recent extragalactic work discussed in Section~\ref{sec:intro}, the surface densities of molecular clouds are observed to vary in different galactic environments and Equation~\ref{eq:scaling-virial} has become a key diagnostic for the dynamical state of gas in galaxies.

The derivation above represents a highly idealized view. In reality, the molecular ISM has complex structure and we do not expect spherically symmetric clouds with simple density profiles. Nevertheless, if cloud substructure -- here parameterized by $f$ -- does not vary significantly, we can assume a constant value of $f$ and obtain meaningful relative measurements of the dynamical state of the molecular gas, even if the absolute value of \alphavir\ remains uncertain. This comparative approach has often been used in the extragalactic literature \citep[e.g.,][]{Rosolowsky_Leroy_2006, Bolatto_etal_2008}, and this is the view we adopt in this work.

When contributions to the gravitational potential other than self-gravity become significant, simply comparing $K$ and $U_g$ does not provide a full description of a cloud's dynamical state. However, insight can still be gained by examining the deviation of cloud line widths and comparing the observed line widths to the expectation for an isolated, self-gravitating cloud. 

In particular, the role of external pressure (\Pext) on cloud line widths has been emphasized in recent studies of Galactic and extragalactic clouds \citep[e.g.,][]{Heyer_etal_2001, Field_etal_2011, Schruba_etal_2018}. For a fixed dynamical state (e.g., virialized, marginally bound) and density profile, we expect a cloud subjected to a high surface pressure to show a larger line width \sig\ than the same cloud without surface pressure. At fixed \alphavir , $R$, and \Pext , the detailed shape of the $\sig$-$\Sig$ relation depends on the sub-cloud density profile \citep[e.g.,][]{Field_etal_2011, Meidt_2016}. We expect $\sig$ to be nearly flat as a function of $\Sig$ near the surface density where $\Pext \sim 0.5 \pi G \Sig^2$, that is, where the confinement due to self-gravity and external pressure are comparable in strength. At higher surface densities, the external pressure plays only a modest role, and we expect Equation~\ref{eq:scaling-virial} to still hold, perhaps with a slightly shallower slope.

At low cloud surface densities ($\Pext > 0.5 \pi G \Sig^2$), the external pressure exceeds the cloud's self-gravitational pressure. If we assume such a cloud to be in pressure equilibrium, then its internal kinetic energy density (or equivalently, internal turbulent pressure \Pturb) scales with the external pressure \Pext\ \citep{Hughes_etal_2013a}. Therefore, we expect the $\sig$-$\Sig$ relation to asymptote to an isobaric relation with $\Pturb \approx \rho \sig^2 \approx \Pext$. For gas clouds with a line of sight depth $\sim 2R$, the expected scaling is:

\begin{equation}\label{eq:scaling-pe}
\begin{split}
\sig &\approx \Pext^{0.5} \rho^{-0.5} \\
     &\approx (2\Pext)^{0.5} R^{0.5} \Sig^{-0.5}.
\end{split}
\end{equation}

\noindent In this case, $\sig$ and $\Sig$ are inversely correlated because for the same kinetic energy density (thus gas pressure), denser gas should have lower velocity dispersion than less dense gas. 

Though we motivate Equation~\ref{eq:scaling-pe} by considering clouds confined by external pressure, this isobaric relation should be a general limit. Whenever the ambient pressure in a medium significantly exceeds self-gravity, we can expect our ``cloud'' to move towards pressure equilibrium with the surrounding gas. In a realistic cold ISM, this should be the case when the ambient pressure in the disk becomes high relative to the self-gravity of the cold molecular clouds. Then we expect the clouds to follow an isobaric relation \citep[defined by the local ambient pressure as a ``pressure floor'';][]{Keto_Myers_1986,Elmegreen_1989,Field_etal_2011,Schruba_etal_2018}, specifying the relationship between the cloud surface density and velocity dispersion. Moreover, the ambient pressure is expected to vary as a function of location in the galaxy in response to the distribution of gas and the potential of the galaxy \citep[e.g.,][]{Elmegreen_1989, Wolfire_etal_2003, Ostriker_etal_2010,Herrera-Camus_etal_2017,Meidt_etal_2018a}. Given this, the low surface density, ``pressure-dominated'' limit should not be a single $\sig$-$\Sig$ relation with $-0.5$ slope across the entire galaxy, but rather a group of curves, each defined by the local ambient pressure value.

\subsection{Additional Expectations Under the Fixed-Scale Analysis~Framework}

We measure $\Sig$ and $\sig$ from data cubes convolved to a common spatial resolution corresponding to the size of a typical Galactic GMC ($2R = 45$-$120$~pc). We view these measurements as characterizing the molecular ISM at a scale comparable to these observing beam sizes, \rbeam , and thus do not engage in any further structure-finding. Following this logic, we will mostly discuss our results equating each individual beam to a molecular cloud. For a fixed $R = \rbeam$, the expected $\sig$-$\Sig$ correlation from Equation \ref{eq:scaling-virial} should then be:

\begin{equation}\label{eq:scaling0.5}
\sig \propto \alphavir^{0.5}\,\Sig^{0.5}
\end{equation}

\noindent with $R = \rbeam$ now part of the coefficient.

There are a few caveats that should be kept in mind when interpreting our measurement: the unknown line of sight depth of the emission, the possible coincidence of physically unassociated structures along the line of sight, and the effect of any mismatch between the beam size and the size of physical structures. Several of these concerns are not exclusively associated with the fixed-scale approach, but relevant to most studies using position-position-velocity data cubes with moderate resolution.

{\em Line of sight depth:} For our fixed-scale measurements, the sampled size scale on the sky is known by construction -- it is the spatial scale to which we convolve the data. However, we do not have an independent constraint on the line of sight depth $l$. In this paper, we take $l \sim 2\rbeam$; that is, we assume that the spatial scale sampled along the line of sight is comparable to the scale that we study on the sky. 

Because the size $R$ in Equation~\ref{eq:scaling-virial} represents the geometric mean of the size in each of three dimensions, $R \sim (\rbeam^2 l)^{1/3}$, we have $\sig \propto l^{1/6} \Sig^{0.5}$, which depends weakly on $l$. Variations in $l$ would add noise or a mild systematic to the slope of our measured $\sig$-$\Sig$ relation. However, the effect is expected to be small: a factor of $2$ variation in $l$ across a decade in $\Sig$ would imply a change in slope of $\sim 0.05$.

{\em Capturing unassociated structures:} Line of sight depth variations alone have a mild impact, but in the case that the beam samples an ensemble of unassociated objects aligned along the line of sight, we expect larger effects. In this case, we would measure a higher surface density, since the emission from multiple structures is added together. If the line width responds to a larger scale potential well, then we would expect the measured line width to be much larger as well. For a sightline through a heavily populated extended gas disk, we would at a minimum capture any ``inter-cloud'' velocity dispersion \citep[e.g.,][]{Solomon_deZafra_1975,Stark_1984,Wilson_etal_2011, CalduPrimo_etal_2016}, reflecting either the broader turbulent cascade or the motion of clouds in the larger galactic potential. 

In regions with strong velocity gradients, we further expect increased velocity dispersion due to ``beam smearing'' or velocity fields unresolved by the beam \citep[e.g.,][]{Colombo_etal_2014b,Meidt_etal_2018a}. Such a situation could lead to complex line profiles showing multiple components. We expect this situation to arise most often in the dense inner parts of galaxies, and likely in bars and spiral arms as well, where non-circular motions and shocks are strong and the chance of capturing multiple unassociated structures along one sightline is highest. This situation is also more likely to occur for observations of highly inclined galaxies.

{\em Contamination of bright PSF wings:} A related concern arises when the beam samples the edge of a large cloud or the extended ``wings'' of a beam dominated by a nearby bright object. In both cases, we might expect the measured \sig\ to remain larger, still indicative of the gravitational potential of the whole cloud or the \sig\ value found in the bright, nearby source. However, we do not expect to see molecular clouds with sizes vastly larger than our beam size ($45$-$120$~pc), so the main sense of this bias in our data will be related to the wings of the PSF. We might expect to find a mild inflation in \sig\ along faint sightlines near isolated bright sightlines. Given the molecular gas rich environment for most of our targets, and the observed tight correlation between \Sig\ and \sig , we do not expect this bias to play a major role.

{\em Individual unresolved clouds:} At the other extreme, we can imagine isolated gas structures on scales much smaller than the beam size ($R \ll \rbeam$). In the case of a single small cloud within the beam, the beam-averaged surface density no longer traces the cloud's surface density. However, the line width and total gas mass, $M = \Sig A_{\rm beam}$, are still faithful measurements of the cloud's properties. Thus, from the first half of Equation~\ref{eq:scaling-virial} we can derive the expected relation for individual unresolved, virialized clouds:

\begin{equation} \label{eq:scalingless0.5}
\sig \propto R^{-0.5} M^{0.5}
       \propto R^{-0.5} \Sig^{0.5}.
\end{equation}

\noindent Here $R$ is the radius of the cloud, not the beam. We expect $R$ to be positively correlated with $M$ and thus $\Sig = M/A_{\rm beam}$. We therefore expect the slope of the $\sig$-$\Sig$ relation to be shallower than $0.5$ when $R < \rbeam$. Moreover, when inferring \alphavir\ from Equation~\ref{eq:alphavir}, if we are still substituting $R$ with \rbeam\ in this case, we will overestimate \alphavir\ by $\sim \rbeam/R$. This situation is most relevant in low gas density regions where molecular clouds are small and sparse.

{\em Synthesis:} We do not expect significant impact on the measured scaling relation due to variations in the line of sight depth. It is also unlikely that the contamination from bright sightlines through PSF wings is significant in our sample. We do expect beam smearing to be important in high density regions and unresolved structures to be prevalent in low density regions. Both \Sig\ and \sig\ could be overestimated in the former case, and \Sig\ could be underestimated in the latter case.

\section{Results}\label{sec:result}

\begin{deluxetable*}{lcccccccc}
\colnumbers
\tabletypesize{\footnotesize}
\tablecaption{Cloud-Scale Molecular Gas Measurements for All 15 Galaxies\label{tab:data}}
\tablewidth{0pt}
\tablehead{\colhead{Name} & 
\colhead{Resolution} & 
\colhead{$\Tpeak$} & 
\colhead{$\Sig$} & 
\colhead{$\sig$} & 
\colhead{$\alphavir$} & 
\colhead{$\Pturb/k_\mathrm{B}$} & 
\colhead{Center} & 
\colhead{Complete} \\
\colhead{} &
\colhead{[pc]} &
\colhead{[K]} &
\colhead{[\Msun~pc$^{-2}$]} &
\colhead{[\kms]} &
\colhead{} &
\colhead{[K~cm$^{-3}$]} &
\colhead{} &
\colhead{}
}
\startdata
NGC0628 & 45 & 4.37E-01 & 1.67E+01 & 2.25E+00 & 3.1E+00 & 9.2E+03 & True & False \\
NGC0628 & 45 & 5.28E-01 & 2.05E+01 & 2.25E+00 & 2.5E+00 & 1.1E+04 & True & False \\
NGC0628 & 45 & 7.31E-01 & 3.37E+01 & 2.75E+00 & 2.3E+00 & 2.8E+04 & True & False \\
NGC0628 & 45 & 4.28E-01 & 2.35E+01 & 3.34E+00 & 4.9E+00 & 2.9E+04 & True & False \\
NGC0628 & 45 & 5.27E-01 & 2.94E+01 & 3.40E+00 & 4.0E+00 & 3.7E+04 & True & False \\
NGC0628 & 45 & 4.26E-01 & 2.35E+01 & 3.36E+00 & 4.9E+00 & 2.9E+04 & True & False \\
NGC0628 & 45 & 7.04E-01 & 3.68E+01 & 3.16E+00 & 2.8E+00 & 4.0E+04 & True & False \\
NGC0628 & 45 & 5.83E-01 & 3.02E+01 & 3.13E+00 & 3.3E+00 & 3.2E+04 & True & False \\
NGC0628 & 45 & 5.60E-01 & 3.24E+01 & 3.54E+00 & 4.0E+00 & 4.4E+04 & True & False \\
NGC0628 & 45 & 7.50E-01 & 3.95E+01 & 3.19E+00 & 2.6E+00 & 4.4E+04 & True & True \\
... & ... & ... & ... & ... & ... & ... & ... & ... \\
\enddata
\tablecomments{Fixed spatial scale measurements for all 15 targets at 45, 80 and 120~pc resolutions, with each row corresponding to one (Nyquist-sampled) sightline. For each sightline we report: 
(1) host galaxy name;
(2) spatial resolution of the measurement (beam full width at half maximum);
(3) brightness temperature at the CO line peak (also see Appendix~\ref{apdx:sig-tpeak});
(4) molecular gas surface density;
(5) molecular gas velocity dispersion;
(6) inferred virial parameter (see Section~\ref{sec:alphavir});
(7) inferred internal gas turbulence pressure (see Section~\ref{sec:Pturb});
(8) if the sightline is located in the central region of the host galaxy (see Section~\ref{sec:dist:region}); and
(9) if the CO detection is above the completeness threshold (see Section~\ref{sec:scaling:disk}).
\\
Only a portion of this table is shown here to demonstrate its form and content. A machine-readable version of the full table is available.}
\end{deluxetable*}

In Appendix \ref{apdx:maps}, we show our surface density and velocity dispersion maps for all 15 galaxies at 120~pc resolution (Figure \ref{fig:maps}). Table~\ref{tab:data} presents these measurements in tabular form for all targets at three resolutions ($45$, $80$, $120$~pc). We report values for all sightlines with significant CO detections. At $120$~pc ($\rbeam=60$~pc), this sample corresponds to nearly $30,000$ independent beams across our sample. This is by far the largest set of measured surface densities and line widths at the scale of individual GMCs. As Table~\ref{tab:stats} shows, these sightlines capture most of the flux in most of our galaxies. The masses and surface densities that we derive for most sightlines agree well with those found for Galactic GMCs \citep[][and references therein]{Heyer_Dame_2015}.

\subsection{Distributions of Mass by Surface Density and Velocity Dispersion}\label{sec:dist}

\begin{figure*}[p]
\fig{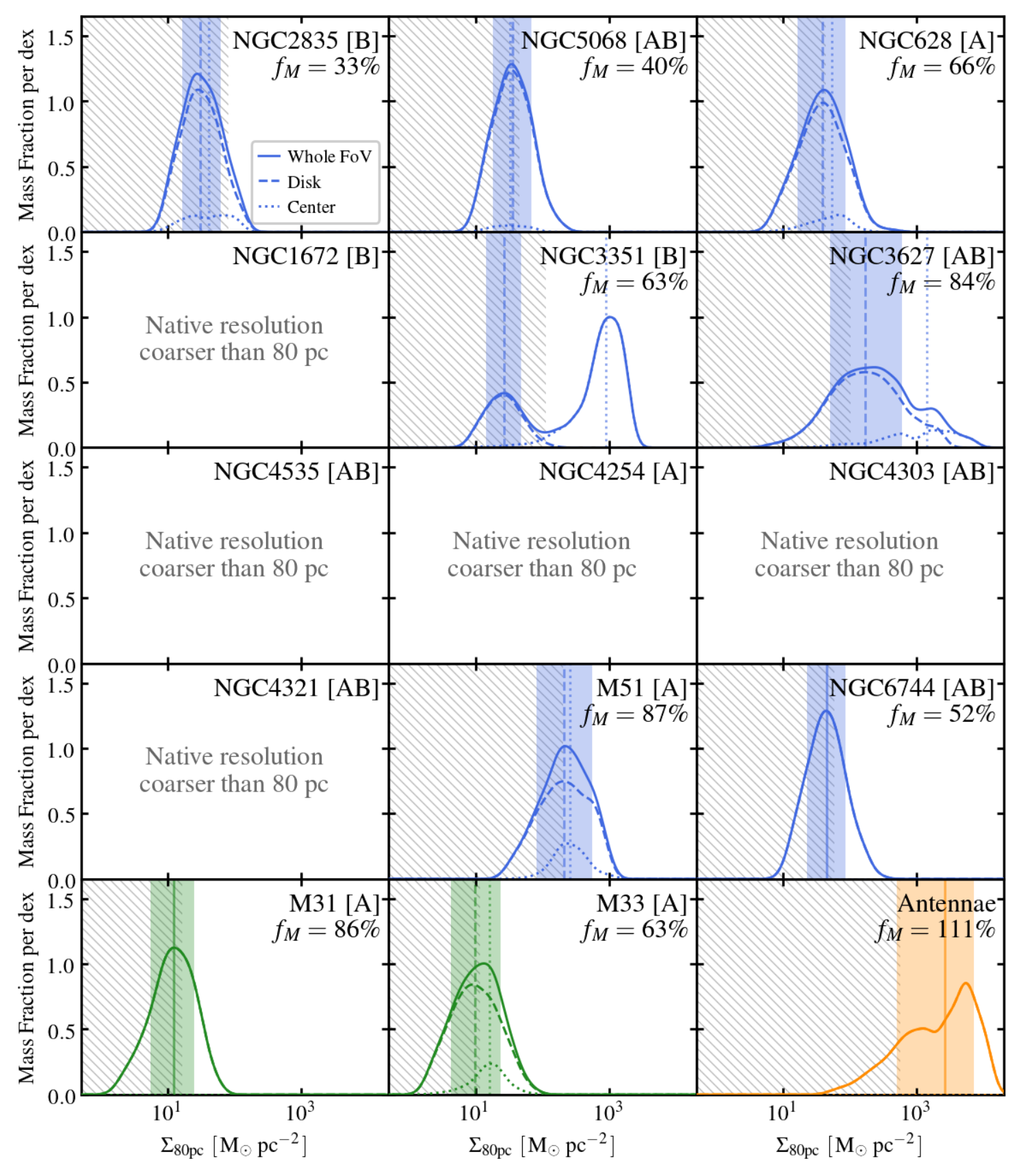}{0.85\textwidth}{}
\vspace{-2em}
\caption{Distribution of molecular gas mass as a function of surface density, \Sig , measured at 80 pc resolution. Each panel shows results for one galaxy, with the target name, bar type (in the brackets), and CO flux recovery fraction $f_M$ indicated in the top right corner. We order the targets from left to right, then top to bottom, following the scheme in Table \ref{tab:stats}. Galaxies in the ``main sample'' (PHANGS-ALMA targets plus M51) are represented by blue color, the Local Group targets by green color, and the Antennae by orange color (this color scheme is used consistently throughout this paper). All curves show Gaussian kernel density estimators (KDE) generated from the data with bandwidth of $0.1$~dex in logarithmic space. The solid curve shows the distribution for all sightlines with CO detections across the whole field of view (FoV). The dashed/dotted curves show the distribution of mass for galaxy disk/central regions (usually defined as outside/inside the $r_{\rm gal} = 1$~kpc boundary), respectively. The vertical dashed line and color shaded region show the mass weighted median value and $16$-$84\%$ range of \Sig\ for the ``disk'' population, while the vertical dotted line shows the median \Sig\ for the ``center''. The hatched region has less than 100\% completeness due to limited sensitivity of the data. Individual galaxy disks typically have most of their molecular gas spread over a $0.5$-$1.0$~dex range of \Sig , and both the median of this distribution and its width vary from galaxy to galaxy. Our strongly barred targets show significantly different distributions for the disk and central regions, see NGC~3351, NGC~3627, and more examples in Figure \ref{fig:Sig-120pc}.}
\label{fig:Sig-80pc}
\end{figure*}

\begin{figure*}[p]
\fig{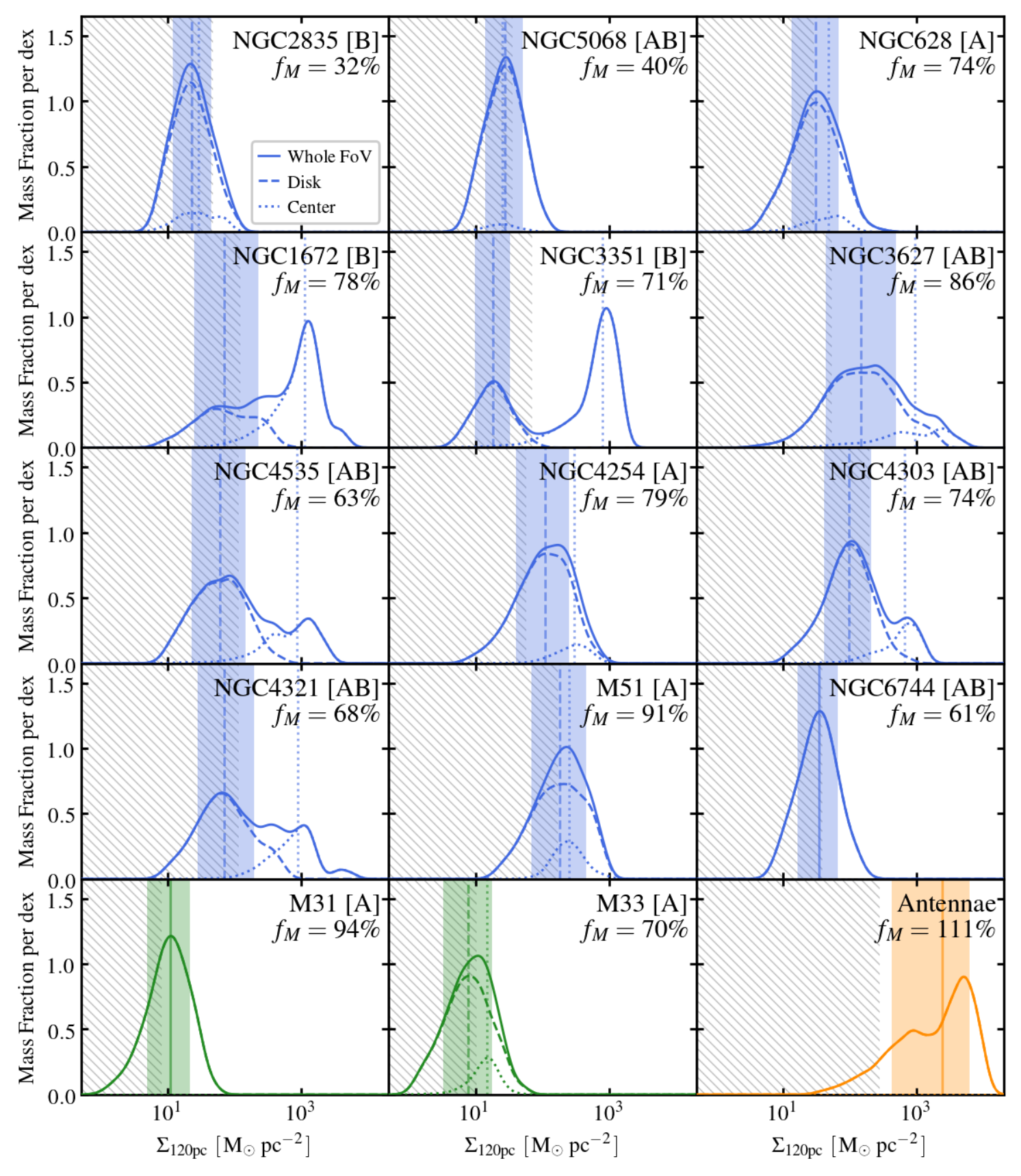}{0.85\textwidth}{}
\vspace{-2em}
\caption{As in Figure~\ref{fig:Sig-80pc}, but here showing the \Sig\ distribution at $120$~pc resolution for all 15 targets. The general sense of galaxy-by-galaxy variations is more clearly revealed in this figure: higher mass star-forming galaxies tend to keep more gas at high \Sig\ (note that the CO map of NGC~6744 does not cover the central region, which might be the reason of this target being an outlier from the general trend). Note that all strongly barred galaxies (NGC~1672, NGC~3351, NGC~3627, NGC~4535, NGC~4303, and NGC~4321) demonstrate significant disk/center dichotomies in their \Sig\ distribution.}
\label{fig:Sig-120pc}
\end{figure*}

\begin{figure*}[p]
\fig{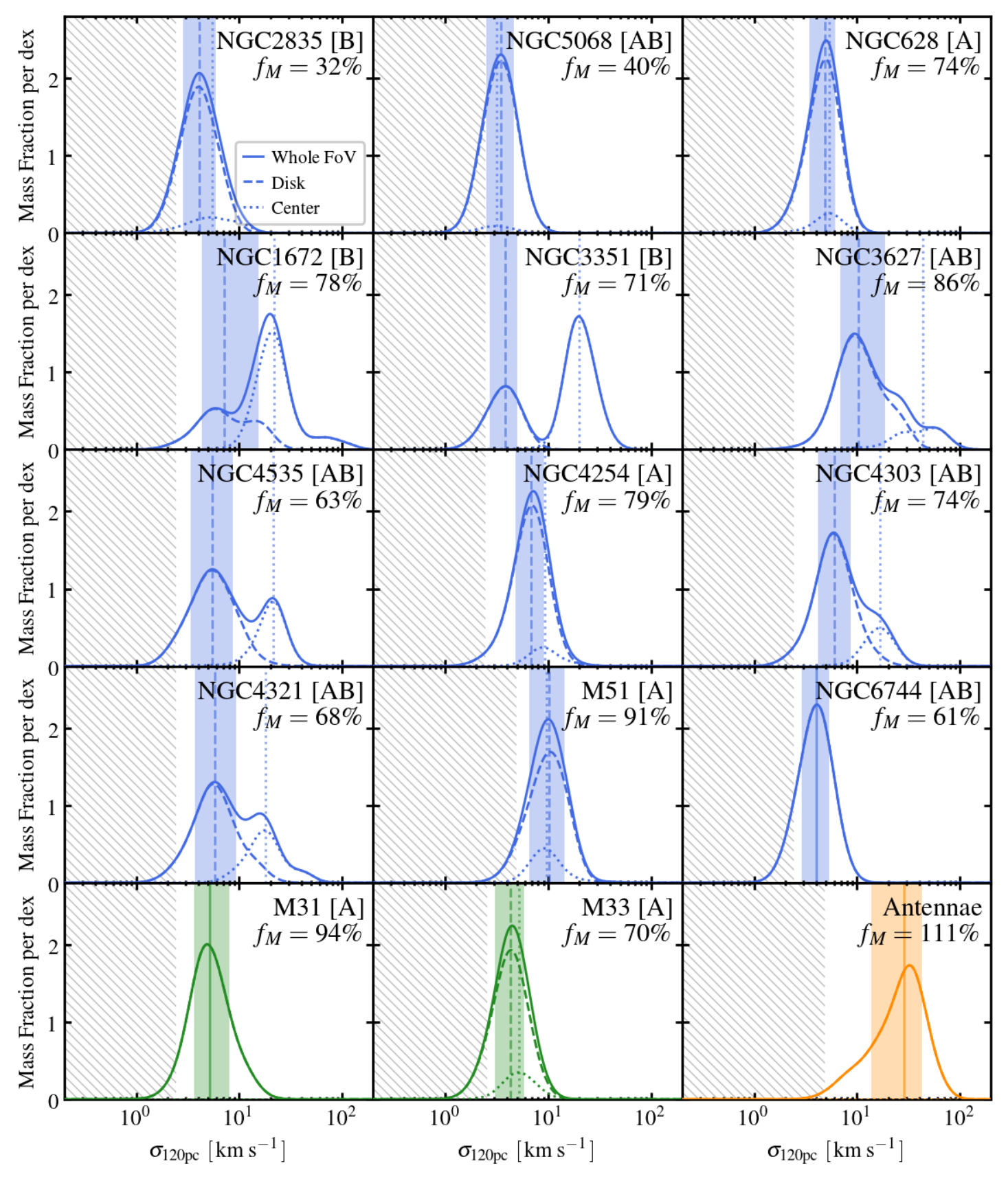}{0.85\textwidth}{}
\vspace{-2em}
\caption{Distribution of molecular gas mass as a function of velocity dispersion, \sig , measured at $120$~pc resolution. All curves are Gaussian KDE generated from the data with bandwidth of $0.1$~dex in logarithmic space. Labels and line-styles have the same meanings as in Figure~\ref{fig:Sig-80pc} and \ref{fig:Sig-120pc}, except that the hatched region here shows the \sig\ range close to or below the spectral resolution limit.
Galaxies show distributions of mass as a function of \sig\ similar to their \Sig\ distributions (Figure \ref{fig:Sig-80pc}), but the dynamical range in \sig\ is only about half of that seen in Figure \ref{fig:Sig-120pc}.}
\label{fig:sig}
\end{figure*}

Figures~\ref{fig:Sig-80pc} and \ref{fig:Sig-120pc} show the distribution of molecular gas mass (as inferred from CO flux) for each galaxy as a function of molecular gas surface density, \Sig , measured at $80$ and $120$~pc resolution respectively. Figures~\ref{fig:sig} shows the corresponding distributions as a function of velocity dispersion, \sig , measured at $120$~pc resolution. For each galaxy, we include all sightlines with detected CO emission across the whole field of view (FoV).

For most of the molecular gas properties considered in this work, their median values often show systematic variation across the spatial scales that we consider, while the shape of their distribution functions and the rank order of galxies barely varies. Therefore in the following part of this section, most of the tables report the median values and widths of all the relevant measurements at 45, 80, and 120~pc scales, while most of the figures only illustrate results at 120~pc scale (where we have available data for all targets).

\begin{deluxetable*}{lccccccccc}
\tabletypesize{\footnotesize}
\tablecaption{Properties of the \Sig\ Distribution Function \label{tab:Sig}}
\tablewidth{0pt}
\tablehead{
\colhead{Galaxy} & 
\multicolumn{3}{c}{at 45 pc resolution} &
\multicolumn{3}{c}{at 80 pc resolution} &
\multicolumn{3}{c}{at 120 pc resolution} \\
\cmidrule(lr){2-4} \cmidrule(lr){5-7} \cmidrule(lr){8-10}
\colhead{} &
\colhead{disk} &
\colhead{disk} &
\colhead{center} &
\colhead{disk} &
\colhead{disk} &
\colhead{center} &
\colhead{disk} &
\colhead{disk} &
\colhead{center}\\
\colhead{} &
\colhead{median} &
\colhead{16-84\%} &
\colhead{median} &
\colhead{median} &
\colhead{16-84\%} &
\colhead{median} &
\colhead{median} &
\colhead{16-84\%} &
\colhead{median} \\
\colhead{} &
\colhead{$\log_{10}\Sig$} &
\colhead{width} &
\colhead{$\log_{10}\Sig$} &
\colhead{$\log_{10}\Sig$} &
\colhead{width} &
\colhead{$\log_{10}\Sig$} &
\colhead{$\log_{10}\Sig$} &
\colhead{width} &
\colhead{$\log_{10}\Sig$}
}
\startdata
NGC~2835 & 1.75 & 0.56 & 1.91 & 1.49 & 0.59 & 1.61 & 1.36 & 0.60 & 1.46 \\
NGC~5068 & 1.78 & 0.52 & 1.79 & 1.55 & 0.59 & 1.51 & 1.44 & 0.57 & 1.40 \\
NGC~628 & 1.78 & 0.64 & 1.83 & 1.58 & 0.72 & 1.72 & 1.48 & 0.71 & 1.67 \\
NGC~1672 & -- & -- & -- & -- & -- & -- & 1.84 & 0.96 & 3.04 \\
NGC~3351 & -- & -- & -- & 1.42 & 0.53 & 2.95 & 1.26 & 0.54 & 2.89 \\
NGC~3627 & -- & -- & -- & 2.23 & 1.08 & 3.15 & 2.17 & 1.06 & 2.97 \\
NGC~4535 & -- & -- & -- & -- & -- & -- & 1.78 & 0.82 & 2.93 \\
NGC~4254 & -- & -- & -- & -- & -- & -- & 2.03 & 0.81 & 2.47 \\
NGC~4303 & -- & -- & -- & -- & -- & -- & 1.98 & 0.71 & 2.82 \\
NGC~4321 & -- & -- & -- & -- & -- & -- & 1.84 & 0.86 & 2.94 \\
M51 & 2.43 & 0.77 & 2.47 & 2.33 & 0.83 & 2.41 & 2.26 & 0.84 & 2.39 \\
NGC~6744 & -- & -- & -- & 1.64 & 0.59 & -- & 1.53 & 0.60 & -- \\
M31 & 1.24 & 0.65 & -- & 1.09 & 0.68 & -- & 1.03 & 0.65 & -- \\
M33 & -- & -- & -- & 0.99 & 0.76 & 1.21 & 0.89 & 0.73 & 1.16 \\
Antennae & -- & -- & -- & 3.41 & 1.16 & -- & 3.37 & 1.18 & --\\
\enddata
\tablecomments{For each galaxy at each resolution, we report:
(1) median $\log_{10}\Sig$ value {\it by gas mass} for the ``disk'' population (in units of $\rm\Msun\,pc^{-2}$);
(2) full width of the $16$-$84\%$ gas mass range of \Sig\ distribution for the ``disk'' population (in units of dex); and
(3) median $\log_{10}\Sig$ value {\it by gas mass} for the ``center'' population (in units of $\rm\Msun\,pc^{-2}$).
}
\end{deluxetable*}

\begin{deluxetable*}{lccccccccc}
\tabletypesize{\footnotesize}
\tablecaption{Properties of the \sig\ Distribution Function \label{tab:sig}}
\tablewidth{0pt}
\tablehead{
\colhead{Galaxy} & 
\multicolumn{3}{c}{at 45 pc resolution} &
\multicolumn{3}{c}{at 80 pc resolution} &
\multicolumn{3}{c}{at 120 pc resolution} \\
\cmidrule(lr){2-4} \cmidrule(lr){5-7} \cmidrule(lr){8-10}
\colhead{} &
\colhead{disk} &
\colhead{disk} &
\colhead{center} &
\colhead{disk} &
\colhead{disk} &
\colhead{center} &
\colhead{disk} &
\colhead{disk} &
\colhead{center}\\
\colhead{} &
\colhead{median} &
\colhead{16-84\%} &
\colhead{median} &
\colhead{median} &
\colhead{16-84\%} &
\colhead{median} &
\colhead{median} &
\colhead{16-84\%} &
\colhead{median} \\
\colhead{} &
\colhead{$\log_{10}\sig$} &
\colhead{width} &
\colhead{$\log_{10}\sig$} &
\colhead{$\log_{10}\sig$} &
\colhead{width} &
\colhead{$\log_{10}\sig$} &
\colhead{$\log_{10}\sig$} &
\colhead{width} &
\colhead{$\log_{10}\sig$}
}
\startdata
NGC~2835 & 0.54 & 0.31 & 0.71 & 0.58 & 0.31 & 0.72 & 0.61 & 0.33 & 0.74 \\
NGC~5068 & 0.44 & 0.28 & 0.39 & 0.50 & 0.28 & 0.42 & 0.54 & 0.28 & 0.50 \\
NGC~628 & 0.61 & 0.31 & 0.64 & 0.66 & 0.29 & 0.70 & 0.69 & 0.26 & 0.73 \\
NGC~1672 & -- & -- & -- & -- & -- & -- & 0.85 & 0.57 & 1.33 \\
NGC~3351 & -- & -- & -- & 0.53 & 0.29 & 1.25 & 0.58 & 0.28 & 1.30 \\
NGC~3627 & -- & -- & -- & 0.97 & 0.47 & 1.59 & 1.01 & 0.44 & 1.64 \\
NGC~4535 & -- & -- & -- & -- & -- & -- & 0.74 & 0.41 & 1.33 \\
NGC~4254 & -- & -- & -- & -- & -- & -- & 0.83 & 0.29 & 0.96 \\
NGC~4303 & -- & -- & -- & -- & -- & -- & 0.78 & 0.32 & 1.22 \\
NGC~4321 & -- & -- & -- & -- & -- & -- & 0.76 & 0.40 & 1.26 \\
M51 & 0.92 & 0.40 & 0.88 & 0.97 & 0.37 & 0.93 & 1.00 & 0.34 & 0.98 \\
NGC~6744 & -- & -- & -- & 0.57 & 0.30 & -- & 0.61 & 0.29 & -- \\
M31 & 0.53 & 0.36 & -- & 0.64 & 0.36 & -- & 0.71 & 0.35 & -- \\
M33 & -- & -- & -- & 0.58 & 0.30 & 0.65 & 0.63 & 0.30 & 0.72 \\
Antennae & -- & -- & -- & 1.41 & 0.51 & -- & 1.45 & 0.50 & --\\
\enddata
\tablecomments{For each galaxy at each resolution, we report:
(1) median $\log_{10}\sig$ value {\it by gas mass} for the ``disk'' population (in units of $\kms$);
(2) full width of the $16$-$84\%$ gas mass range of \sig\ distribution for the ``disk'' population (in units of dex); and
(3) median $\log_{10}\sig$ value {\it by gas mass} for the ``center'' population (in units of $\kms$).
}
\end{deluxetable*}

\subsubsection{Central and Disk Distributions}\label{sec:dist:region}

For many of the galaxies with the widest range of \Sig\ and \sig , we observe multiple peaks in the \Sig\ and \sig\ distributions (e.g., NGC~3351, NGC~3627, NGC~1672, NGC~4535, NGC~4303, and NGC~4321). By visually inspecting the maps in Figure \ref{fig:maps}, we see that the high value peak(s) of \Sig\ and \sig\ tend to arise from bright structures in the innermost regions of these targets. It has long been known that molecular gas in the central region of disk galaxies has different properties compared to the gas in the disk \citep[e.g.,][among many others]{Oka_etal_2001, Regan_etal_2001, Jogee_etal_2005, Shetty_etal_2012, Kruijssen_Longmore_2013, Colombo_etal_2014a, Leroy_etal_2015, Freeman_etal_2017}. Especially in galaxies with strong bars, the inner parts of galaxies often harbor high gas surface densities and complex structures such as starburst rings \citep{Kenney_etal_1992,Sakamoto_etal_1999,Sheth_etal_2002,Kormendy_Kennicutt_2004,Jogee_etal_2005}.

To illustrate the impact of nuclear gas concentrations, we define a central region for each galaxy. In Figures~\ref{fig:Sig-80pc}--\ref{fig:sig}, we plot the distribution for gas in this central region and the outer disk as separate histograms (dotted and dashed lines). For most galaxies, we define the center as the region within $1$~kpc of the galaxy nucleus. For NGC~3351, we slightly expand the defined radius to $1.5$~kpc, so that the visually distinct inner disk is entirely designated as central. The central regions of M31 and NGC~6744 are not included in our CO data, and the CO map of the Antennae only covers the interacting region. Therefore we do not plot any separate histograms for these galaxies.

Comparing the ``center'' distributions to the ``disk'' distributions, we find that for the strongly barred galaxies (NGC~3351, NGC~3627, NGC~1672, NGC~4535, NGC~4303, and NGC~4321) the peaks of the distribution at large values of \Sig\ and \sig\ are often predominantly tracing gas in the central region. In galaxies without bar-driven inner structures, gas in the inner part of the galaxy also tends to have higher \Sig\ and \sig\ than the galaxy average, but the effect is much weaker and the fraction of emission arising from the galaxy center tends to be smaller.

In addition to these radial variations, the maps in Figure~\ref{fig:maps} show significant azimuthal variations. At fixed galactocentric radius, spiral arms and bars show clear enhancements in both \Sig\ and \sig . Such variations have been emphasized in individual galaxies before \citep[e.g.,][]{Koda_etal_2009, Colombo_etal_2014a}. They manifest in the histograms as broad, non-Gaussian shapes of the mass distributions in galaxies with prominent dynamical features. For example, NGC~1672, NGC~3627, NGC~4535 and M51 all show distributions skewed towards high \Sig . In the Antennae, the ``superclouds'' created by the interaction stand out from the rest of the gas \citep{Wilson_etal_2003, Wei_etal_2012}. Obtaining a quantitative mapping between a galaxy's dynamical features and the distributions of \Sig , \sig , and $\alpha_{\rm vir}$ in the molecular gas reservoir is a main goal of the next set of PHANGS papers.

\subsubsection{Width and Median of the Distributions}\label{sec:dist:shape}

For each galaxy disk (i.e., excluding the central region), we measure the median \Sig\ and \sig\ (by gas mass; shown as vertical dashed lines in the Figures) as well as the $16$-$84\%$ width of the distribution (color shaded region). These results are reported in Tables~\ref{tab:Sig} and \ref{tab:sig}, with galaxies ordered in the same way as Figures~\ref{fig:Sig-80pc}--\ref{fig:sig} for easy comparison.

Figures~\ref{fig:Sig-80pc}, \ref{fig:Sig-120pc} and Table~\ref{tab:Sig} reveal a range of surface densities at $80$ and $120$~pc resolution. Within a galaxy disk, the $16$-$84\%$ width of the distribution in \Sig\ is typically $0.55$-$1.10$~dex. The $16$-$84\%$ width for the \sig\ distribution is typically $0.25$-$0.60$~dex, about half the width of the \Sig\ distribution. Galaxy-to-galaxy variations in \sig\ are also about half of those found for \Sig\ (in logarithmic space). This is what we would expect from a scaling relation of $\sig \propto \Sig^{0.5}$, i.e., fixed \alphavir\ gas (Section \ref{sec:expectation}). Section~\ref{sec:scaling} presents the observed scaling relation and explores this result further.

The estimated statistical error on our \Sig\ and \sig\ measurements is almost always less than $0.1$~dex. Thus the observed ranges of \Sig\ and \sig\ reflect real, significant variations in the surface brightness and line width of CO emission within galaxies.

Note that the distributions in Figures~\ref{fig:Sig-80pc}, \ref{fig:Sig-120pc} and Table~\ref{tab:Sig} are calculated from all {\it detected} CO emission. As reported in Table \ref{tab:stats}, the sensitivity limit of the data prevents us from characterizing all CO emission within the field of view, and the amount of excluded emission can be significant in the low mass PHANGS-ALMA targets (e.g., NGC~2835, NGC~5068). Because the non-detected emission often lies at low surface density, our calculated distribution function becomes incomplete at the low \Sig\ end. For galaxies with high CO emission recovery fraction (e.g., NGC~3627, M31, M51, and the Antennae), we expect to recover their true distribution fairly well, extending down to our (relatively low) detection limit. However, for targets like NGC~2835 and NGC~5068, we likely only capture the high end of the intrinsic distribution. In this case our reported median values will be biased high and the distribution width will be underestimated.

We also measure the median values of \Sig\ and \sig\ for each galaxy center (vertical dotted lines) whenever possible. In the strongly barred galaxies, the median \Sig\ value in the central region can be $0.8$-$1.7$~dex higher than in the disk. This offset is significant compared to the width of the \Sig\ distribution in the disks. The \sig\ distributions follow a similar pattern, with the center population showing $0.4$-$0.8$~dex excess in median \sig\ relative to the disk population.

The magnitude of the disk-center distinction does depend on our treatment of the CO-to-H$_2$ conversion factor, \alphaco . \citet{Sandstrom_etal_2013} found a factor of $\sim 2$ decrease in \alphaco\ in the centers of star-forming galaxies relative to their disks. Lower \alphaco\ are also reported in the central regions of other types of galaxies \citep[see][and references therein]{Bolatto_etal_2013}. We use a fixed \alphaco\ in this paper and so may expect to somewhat overestimate the difference in  median \Sig\ between galaxy disks and centers. However, the factor of 2 (or $\sim 0.3$ dex) change in \alphaco\ found by \citet{Sandstrom_etal_2013} is still not strong enough to explain the $0.8-1.7$ dex separation between the two \Sig\ peaks in our strongly barred targets. Therefore, it is unlikely that the bimodal \Sig\ distribution function shapes merely reflect \alphaco\ variation. \alphaco\ variations should have less impact on the observed \sig , though they could change amount of mass associated with different parts of the histogram.

Compared to the intra-galaxy distribution width, the median \Sig\ and \sig\ show as strong or even stronger inter-galaxy variations. In Figure~\ref{fig:Sig-80pc}--\ref{fig:sig}, we arrange the panels so that the 11 PHANGS-ALMA targets and M51 (our ``main sample'') are ordered in terms of stellar mass from left to right, top to bottom. M31, M33 and Antennae instead appear in the bottom row. Table~\ref{tab:Sig} and \ref{tab:sig} also follow this sorting scheme. Among the 12 galaxies in the main sample, generally low mass star-forming galaxies (NGC~2835, NGC~5068, and NGC~628) have low median \Sig\ of $10$-$30 ~\Msun~{\rm pc}^{-2}$ and low median \sig\ of $3$-$5~\kms$. High mass star-forming galaxies (NGC~3627, NGC~4303, NGC~4321, M51) have median \Sig\ as high as $100$-$200~\Msun~{\rm pc}^{-2}$ and median \sig\ of $5$-$10~\kms$. NGC~6744 is an outlier from this trend. We note that it is an early-type spiral galaxy compared to the other high mass star-forming galaxies in our sample, and that its CO map does not cover the entire inner part of the galaxy.

More massive star-forming galaxies are known to harbor more massive molecular reservoirs \citep[e.g.,][]{Young_etal_1989,Young_Scoville_1991,Saintonge_etal_2011}, with a good match between the distribution of stars and gas \citep[e.g.,][]{Young_etal_1995,Regan_etal_2001,Leroy_etal_2008}. Figures~\ref{fig:Sig-80pc}--\ref{fig:sig} show that for our targets in the main sample, the apparent cloud-scale molecular gas surface density and velocity dispersion also correlate with stellar mass. We discuss the possible physical origin of such correlation in more detail in Section~\ref{sec:discussion}.

Finally, among our supplementary targets, M31 and M33 both show extraordinarily low \Sig\ values but normal \sig\ values. The Antennae galaxies show high values of both \Sig\ and \sig . A more detailed discussion about these three targets is presented in Section~\ref{sec:scaling:high-end} and \ref{sec:scaling:low-end}.

\subsubsection{Scale Dependence and Clumping}\label{sec:dist:clumping}

We characterize the emission at a set of fixed spatial scales. By comparing results obtained at different scales, we can investigate the structure and ``clumping'' of the molecular ISM. In Tables~\ref{tab:Sig} and \ref{tab:sig}, we report the median value and $16$-$84\%$ width of \Sig\ and \sig\ at each available resolution for each galaxy. For galaxies with multi-resolution measurements, the width of their \Sig\ and \sig\ distributions show little variation across physical scales, but the median \Sig\ systematically decreases, and the median \sig\ increases, as the measurements are obtained using data with coarser resolution.

These trends match expectations for a highly-structured turbulent medium. We expect smaller velocity dispersions, \sig , when sampling a turbulent medium at smaller spatial scales, a manifestation of the size-line width relation \citep{Larson_1981}. Meanwhile, clumpy substructures will suffer from less beam dilution at better resolution so we also expect to find higher surface densities.

Our sensitivity changes as a function of angular resolution because the noise level improves with spatial averaging. Therefore the change in the median \Sig\ across resolutions will be somewhat exaggerated by observational selection effects. We intend to revisit the specific topic  of molecular gas clumping factor in a future PHANGS-ALMA paper.

\subsection{Line Width--Surface Density Scaling Relation}\label{sec:scaling}

\begin{figure*}[htb]
\fig{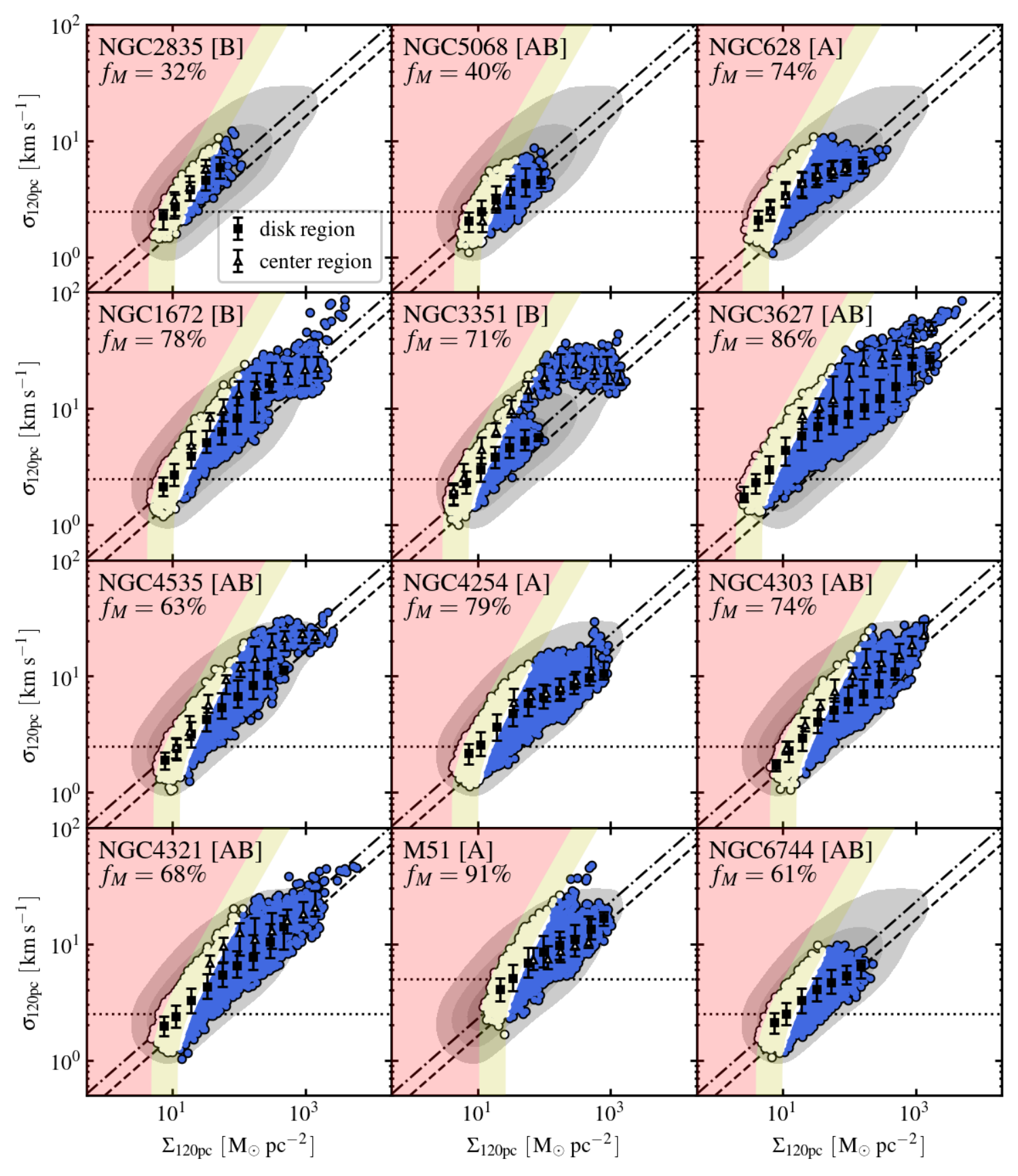}{0.85\textwidth}{}
\vspace{-2em}
\caption{$\sig$-$\Sig$ relation measured at $120$~pc resolution across our main sample. In each panel, we show all detected sightlines in one galaxy as blue or white filled circles. The black filled squares (resp. open triangles) represent the median \sig\ in each \Sig\ bin for sightlines in the disk (resp. center) region, with their associated error bars showing $1\sig$ scatter. The gray contours in the background show data density for measurements in all targets in the main sample available at this resolution (the 3 levels include 60\%, 90\% and 99\% of data points, respectively). The dashed line (resp. dot-dash line) shows the ${\rm slope} = 0.5$ prediction for $100\%$ beam-filling spherical clouds with virial parameter $\alphavir = 1$ (resp. $\alphavir = 2$). The red and yellow shaded regions show the sensitivity limits of our CO emission identification strategy, signifying that the completeness is not 100\% inside the yellow region, and it drops to zero inside the red region (see Section~\ref{sec:scaling:disk}). We take this selection function into account when fitting the average relation between \sig\ and \Sig . The horizontal dotted lines show the velocity resolution for each data cube, far below which the measured \sig\ values become increasingly less robust. We observe strong positive correlations between \sig\ and \Sig\ in all galaxies. The measured slopes $\beta$ for the disk populations are close to the expected $\beta = 0.5$ value for resolved, self-gravitating clouds. Data from the centers of the strongly barred targets (NGC~1672, NGC~3351, NGC~3627, NGC~4535, NGC~4303, and NGC~4321) show higher \Sig\ and elevated \sig\ at fixed \Sig\ relative to disk population.}
\label{fig:scaling}
\end{figure*}

\begin{deluxetable*}{lccccccccc}
\tabletypesize{\footnotesize}
\tablecaption{Best Fit Parameters of the $\sig$-$\Sig$ Scaling Relation in Galaxy Disks \label{tab:fit}}
\tablewidth{0pt}
\tablehead{
\colhead{Galaxy} & 
\multicolumn{3}{c}{at 45 pc resolution} &
\multicolumn{3}{c}{at 80 pc resolution} &
\multicolumn{3}{c}{at 120 pc resolution} \\
\cmidrule(lr){2-4} \cmidrule(lr){5-7} \cmidrule(lr){8-10}
\colhead{} &
\colhead{$\beta$} &
\colhead{$A$} &
\colhead{$\Delta_\mathrm{intr}(\Delta_\mathrm{tot})$} &
\colhead{$\beta$} &
\colhead{$A$} &
\colhead{$\Delta_\mathrm{intr}(\Delta_\mathrm{tot})$} &
\colhead{$\beta$} &
\colhead{$A$} &
\colhead{$\Delta_\mathrm{intr}(\Delta_\mathrm{tot})$}
}
\startdata
NGC~628 & $0.50^{*}$ & $0.63^{*}$ & $0.00$($0.10$) & $0.38^{*}$ & $0.74^{*}$ & $0.03$($0.11$) & $0.34^{*}$ & $0.79^{*}$ & $0.02$($0.11$) \\
NGC~1672 & -- & -- & -- & -- & -- & -- & $0.63^{*}$ & $0.91^{*}$ & $0.08$($0.12$) \\
NGC~2835 & $0.51^{*}$ & $0.60^{*}$ & $0.00$($0.09$) & $0.49^{*}$ & $0.74^{*}$ & $0.00$($0.09$) & $0.45^{+0.09}_{-0.09}$ & $0.83^{*}$ & $0.00$($0.08$) \\
NGC~3351 & -- & -- & -- & $0.56^{*}$ & $0.76^{*}$ & $0.00$($0.09$) & $0.41^{+0.06}_{-0.06}$ & $0.82^{*}$ & $0.00$($0.09$) \\
NGC~3627 & -- & -- & -- & $0.44^{*}$ & $0.84^{*}$ & $0.10$($0.14$) & $0.42^{*}$ & $0.92^{*}$ & $0.10$($0.15$) \\
NGC~4254 & -- & -- & -- & -- & -- & -- & $0.34^{*}$ & $0.81^{*}$ & $0.08$($0.11$) \\
NGC~4303 & -- & -- & -- & -- & -- & -- & $0.41^{*}$ & $0.79^{*}$ & $0.08$($0.10$) \\
NGC~4321 & -- & -- & -- & -- & -- & -- & $0.50^{*}$ & $0.82^{*}$ & $0.07$($0.10$) \\
NGC~4535 & -- & -- & -- & -- & -- & -- & $0.53^{*}$ & $0.83^{*}$ & $0.07$($0.10$) \\
NGC~5068 & N/A\tablenotemark{$\dagger$} & N/A\tablenotemark{$\dagger$} & N/A\tablenotemark{$\dagger$} & N/A\tablenotemark{$\dagger$} & N/A\tablenotemark{$\dagger$} & N/A\tablenotemark{$\dagger$} & $0.50^{*}$ & $0.74^{*}$ & $0.04$($0.11$) \\
NGC~6744 & -- & -- & -- & $0.50^{*}$ & $0.69^{*}$ & $0.05$($0.09$) & $0.47^{*}$ & $0.76^{*}$ & $0.04$($0.10$) \\
M51 & $0.49^{*}$ & $0.69^{*}$ & $0.09$($0.11$) & $0.40^{*}$ & $0.83^{*}$ & $0.10$($0.11$) & $0.36^{*}$ & $0.90^{*}$ & $0.10$($0.11$) \\
M31 & $0.4^{+0.1}_{-0.1}$ & $0.66^{*}$ & $0.00$($0.23$) & $0.4^{+0.1}_{-0.1}$ & $0.77^{+0.05}_{-0.05}$ & $0.00$($0.25$) & $-0.0^{+0.2}_{-0.3}$ & $0.7^{+0.1}_{-0.1}$ & $0.00$($0.18$) \\
M33 & -- & -- & -- & $0.5^{+0.2}_{-0.2}$ & $0.79^{*}$ & $0.00$($0.31$) & $0.4^{+0.3}_{-0.4}$ & $0.9^{+0.1}_{-0.1}$ & $0.00$($0.24$) \\
Antennae & -- & -- & -- & $0.50^{*}$ & $0.71^{*}$ & $0.12$($0.14$) & $0.47^{*}$ & $0.81^{*}$ & $0.13$($0.15$) \\
\cmidrule(lr){1-10}
PHANGS+M51 & $0.48^{*}$ & $0.66^{*}$ & $0.07$($0.11$) & $0.47^{*}$ & $0.85^{*}$ & $0.10$($0.14$) & $0.37^{*}$ & $0.85^{*}$ & $0.12$($0.13$) \\
\enddata
\tablecomments{For each galaxy at each resolution level, we report: 
(1) the best fit power-law slope $\beta$ (see Equation~\ref{eq:scaling}), 
(2) the best fit power-law normalization $A$, and 
(3) scatter in \sig\ at fixed \Sig\ around the best fit scaling relation (in unit of dex; we report both the estimated intrinsic scatter, $\Delta_\mathrm{intr}$, and the total observed rms scatter, $\Delta_\mathrm{tot}$).
The last row show the best fit values when combining the main sample together (see Section~\ref{sec:scaling:disk} and Appendix~\ref{apdx:mcmc}).}
\tablenotetext{*}{The MCMC sampling suggests that the statistical uncertainties on these $\beta$ and $A$ estimates are smaller than 0.05 (i.e., insignificant compared to the systematic uncertainties). We suggest adopting a total uncertainty of 0.05 for these $\beta$ and $A$ estimates (see Section~\ref{sec:scaling:disk}).}
\tablenotetext{\dagger}{``N/A'' means the MCMC sampling does not converge.}
\end{deluxetable*}

In Section~\ref{sec:expectation}, we argued that the scaling relation between surface density, \Sig , and line width, \sig , reflects the physical state of the gas structure in each beam. If molecular gas in galaxies tends towards a universal dynamical configuration, then we expect to observe a correlation (e.g., fixed \alphavir ) or anti-correlation (fixed \Pturb ) between these two quantities. The gas maps (Figure~\ref{fig:maps}) and distribution functions that we present already reveal some similarities between the distributions of \Sig\ and \sig , indicating the existence of a positive correlation.

In Figure~\ref{fig:scaling}, we show the line width, \sig , as a function of surface density, \Sig , at $120$~pc resolution across our main sample. In each panel, we plot all measurements in one galaxy as colored or white filled circles. For reference to the larger population, we also plot all measurements across this main sample as gray contours (showing data density levels including 60\%, 90\% and 99\% of points) in the background.

To highlight the difference between the central regions and the disks, we separate sightlines into disk and center populations. We calculate the median and rms scatter in \sig\ in $0.2$~dex wide bins of fixed \Sig\ for both populations, and show them as black filled squares or open triangles with error bars.

Our selection criteria introduce a bias into the analysis (Section~\ref{sec:method:procedure}). Since detection depends on the S/N in a channel-by-channel sense, we preferentially pick out sightlines with narrow line profiles at fixed line-integrated intensity. This can affect our measurement at low \Sig , near the sensitivity limit. In Figure~\ref{fig:scaling}, we plot the sensitivity limits imposed by our selection criteria as yellow and red shaded regions\footnote{The curvature at the low \Sig\ end arises because the channel-width-subtracted \sig\ deviates from the uncorrected value near the velocity resolution limit. We show the corrected values in the plots, while we note that the selection criteria operate directly on uncorrected values. See Equation~\ref{eq:resolution} for the relation between the two.}. These represent the 2-consecutive-5$\sigma$-channel and 2-consecutive-2$\sigma$-channel criteria, respectively (see Section~\ref{sec:method:procedure}). 

Our completeness will be less than 100\% throughout the yellow hatched region, and it rapidly drops to zero inside the red region. As discussed above, in most targets the 2-at-2$\sigma$ represents the relevant case over most of the area in most of our targets, and completeness is still reasonably high in the yellow region. Thus there is only a sharp edge at the boundary of the red region. But for NGC~2835 and NGC~5068, the 2-at-5$\sigma$ criterion is more restrictive and completeness through the yellow region is lower.

We also label the $\sig_{\rm measured} = \Delta v$ threshold of each data set as a horizontal dotted line. Although we account for broadening of the line due to the finite channel width and spectral response curve via Equation~\ref{eq:resolution}, we do not expect \sig\ to be reliable much below this value.

Overall, Figure~\ref{fig:scaling} shows the equivalent of the $\sig^2/r \propto \Sig$ scaling relation \citep[][Section~\ref{sec:expectation}]{Heyer_etal_2009} for around $30,000$ independent beams spanning $12$ nearby star-forming galaxies. Although this relationship is usually studied for individual clouds or sub-cloud structure, Figure~\ref{fig:scaling} shows that a version of this $\sig$-$\Sig$ relation holds sightline-by-sightline across our main sample. The relation spans $3$-$4$ orders of magnitude in surface density and two decades in line width, and appears to be a fundamental property of the molecular ISM at cloud scales.

\subsubsection{Star-forming Galaxy Disks}\label{sec:scaling:disk}

In Figure~\ref{fig:scaling}, the diagonal lines show the $\sig \propto \Sig^{0.5}$ expectation for gas structures with $\alphavir = 1$ (dashed) and $2$ (dash-dot), assuming our fiducial sub-beam geometries, no surface pressure term, and taking the beam size as the relevant size scale. By eye, the disk population of these star-forming galaxies all show scaling relations that are approximately parallel to these lines, which implies a roughly constant \alphavir\ within each galaxy disk region. 

To provide a more quantitative description of the observed scaling relations, we model the data for each galaxy disk at each resolution with a power law of the form

\begin{equation}
\label{eq:scaling}
\log_{10} \left( \frac{\sig}{\kms} \right)
= \beta \log_{10} \left( \frac{\Sig}{10^2\;\Msun\,\rm pc^{-2}} \right) + A \, ,
\end{equation}

\noindent where $\beta$ is the power law index and $A$ is the normalization of the fit at $\Sig = 10^2~\Msun\,\rm pc^{-2}$. We also include an intrinsic scatter along the \sig\ direction in our model, for which we denote the rms scatter in the logarithmic space as $\Delta_\mathrm{intr}$.

To take both the selection effect and the measurement uncertainties into account, we find the best fit model parameters and their associated uncertainties using a Markov-Chain Monte Carlo (MCMC) method \citep[as implemented in {\tt emcee} by][]{Foreman-Mackey_etal_2013}. We include only the data points above our 2-consecutive-5-sigma selection criterion (i.e. the blue points in Figure~\ref{fig:scaling}, or the rows flagged as ``complete'' in Table~\ref{tab:data}) as the input for MCMC, and also take into account this truncation in our Bayesian model. We assume that for each galaxy at each resolution, the statistical uncertainties in \sig\ and \Sig , as well as the correlation in their uncertainties, can be described by the estimated values from the Monte Carlo simulation that we describe in the last paragraph of Section~\ref{sec:method:procedure}. We report all derived model parameters in Table~\ref{tab:fit}. A more detailed description of the MCMC setup, definition of the priors and likelihood functions, as well as the distribution-correlation plots are presented in Appendix~\ref{apdx:mcmc}.

At 80~pc resolution, we find best fit power-law slopes of $\beta = 0.34$-$0.63$ in individual galaxy disks. The best fit normalization at $\Sig = 100\rm\;\Msun/pc^2$ is $4$-$8\rm\;\kms$. If we combine the data at 80~pc resolution for all galaxies in the main sample (i.e., the row named ``PHANGS+M51'' in Table \ref{tab:fit}), we find a best-fit $\sig$-$\Sig$ relation of

\begin{equation}
\label{eq:bestfit}
\log_{10} \left( \frac{\sig}{\kms} \right) = 0.47\,\log_{10} \left( \frac{\Sig}{10^2\;\Msun\,\rm pc^{-2}} \right) + 0.85
\end{equation}

\noindent for this whole sample of 12 galaxies. This relation holds for sightlines with molecular gas surface density over  the range $\Sig \sim 20$-$2,000~\Msun\,\rm pc^{-2}$.

Due to the huge number of data points involved, the formal statistical errors on the best fit power-law slope and zero point are often quite small. However, by comparing these MCMC results with those obtained from other fitting strategies (e.g., binning by \Sig\ then fitting, changing weighting schemes), we find that different methods produce results that differ by $\sim 0.05$ for both $\beta$ and $A$. Since the best fit results depend mostly on which fitting scheme we adopt, for comparisons with other studies or applications in theoretical models/numerical simulations as empirical relations, we adopt a typical uncertainty of $0.05$ in $\beta$ and $A$. For the few cases where we find $> 0.05$ statistical errors from the MCMC method, we instead explicitly report these values in Table~\ref{tab:fit}.

We find the intrinsic scatter in \sig\ at fixed \Sig\ to be less than $0.10$~dex for all galaxies at all resolutions. In a few cases (NGC~628 at 45~pc, NGC~2835, NGC~3351, M31, and M33 at all resolutions), the MCMC modeling returns a best fit value of zero for the intrinsic scatter. Given that the measurement uncertainties in \sig\ and \Sig\ are comparable to the total observed rms scatter around the best-fit relation, these results most likely imply that we are over-estimating the measurement uncertainties in \sig\ and/or \Sig\ for these galaxies. For clarity, in Table~\ref{tab:fit} we also report the total rms scatter around the best-fit relation, which includes both the intrinsic scatter and the measurement uncertainties.

Our power-law fits suggest that molecular gas has similar dynamical properties in star-forming galaxy disks. Considering the uncertainty in $\beta$, about half of the best fit values are consistent with $0.5$, as expected for resolved, self-gravitating structures (see Section~\ref{sec:expectation}). For the remaining targets, most show shallower slopes and high \sig\ in the low \Sig\ regime. We discuss the possible origin of such behavior in Section~\ref{sec:scaling:discuss}.

\subsubsection{Star-forming Galaxy Centers}\label{sec:scaling:center}

In Section~\ref{sec:dist:region}, we find that molecular gas in the central regions of the strongly barred galaxies (NGC~1672, NGC~3351, NGC~3627, NGC~4303, NGC~4321 and NGC~4535) shows higher \Sig\ and \sig\ relative to molecular gas in the disk. By inspecting the corresponding panels in Figure~\ref{fig:scaling}, we can further conclude that their center populations also show higher \sig\ at a given \Sig\ compared to the relation inferred from their disk population. NGC~2835 may show a similar trend, but the number of data points is small due to its faint CO emission.

These enhanced line widths may reflect differences in the dynamical state of gas in the central region compared to the disk. Despite their high \Sig , molecular gas in central regions appears less strongly bound by its own self-gravity. At our spatial resolutions, every beam in our data captures a mixture of more and less bound gas. Therefore, our measurements might be interpreted as follows: in the central parts of strongly barred galaxies, the CO emission tends to be dominated by apparently less bound material. More strongly bound structures may still exist within this medium, and may become distinguishable at higher spatial resolution \citep[e.g., as seen in NGC 253 by][and possibly associated with higher \alphaco]{Leroy_etal_2015}. 

Our conclusion that there is an offset between disk and center population in the \sig-\Sig\ space is robust against several systematic effects. The filling factor of CO emission is expected to increase towards the gas rich central regions, and \alphaco\ should become lower there \citep{Sandstrom_etal_2013}. The former effect is preferentially biasing our disk \Sig\ measurements to lower values, whereas the latter one is elevating the \Sig\ values for the center population. If we were to correct for these systematics, the disk-center offset in the \sig-\Sig\ space would be further increased. However, the rapidly rising rotation curves in the central regions imply that the possibility of capturing unassociated structures moving at different rotational velocities within the beam (``beam smearing'') is higher there. We expect to be able to test the influence of beam smearing on our measurements once we have PHANGS-ALMA rotation curves (\citeinprep{P. Lang et al. in prep.}).

In our other targets -- NGC~5068, NGC~628, M51, and NGC~4254 -- the excess in line width at the galaxy centers is more subtle or absent. NGC~628 and NGC~5068 are low mass spirals without strong bars and do not show a clear separation between the disk and centers in their distributions of \Sig\ and \sig\ (Section~\ref{sec:dist}). M51 and NGC~4254 are more massive spirals that also lack strong, large-scale bars. They both show concentrations of molecular gas in their central region, but their \Sig\ and \sig\ distributions for center and disk sightlines overlap. We note, though, that M51 shows evidence for a population of high line width sightlines in both the disk and the central region. These correspond to both the very center \citep[$\rgal < 100$~pc;][]{Querejeta_etal_2016} and particular regions in the spiral arms \citep[][]{Meidt_etal_2013, Colombo_etal_2014a, Leroy_etal_2017}.

\subsubsection{High Surface Density Regime: the~Antennae~Galaxies}\label{sec:scaling:high-end}

\begin{figure*}[htb]
\fig{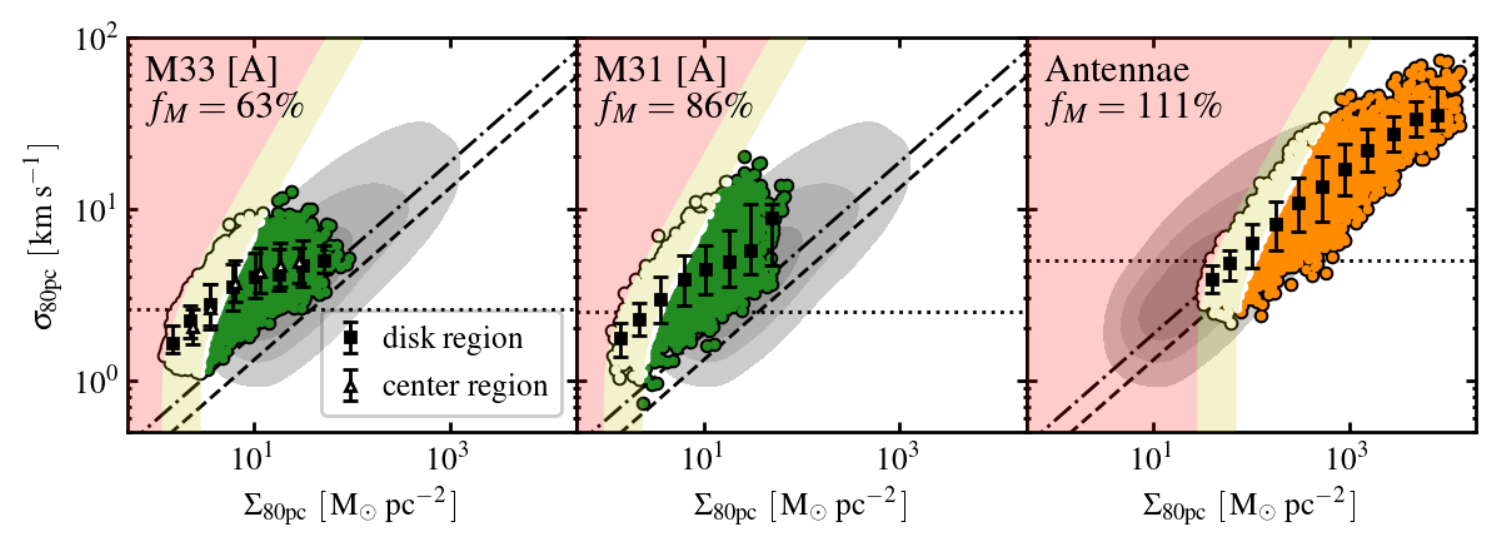}{0.85\textwidth}{}
\vspace{-2em}
\caption{$\sig$-$\Sig$ relation at $80$~pc resolution measured for the two Local Group targets M33 (left panel) and M31 (middle panel), and the nearby major merger, the Antennae galaxies (right panel). Gray contours in the background show the data density for measurements across the main sample as reference. Both M31 and M33 have no sightlines showing \Sig\ higher than $100\rm\;\Msun\,pc^{-2}$. They instead have many more detected sightlines showing \Sig\ lower than $10\rm\;\Msun\,pc^{-2}$, among which the majority also show higher \sig\ at given \Sig\ compared to the main sample population. In contrast, the Antennae galaxies have most detected sightlines showing high \Sig\ and high \sig , but their ratio is roughly consistent with the extrapolation of the average \sig-\Sig\ relation of the main sample.}
\label{fig:scaling-ext}
\end{figure*}

Our main sample, consisting of 11 PHANGS-ALMA targets and M51, emphasizes relatively massive galaxies ($M_\star = 10^{10}$-$10^{11}\,\Msun$) on the star-forming main sequence. These galaxies represent the typical environment for star formation in the local universe. Nevertheless, our ultimate goal is to achieve a quantitative, homogeneously analyzed picture that covers the full range of conditions found in galaxies, from molecule-poor outer disks and dwarf galaxies to gas-rich turbulent disks at high redshift. With the aim of extending our sample towards these extremes, we included the Antennae galaxies, the nearest major merger, M31, a quiescent massive spiral, and M33, a star-forming dwarf spiral. These targets offer clues about the behavior of the scaling relations that we measure outside the active regions of disk galaxies.

In the right panel of Figure~\ref{fig:scaling-ext}, we show the \sig-\Sig\ relation measured in the interacting region of the Antennae galaxies (orange and white filled circles), on top of the main sample population (gray contours in the background). Molecular gas in this regime mostly populates the upper right corner in the $\sig$-$\Sig$ parameter space. Such extraordinarily high surface density and velocity dispersion are expected and have been noted before, as the galaxy merger brings a huge amount of gas into the interacting region and concentrates it into a small area \citep{Wilson_etal_2003}. The complex kinematics of the collision can create large velocity dispersions \citep{Wei_etal_2012}.

The measured $\sig$-$\Sig$ scaling relation for the Antennae lies close to the extrapolation of the average relation in the disks of star-forming galaxies, albeit with a slightly larger scatter ($0.12$-$0.13$~dex; see the last row in Table~\ref{tab:fit}). In other words, the measured high line widths match the expectation given the high surface densities and approximately fixed \alphavir\ to first order \citep[see also][]{Leroy_etal_2016}. In this case, the main effect of the major merger may be to drive the internal pressure of the gas to high values (see Section \ref{sec:Pturb} below), while not substantially altering its observed dynamical state. 

We caution that additional caveats, including the CO-to-H$_2$ conversion factor and line ratios \citep{Zhu_etal_2003, Wilson_etal_2003, Schulz_etal_2007}, might have stronger impact on the Antennae galaxies than our other targets. \citet{Zhu_etal_2003} suggested that \alphaco\ could be a factor of $2$-$4$ smaller than the Galactic value in this interacting region \citep[but see e.g.,][]{Wilson_etal_2003}. If \alphaco\ is indeed smaller that the Galactic value that we adopt, then it means that the true \Sig\ values could be lower than our estimates and the points could shift leftwards in the \sig-\Sig\ space. Moreover, in such interacting system, many of the brightest regions show complex, multi-component line profiles \citep[][]{Herrera_etal_2012, Johnson_etal_2015}, which makes the CO line width measurements trickier. Our effective width approach de-emphasizes the component-to-component width of the line, but some spectral decomposition is likely necessary in future works.

\subsubsection{Low Surface Density Regime: M31~and~M33}
\label{sec:scaling:low-end}

\begin{figure*}[htb]
\fig{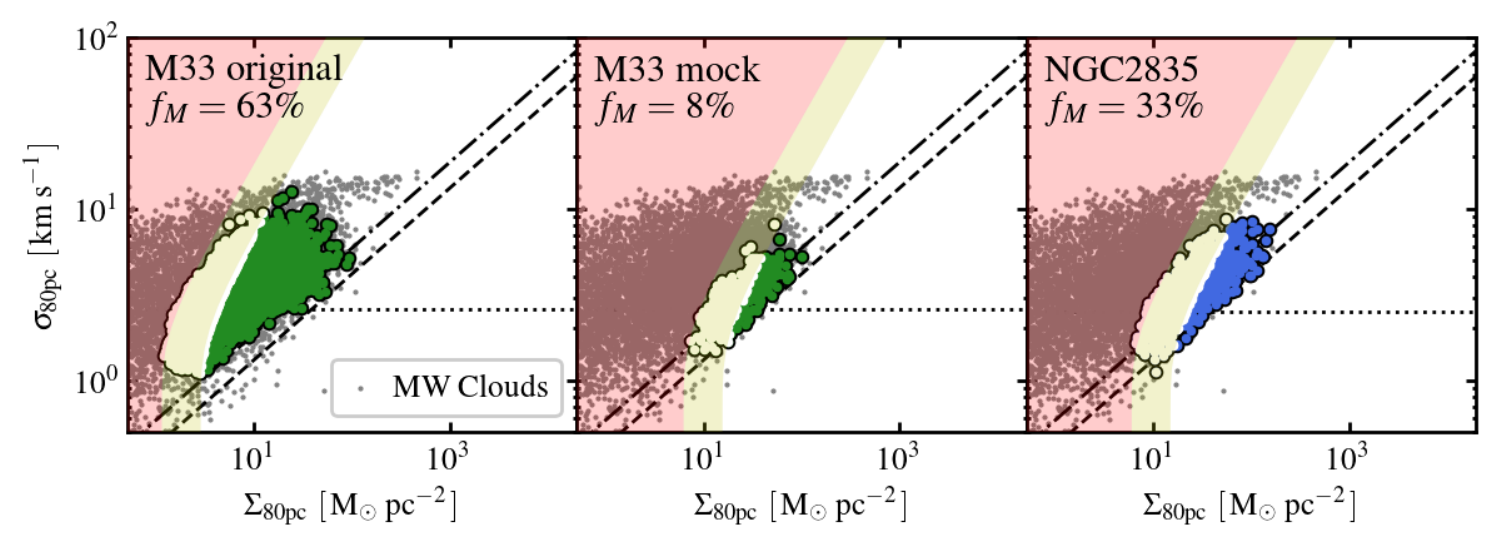}{0.85\textwidth}{}
\vspace{-2em}
\caption{{\it left panel}: Similar to the left panel in Figure \ref{fig:scaling-ext}, but here we show in the background the distribution of molecular clouds in the Milky Way as gray dots. We use the Milky Way molecular cloud catalog provided by \citet{MivilleDeschenes_etal_2017}, and apply additional beam dilution accordingly for clouds smaller than our beam size, in order to mimic the same effect in our work. {\it middle panel}: Measurement results in M33 on mock data cubes with their noise level elevated to match that in NGC~2835. The detected molecular gas mass fraction $f_M \approx 8\%$ in this case, even lower than the $f_M \approx 33\%$ in NGC~2835. The low detection fraction means that only the low \sig\ sightlines at highest \Sig\ are identified, pushing the derived \sig-\Sig\ relation close to the $\alphavir = 2$ line. {\it right panel}: Similar to the panel for NGC~2835 in Figure \ref{fig:scaling}.}
\label{fig:scaling-mock}
\end{figure*}

M31 and M33 represent low gas surface density, {\sc Hi} dominated environments. As illustrated in the left two panels of Figure~\ref{fig:scaling-ext}, both of these galaxies have sightlines showing much lower \Sig\ than targets in the main sample (also see Table~\ref{tab:Sig}), as well as higher \sig\ compared to an extrapolation of the $\sig$-$\Sig$ trend for the main sample. Such behavior suggests that the fixed-scale $\sig$-$\Sig$ relation deviates at low \Sig\ from the fixed-\alphavir\ lines.

The distinction between these two targets and other targets in our sample is largely due to the difference in their CO map depth. Because of the proximity of the Local Group targets, CO emission has been surveyed at relatively high sensitivity. Comparing the position of the yellow/red hatched, regions in the M31 or M33 panels in Figure~\ref{fig:scaling-ext} to any of the panels in Figure~\ref{fig:scaling}, one can see that the M31 and M33 sightlines probe a part of the \sig-\Sig\ parameter space that is inaccessible for our other targets at typically $10$-$20$ times the distance.

As a more quantitative way to illustrate this sensitivity effect, we degraded the M33 data to the sensitivity of our more distant targets. We add random noise into the M33 data cube to artificially degrade it, so that its noise level matches that in the data cube for NGC~2835, a galaxy with similar global properties. We then repeat our whole analysis (including regenerating the new mask for signal identification) on these degraded data cubes, and measure \sig\ and \Sig\ for all identified sightlines.

We show the result for these mock measurements in Figure \ref{fig:scaling-mock}. After we elevate the noise level, the recovered molecular gas mass faction in M33 drastically drops from 63\% to 8\%. This low detection rate agrees with the results from previous observations, in which case low sensitivity interferometer observations only recovered a small fraction (20\%) of the CO flux in the galaxy \citep[see][and reference therein]{Rosolowsky_etal_2007}.

The sightlines that we do detect come from a handful of dense, bright regions. The sensitivity limit imposes a selection bias against low \Sig\ and high \sig\ sightlines, and thus the only remaining detections are those with low \sig\ at the highest \Sig . Consequently, at this lower sensitivity, only the sightlines including the most apparently bound gas structures enter the analysis, and the \sig-\Sig\ relation becomes close to the $\alphavir = 2$ line.

As the low sensitivity and low detection rate could bias the inferred average \alphavir\ value, we expect that similar issues also affect our measurements of NGC~2835 and NGC~5068, the two PHANGS-ALMA objects with the faintest CO emission and lowest CO flux recovery fractions ($30$-$40$\% at 80~pc). Though we do not know for certain where the non-detected emission in these targets lies in the \sig-\Sig\ space, the most natural expectation would be that it occupies a similar part of the parameter space as it does in M31 and M33. We believe that the systematic deviation of M31 and M33 in the \sig-\Sig\ parameter space, as compared to the PHANGS-ALMA targets, implies a flattening of the $\sig \propto \Sig^{0.5}$ scaling that is common in low molecular gas density environments ($\Sig \lesssim 30\rm\;M_\odot\,pc^{-2}$). Limited sensitivity prevents such detections outside the Local Group.

Existing observations in the Milky Way do support this prediction. To show this, we take the molecular cloud catalog published by \citet{MivilleDeschenes_etal_2017} as a reference sample that is observed at much better sensitivity. To replicate the effect of beam dilution in our sample, for all Milky Way clouds that have their size $R$ smaller than our beam radius $r_{\rm beam} = 80$~pc, we calculate their ``beam-averaged'' surface density as $\Sig' = \Sig\;R^2 / r_{\rm beam}^2$, and plot their \sig-$\Sig'$ relation in Figure~\ref{fig:scaling-mock} as gray dots. These beam-diluted Milky Way data suggest that we would have also seen a large population of high $\sigma$ low $\Sigma$ measurements if we were to observe our own Galaxy at high sensitivity but fixed 80 pc spatial resolution. Note that this exercise does not take into account the possibility of catching multiple clouds in a beam, and thus the beam dilution effect might be slightly stronger than reality.

\subsubsection{Interpretations of \sig\ Excess at Low \Sig }\label{sec:scaling:discuss}

We observe an excess in \sig\ at low \Sig\ relative to the expected relation for self-gravity dominated gas. This behavior leads to a shallower $\sig$-$\Sig$ relation in several galaxies. It appears strongest in the Local Group galaxies, where most measurements appear at the low \Sig\ end and clearly deviate from an $\sig \propto \Sig^{0.5}$ relation extrapolated from the high \Sig\ regime. What mechanisms are responsible for enhancing \sig\ in low \Sig\ environments? We consider several possibilities:
1) the beam filling factor for CO emission might be lower in the low \Sig\ regime,
2) at low \Sig\ we might be underestimating the mass that sets the local gravitational potential, and
3) gas structures in this regime may be more susceptible to external pressure originating from the ambient medium and/or motions due to the galaxy potential.

{\it Low filling factor:} In low density environments, the overall number density of gas concentrations should be lower, and gas structures are also expected to be more compact (due to less shielding). This means that bright CO emission may fill a small fraction the beam, an effect that is commonly referred to as ``beam dilution''. As discussed in Section~\ref{sec:expectation}, this should lead to a shallow $\sig$-$\Sig$ relationship with $\beta < 0.5$ and a lower \Sig\ at a given \sig. In the limit of each beam capturing a single unresolved compact structure, the beam-averaged CO intensity (or gas surface density) may encode little or no information on the true surface density of the structure, but only reflect its total mass instead.

The fact that we see CO emission from these regions is itself supporting this interpretation. As discussed by \citet{Leroy_etal_2016}, given these \Sig\ values, the implied volume densities in these targets are low compared to the density needed to excite CO emission. This fact strongly implies that a substantial amount of sub-beam clumping must be present.

{\it Missing gas relevant to the gas self-gravity}:
Low density regions may preferentially harbor ``hidden'' molecular gas mass contributing to the gravitational potential but not captured by the observed CO emission. In this case the CO-to-H$_2$ conversion factor \alphaco\ should be correspondingly higher. We expect a higher \alphaco\ in low density regions because a larger portion of the gas sits in poorly shielded envelopes \citep[see e.g.,][]{Bolatto_etal_2013}. We do not attempt to correct for possible \alphaco\ variations in this work, and it might lead to an underestimation of \Sig\ preferentially in the low \Sig\ regime. Nevertheless, it is worth noting that the poor shielding might also lead to a higher gas temperature, which could lead to changes in \alphaco\ in the opposite direction \citep{Maloney_Black_1988}.

``CO-dark'' H$_2$ might not be the only missing part of the mass budget relevant to the local gravitational potential. Atomic gas could be well-mixed with molecular gas, at least near the edges of the molecular gas concentrations. This has been observed in at least one Galactic molecular cloud \citep[W43;][]{Bihr_etal_2015}, and is also naturally expected at least in M31 and M33 given their rich atomic gas content. 

{\it External pressure}: Kinetic pressure from the ambient ISM can increase the molecular gas velocity dispersion. This should happen when the external kinetic pressure becomes significant compared to the molecular gas's self-gravitational pressure (see Section~\ref{sec:expectation}). In this case, the gas structure in question may come to resemble a small part of a larger medium, approaching pressure equilibrium with its surroundings. The ambient gas pressure in the disk will define an isobar that our measurements will follow in the $\sig$-$\Sig$ space. In theory, this situation could occur in either molecular or atomic dominated regions, with diffuse gas of either type forming the ambient medium. In practice, we know that atomic gas in galaxy disks has a high volume filling factor and its typical surface density is $\sim 10\rm\;M_\odot\;pc^{-2}$, and we expect the pressure in this atomic gas to set some floor below which the molecular gas pressure should not fall.

The gas budget in M31 is dominated by atomic gas at all radii \citep{Braun_etal_2009}, while the ISM in M33 may be dominated by molecular gas only within the inner kpc \citep{Druard_etal_2014}. Both galaxies have atomic gas surface densities and line widths that reach values comparable to what we see in the molecular gas \citep{Braun_2012,Druard_etal_2014}. Given that the \sig\ and \Sig\ values we measure resemble those typically found for {\sc Hi}, the internal pressure in the molecular gas is likely comparable to the ambient medium pressure in these galaxies. Reinforcing this view, the pressures implied by our measurements approach the thermal pressures found by \citet{Herrera-Camus_etal_2017} in the {\sc Hi} dominated parts of KINGFISH galaxies. They showed that these thermal pressures are typically a factor of $\sim 3$ lower than the ambient gas pressure. This sets a strong expectation for the ambient kinetic pressure in these galaxies, and our molecular gas measurements approach this ``pressure floor.''

Similar behavior has been discovered in the outer Milky Way \citep{Heyer_etal_2001}, where the ISM is also diffuse and predominately atomic, and molecular gas appear tenuous. In a paper utilizing these same M31 and M33 data sets used here, \citet{Schruba_etal_2018} explicitly compare the inferred hydrostatic pressure in these galaxies to the measured molecular gas properties and argue that ambient pressure indeed plays a key role setting the dynamical state of molecular clouds in M31 and M33.

The gravitational potential associated with the stellar disk can be another possible source of the external pressure. \citet{Meidt_etal_2018a} show that for typical-sized clouds inside a galaxy with typical stellar densities, the in-plane and vertical motions due to the galaxy's potential can rival the velocity dispersions expected from a cloud's self-gravity. These motions would broaden the observed line profiles and could be expected to have their largest effects where the stellar density is high and gas density is low. Combined with the PHANGS-ALMA rotation curves (\citeinprep{P. Lang et al., in prep.}), our data should be ideal to test this scenario.

{\it Synthesis:} 
Physically, the relatively high line widths at low \Sig\ likely result from a combination of several effects. First, the low values of \Sig\ imply lower filling fractions of CO emission. For these two targets, the low peak brightness temperatures and stronger resolution dependence of our measurements support the argument that beam dilution must play a large role. 

Second, in low \Sig\ regions, CO emission may not track all of the gas mass relevant to the dynamical state. ``Missing'' gas might reside in either a CO-dark molecular phase or an atomic phase \citep[perhaps opaque, see][]{Braun_2012} that is well-mixed with molecular gas.

Third, the low internal pressure in these low density regions suggests that external pressure may play a strong role. Such external pressure could originate from either kinetic pressure in the ambient medium or the background galaxy potential.

A first step towards disentangling these mechanisms is to directly examine the correlation of cloud-scale gas properties with large-scale environment. This will be presented in the next set of PHANGS-ALMA papers. Comparison to the surrounding atomic gas and the galaxy gravitational potential could help identify the physics that drive this behavior. We refer the reader to the more detailed investigations in \citet{Hughes_etal_2013a}, \citet{Meidt_etal_2018a}, \citet{Jeffreson_Kruijssen_2018}, and \citet{Schruba_etal_2018}, all of which conclude that environmental factors likely play a role in determining the dynamical state of the gas int the low \Sig\ regime.

\subsection{The Virial Parameter}\label{sec:alphavir}

\begin{figure*}[htb]
\fig{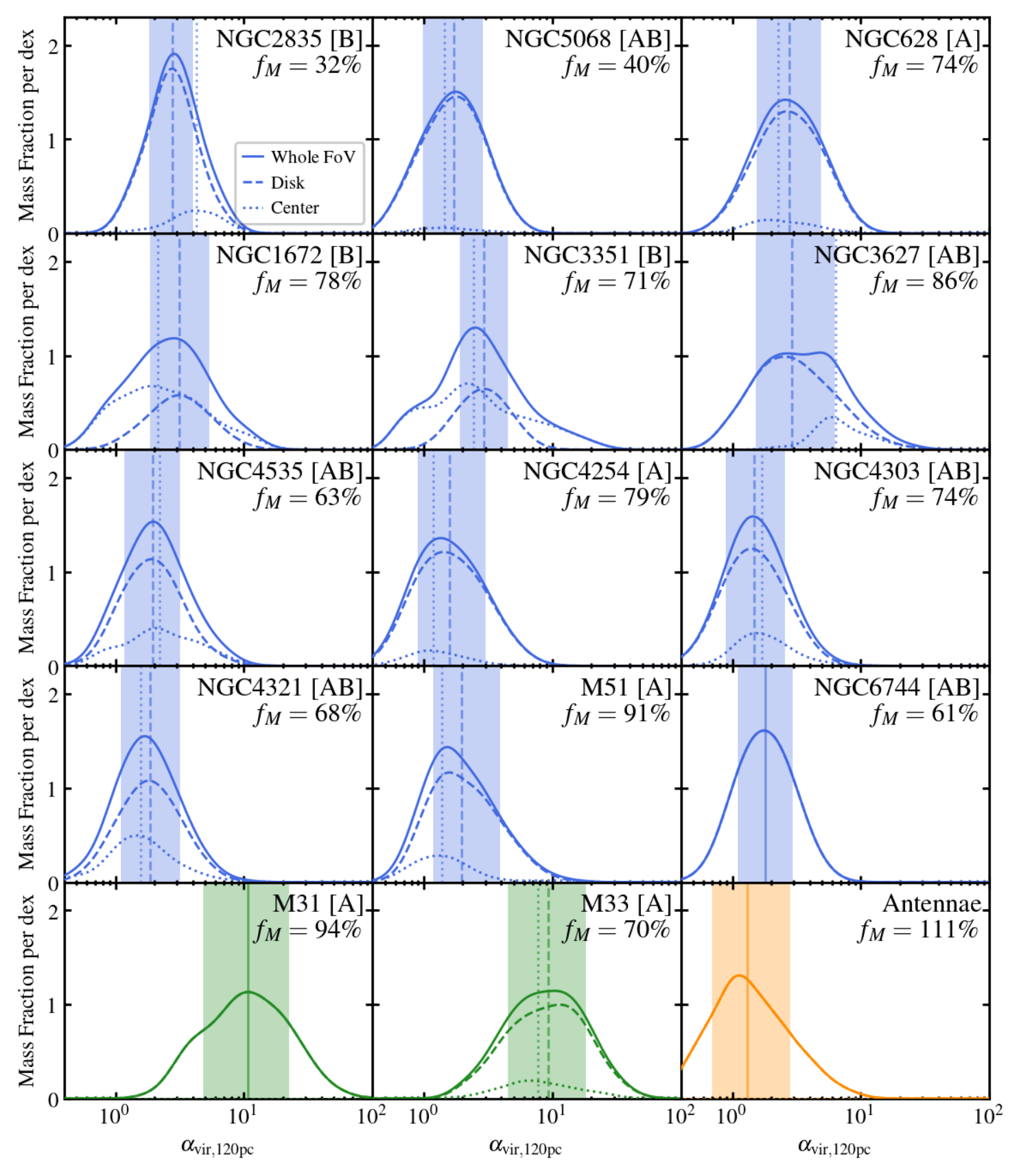}{0.85\textwidth}{}
\vspace{-2em}
\caption{
Molecular gas mass distribution as a function of \alphavir , measured for each galaxy at $120$~pc resolution. All curves are Gaussian KDE generated from the data with bandwidth of 0.1 dex. The solid line represents the distribution for gas in the whole field of view, while the dashed line represents the distribution for sightlines in the disk. The vertical dashed line and color shaded region show the median value and $16$-$84\%$ range of \alphavir\ for the disk distribution. All targets show narrow ($0.3-0.7$~dex) range in \alphavir\ , especially for their disk population. Most of them (except M31 and M33) have mass weighted median \alphavir\ values around $1.5$-$3.0$.}
\label{fig:alphavir}
\end{figure*}

\begin{deluxetable*}{lccccccccc}
\tabletypesize{\footnotesize}
\tablecaption{Properties of the \alphavir\ Distribution Function \label{tab:alphavir}}
\tablewidth{0pt}
\tablehead{
\colhead{Galaxy} & 
\multicolumn{3}{c}{at 45 pc resolution} &
\multicolumn{3}{c}{at 80 pc resolution} &
\multicolumn{3}{c}{at 120 pc resolution} \\
\cmidrule(lr){2-4} \cmidrule(lr){5-7} \cmidrule(lr){8-10}
\colhead{} &
\colhead{disk} &
\colhead{disk} &
\colhead{center} &
\colhead{disk} &
\colhead{disk} &
\colhead{center} &
\colhead{disk} &
\colhead{disk} &
\colhead{center} \\
\colhead{} &
\colhead{median} &
\colhead{16-84\%} &
\colhead{median} &
\colhead{median} &
\colhead{16-84\%} &
\colhead{median} &
\colhead{median} &
\colhead{16-84\%} &
\colhead{median} \\
\colhead{} &
\colhead{$\alphavir$} &
\colhead{width} &
\colhead{$\alphavir$} &
\colhead{$\alphavir$} &
\colhead{width} &
\colhead{$\alphavir$} &
\colhead{$\alphavir$} &
\colhead{width} &
\colhead{$\alphavir$} 
}
\startdata
NGC~2835 & 2.1 & 0.40 & 3.5 & 2.6 & 0.40 & 4.3 & 2.8 & 0.35 & 4.3 \\
NGC~5068 & 1.3 & 0.47 & 1.0 & 1.6 & 0.52 & 1.2 & 1.7 & 0.48 & 1.4 \\
NGC~628 & 2.7 & 0.41 & 3.0 & 2.9 & 0.48 & 2.8 & 2.8 & 0.51 & 2.3 \\
NGC~1672 & -- & -- & -- & -- & -- & -- & 3.2 & 0.47 & 2.1 \\
NGC~3351 & -- & -- & -- & 2.5 & 0.37 & 2.2 & 3.0 & 0.38 & 2.5 \\
NGC~3627 & -- & -- & -- & 3.1 & 0.62 & 6.3 & 2.9 & 0.63 & 6.3 \\
NGC~4535 & -- & -- & -- & -- & -- & -- & 1.9 & 0.45 & 2.2 \\
NGC~4254 & -- & -- & -- & -- & -- & -- & 1.6 & 0.55 & 1.2 \\
NGC~4303 & -- & -- & -- & -- & -- & -- & 1.5 & 0.47 & 1.7 \\
NGC~4321 & -- & -- & -- & -- & -- & -- & 1.9 & 0.47 & 1.6 \\
M51 & 2.4 & 0.47 & 1.9 & 2.1 & 0.51 & 1.6 & 1.9 & 0.53 & 1.4 \\
NGC~6744 & -- & -- & -- & 1.7 & 0.42 & -- & 1.8 & 0.44 & -- \\
M31 & 7.6 & 0.62 & -- & 10.2 & 0.67 & -- & 10.7 & 0.68 & -- \\
M33 & -- & -- & -- & 8.3 & 0.64 & 7.1 & 9.3 & 0.62 & 7.8 \\
Antennae & -- & -- & -- & 1.5 & 0.59 & -- & 1.3 & 0.61 & -- \\
\enddata
\tablecomments{For each galaxy at each resolution, we report:
(1) median \alphavir\ value {\it by gas mass} for the ``disk'' population;
(2) full width of the $16$-$84\%$ gas mass range of \alphavir\ distribution for the ``disk'' population (in units of dex); and
(3) median \alphavir\ value {\it by gas mass} for the ``center'' population.}
\end{deluxetable*}

Our fixed-scale \Sig\ and \sig\ measurements provide access to the dynamical state of molecular gas in each beam, commonly expressed via the virial parameter \alphavir . Figure~\ref{fig:scaling} and Table~\ref{tab:fit} show that, to first order, $\alphavir \sim \sig^2 / \Sig$ varies modestly across our targets. Meanwhile the complementary quantity, $\Pturb \sim \Sig\,\sig^2$ shows an enormous range (see Section~\ref{sec:Pturb}). Both quantities have physical significance, and $\alphavir-\Pturb$ represents a useful parameter space to diagnose the state of molecular gas. Specifically, most modern star formation theories predict a strong dependence of the star formation efficiency per free-fall time on the virial parameter and the turbulent pressure \citep[][among many others]{Kruijssen_2012,Krumholz_etal_2012,Hennebelle_Chabrier_2013,Padoan_etal_2017}. In this section and the next, we examine the distributions of \alphavir\ and \Pturb\ in our sample.

Following the discussion in Section~\ref{sec:expectation}, we take the beam size as the relevant size scale (i.e. $R = \rbeam$, thus $R = 40$ pc at $80$~pc resolution), adopt $f=10/9$ (appropriate for a density profile of $\rho (r) \propto r^{-1}$), and infer \alphavir\ following Equation~\ref{eq:alphavir}

\begin{equation} \label{eq:alphavir-obs}
\begin{split}
\alphavir
   &= \frac{5\,\sig^2\rbeam}{f G\,M}
   = \frac{5}{fG}\frac{\sig^2\rbeam}{\Sig\,A_{\rm beam}} 
   = \frac{5\ln{2}}{\pi fG}\frac{\sig^2}{\Sig\,\rbeam}\\
   &\approx 5.77 \left( \frac{\sig}{\rm km\,s^{-1}} \right)^2 \left( \frac{\Sig}{\rm \Msun\,pc^{-2}} \right)^{-1} \left( \frac{\rbeam}{40\rm\,pc} \right)^{-1}.
\end{split}
\end{equation}

\noindent While our adopted prefactors might be subject to systematic errors, this mainly renders the absolute value of \alphavir\ uncertain. We expect the relative sense of our inferred \alphavir\ values to be relatively robust.

\subsubsection{Disk and Center Distributions}

Figure~\ref{fig:alphavir} shows the distribution of molecular gas mass as a function of \alphavir , measured at 120~pc resolution. For each galaxy, we plot the distribution for gas in the whole field of view (solid curves) as well as that in only the disk region (dashed curves). Table~\ref{tab:alphavir} lists the median value and $16$-$84\%$ range of \alphavir\ in each galaxy.

Most galaxies (excluding M31 and M33) have distributions centered near $\alphavir \approx 1.5$-$3.0$, with $16$-$84\%$ width of $0.4$-$0.65$ dex. Because a fixed \alphavir\ corresponds to a slope $\beta = 0.5$ in Figure~\ref{fig:scaling}, the width of the distributions in Figure~\ref{fig:alphavir} captures a mixture of the scatter about the best fit relation and the deviation from $\beta = 0.5$. As we might expect from the results above, the distribution of \alphavir\ appears more uniform among the high mass star-forming galaxies in the main sample, when focusing only on their disks. In this case, the distribution resembles a log-normal but usually with a mild skew towards higher \alphavir\ values.

Compared to the large range in \Sig , \alphavir\ shows a narrow range of values across our sample. Specifically, while the median value of \Sig\ varies substantially from galaxy to galaxy, the median \alphavir\ varies much less, suggesting that most molecular gas shares a common dynamical state across massive, star-forming disk galaxies. As shown above and in the literature, the variations in \alphavir\ that do exist can be linked to environment. For example, studies by the PAWS survey have shown a link between \alphavir\ and dynamical environment in M51 \citep{Meidt_etal_2013, Colombo_etal_2014a}. But in the disks of most of our target galaxies, the strong correlation between \Sig\ and \sig\ with small scatter implies that the variation in \alphavir\ is much smaller than variations in the surface density, line width, and their combination -- the internal turbulent pressure (see Section~\ref{sec:Pturb}).

We find more variation in \alphavir\ when including the galaxy centers. Figure~\ref{fig:alphavir} shows a wide distribution of \alphavir\ associated with the central regions of NGC~1672 and NGC~3351. Another two barred galaxies, NGC~2835 and NGC~3627, show relatively high \alphavir\ in their central regions compared to disks (by $25\%$ to $100\%$). This captures the displacement of the $\sig$-$\Sig$ relation to higher line width in galaxy centers. As emphasized above, the line width in central regions may include contributions from bulk gas motions like rotation. Therefore interpretation of these high \alphavir\ should be made with some caution. Nevertheless, these line widths are consistent with the idea that the presence of a bar leads to efficient turbulence driving within the inner Lindblad resonance \citep[e.g.,][]{Krumholz_Kruijssen_2015}.

The peak positions of the \alphavir\ distributions lie between $1.5$-$3.0$, given $f = 10/9$ and $R = \rbeam$. As discussed in Section~\ref{sec:expectation}, marginally bound or free-falling gas should have $K \approx U_g$ (i.e., ``energy equipartition''), which implies $\alphavir \approx 2$. Thus our inferred \alphavir\ values indicate $K \sim U_g$ but with a modest excess of turbulent kinetic energy compared to self-gravitational potential energy. However, considering the uncertainties in the pre-factors on \alphavir\ (Equation~\ref{eq:alphavir-obs}), which reflects our limited ability to infer an absolute molecular mass and calculate the true gravitational potential, our derived absolute values of \alphavir\ are correspondingly uncertain. We suggest to read Figure~\ref{fig:alphavir} as showing a remarkable degree of uniformity in the dynamical state of molecular gas across the disks of galaxies and among galaxies, given matched assumptions and treatment. Given the associated uncertainties, our measurement does not distinguish between the various physical states dominated by self-gravity (e.g., marginally bound, spherical free-fall, hierarchical collapse). In Section~\ref{sec:discussion}, we discuss the path towards accurate \alphavir\ estimates.

\subsubsection{Outliers}

Again M31 and M33 stand out. They show most of their mass at high \alphavir\ with large scatter, reflecting the high line widths and shallow scaling relation that we observed for these targets in the previous section. See the discussion in \S \ref{sec:scaling:low-end} for likely explanations for this behavior. In short, we see this high \alphavir\ gas only in these targets because of the high sensitivity of the maps. The high \alphavir\ likely results from a mixture of beam filling effects, ``missing'' gas not included in the CO-based \alphavir\ estimate, and the influence of external pressure.

\subsubsection{Scale Dependence}

\begin{figure}[htb]
\fig{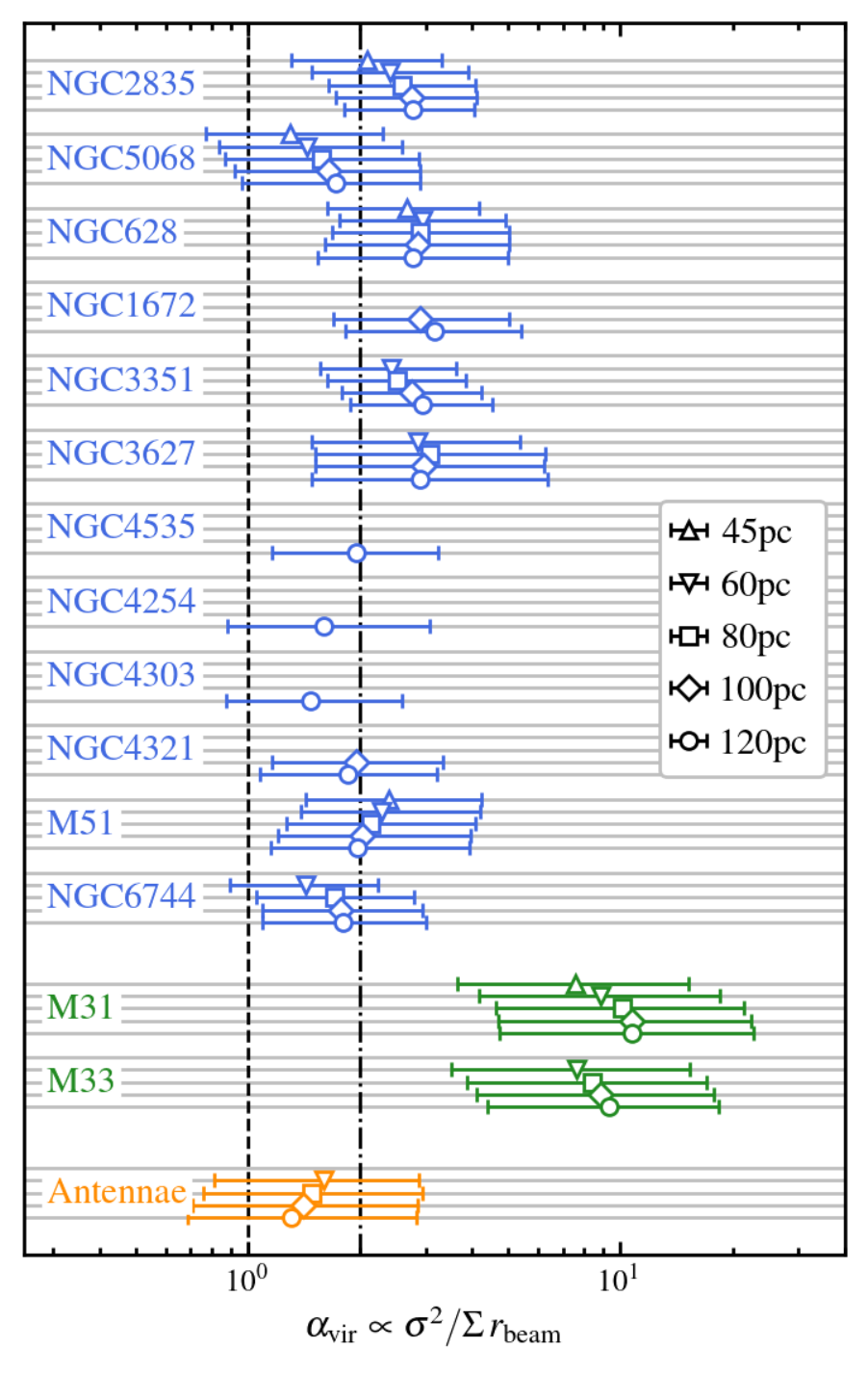}{0.47\textwidth}{}
\vspace{-3em}
\caption{Mass weighted distribution of measured \alphavir\ for all galaxy disks at all of the available resolutions (shown by different symbols). The horizontal position of the symbols and their associated error-bars show the median by mass and $16$-$84\%$ range in \alphavir , respectively. We find the scale dependence of our \alphavir\ measurements to be mild compared to both the distribution width and the factor of 2.7 change in linear resolution.}
\label{fig:scale-alphavir}
\end{figure}

Comparing \alphavir\ at different resolutions can reveal the clumpiness and kinematic structure of molecular gas across physical scales. For example, in a homogeneous turbulent medium with a fiducial velocity dispersion--size relation $\sig \propto r^{0.5}$, we expect \alphavir\ to show little scale dependence (as the \sig\ and \rbeam\ dependences cancel out by each other in Equation~\ref{eq:alphavir-obs}). In the other extreme, i.e. in the case of a single isolated cloud unresolved at any accessible resolution, beam dilution will cause \Sig\ to scale with beam size as $\rbeam^{-2}$, and thus $\alphavir \propto \Sig^{-1} \rbeam^{-1} \propto \rbeam$.

In Figure~\ref{fig:scale-alphavir}, we compare the \alphavir\ distribution in the disks of our targets at different resolutions. We consider each galaxy at $45$, $60$, $80$, $100$, and $120$~pc resolution (when accessible) and take the beam size as the relevant length scale. As in Figure~\ref{fig:alphavir}, we show the median by mass and $16$-$84\%$ mass range of \alphavir . We highlight $\alphavir=1$ and $\alphavir=2$ as dashed and dash-dot vertical lines.

Though both \sig\ and \Sig\ depend on spatial scale, the inferred gas dynamical state varies much more weakly. A few of our targets (NGC~2835, NGC~5068, NGC~3351, NGC~6744, M31, and M33) show higher median \alphavir\ at coarser resolution, while the others show no trend or sometimes an inverse trend. These trends are statistically significant due to the large number of independent measurements. However, the changes in the median \alphavir\ never exceed $0.2$~dex across available resolutions, which is mild compared to the $0.4$~dex change in beam size (from $45$ to $120$~pc). Thus, the observed scale dependence in \alphavir\ is much weaker than the expected $\alphavir \propto \rbeam$ for unresolved isolated clouds, and in at least half of our targets, \alphavir\ shows no or inverse correlation with the sampling scale.

As mentioned above, this analysis is subject to observational selection effects because the fraction of molecular gas mass that enters our calculation decreases as we move from coarser to finer resolution (Table~\ref{tab:stats}). Because the total gas mass being analyzed is not conserved across resolutions, an artificial scale dependence can be introduced by systematically excluding more gas with higher \alphavir\ at finer resolution. In Section~\ref{sec:scaling} we measure the slope of the $\sig$-$\Sig$ relation to be $\beta < 0.5$, which means that gas in low \Sig\ regions tends to have higher \alphavir . Consequently, losing these low \Sig\ sightlines in finer resolution maps can bias the median \alphavir\ to lower values, and thus artificially produce a positive correlation between \alphavir\ and \rbeam .

Taking these observational biases into account, our best estimate is that our observations do not provide strong evidence for a dependence of \alphavir\ on spatial scale. Our measurements appear to capture gas with kinetic energy similar to but slightly greater than the self-gravitational potential energy across the scales that we consider (with the exception of M31 and M33).

\subsection{Internal Turbulent Pressure}\label{sec:Pturb}

\begin{figure*}[htb]
\fig{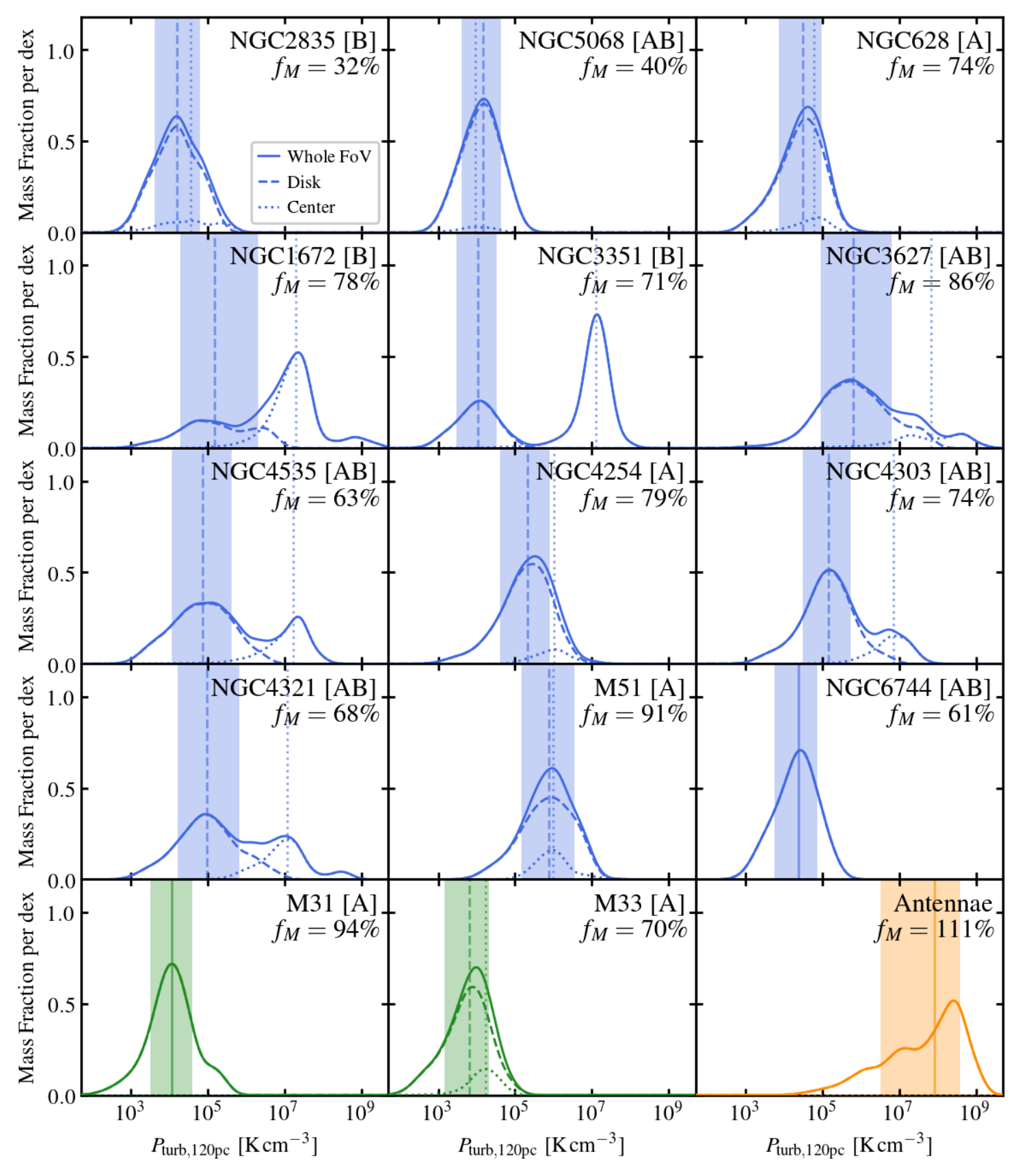}{0.85\textwidth}{}
\vspace{-2em}
\caption{Molecular gas mass distribution as a function of \Pturb , measured for each galaxy at $120$~pc resolution. Curves are Gaussian KDE generated from the data with bandwidth of $0.1$~dex. The solid line represents the distribution for gas in the whole field of view, while the dashed and dotted line represents the distribution for gas in the disk and the central region, respectively. The vertical dashed line and color shaded region show the median value and $16$-$84\%$ range of \Pturb\ for the disk distribution.
The measured \Pturb\ values span decades in each galaxy. NGC~3351 and NGC~3627 show very high \Pturb\ values in their centers, which form distinct peaks at the high end of their \Pturb\ distribution.}
\label{fig:Pturb}
\end{figure*}

\begin{deluxetable*}{lccccccccc}
\tabletypesize{\footnotesize}
\tablecaption{Properties of the \Pturb\ Distribution Function \label{tab:Pturb}}
\tablewidth{0pt}
\tablehead{
\colhead{Galaxy} & 
\multicolumn{3}{c}{at 45 pc resolution} &
\multicolumn{3}{c}{at 80 pc resolution} &
\multicolumn{3}{c}{at 120 pc resolution} \\
\cmidrule(lr){2-4} \cmidrule(lr){5-7} \cmidrule(lr){8-10}
\colhead{} &
\colhead{disk} &
\colhead{disk} &
\colhead{center} &
\colhead{disk} &
\colhead{disk} &
\colhead{center} &
\colhead{disk} &
\colhead{disk} &
\colhead{center}\\
\colhead{} &
\colhead{median} &
\colhead{16-84\%} &
\colhead{median} &
\colhead{median} &
\colhead{16-84\%} &
\colhead{median} &
\colhead{median} &
\colhead{16-84\%} &
\colhead{median} \\
\colhead{} &
\colhead{$\log_{10}\Pturb$} &
\colhead{width} &
\colhead{$\log_{10}\Pturb$} &
\colhead{$\log_{10}\Pturb$} &
\colhead{width} &
\colhead{$\log_{10}\Pturb$} &
\colhead{$\log_{10}\Pturb$} &
\colhead{width} &
\colhead{$\log_{10}\Pturb$}
}
\startdata
NGC~2835 & 4.88 & 1.13 & 5.39 & 4.45 & 1.20 & 4.85 & 4.21 & 1.21 & 4.56 \\
NGC~5068 & 4.70 & 0.99 & 4.66 & 4.34 & 1.07 & 4.18 & 4.16 & 1.03 & 3.97 \\
NGC~628 & 5.05 & 1.22 & 5.18 & 4.70 & 1.21 & 4.94 & 4.49 & 1.14 & 4.77 \\
NGC~1672 & -- & -- & -- & -- & -- & -- & 5.17 & 2.04 & 7.30 \\
NGC~3351 & -- & -- & -- & 4.28 & 1.08 & 7.20 & 4.04 & 1.05 & 7.11 \\
NGC~3627 & -- & -- & -- & 5.95 & 1.93 & 8.11 & 5.81 & 1.85 & 7.82 \\
NGC~4535 & -- & -- & -- & -- & -- & -- & 4.87 & 1.59 & 7.23 \\
NGC~4254 & -- & -- & -- & -- & -- & -- & 5.33 & 1.29 & 6.02 \\
NGC~4303 & -- & -- & -- & -- & -- & -- & 5.16 & 1.26 & 6.85 \\
NGC~4321 & -- & -- & -- & -- & -- & -- & 4.98 & 1.62 & 7.06 \\
M51 & 6.30 & 1.50 & 6.26 & 6.05 & 1.46 & 6.07 & 5.88 & 1.40 & 5.99 \\
NGC~6744 & -- & -- & -- & 4.58 & 1.13 & -- & 4.37 & 1.12 & -- \\
M31 & 4.35 & 1.23 & -- & 4.18 & 1.17 & -- & 4.08 & 1.10 & -- \\
M33 & -- & -- & -- & 3.98 & 1.21 & 4.29 & 3.81 & 1.17 & 4.23 \\
Antennae & -- & -- & -- & 8.02 & 2.11 & -- & 7.92 & 2.10 & -- \\
\enddata
\tablecomments{For each galaxy at each resolution, we report:
(1) median $\log_{10}\Pturb$ value {\it by gas mass} for the ``disk'' population (in units of $\rm\,K\,cm^{-3}$);
(2) full width of the $16$-$84\%$ gas mass range of \Pturb\ distribution for the ``disk'' population, (in units of dex); and
(3) median $\log_{10}\Pturb$ value {\it by gas mass} for the ``center'' population (in units of $\rm\,K\,cm^{-3}$).}
\end{deluxetable*}

Besides the virial parameter, our measurements also allow us to infer the internal turbulent pressure, \Pturb , or equivalently the kinetic energy density, in the molecular gas. Comparing \Pturb\ to self-gravity, the external pressure in the ambient medium, and disk structure offers more insights into the dynamical state of the molecular ISM.

The internal pressure in molecular gas with line of sight depth ${\sim}2R$ can be expressed as

\begin{equation} \label{eq:pressure}
\Pturb \approx \rho \sig^2 \approx \frac{1}{2R} \Sig \sig^2.
\end{equation}

\noindent Under the assumption of $R=\rbeam$, we estimate \Pturb\ from\footnote{We point out that this equation, together with Equation~\ref{eq:alphavir-obs}, means that $\Pturb\propto\alphavir\,\Sigma^2$. Given the observed narrow distribution of \alphavir\ across our main sample, \Pturb\ will mostly reflect the variation in $\Sigma$ in these galaxies. Nevertheless, as described later in this section and in Section~\ref{sec:discussion}, there are several reasons for us to believe that reporting \Pturb\ here is valuable.}

\begin{equation} \label{eq:p-parameter}
\Pturb/k_B \approx 61.3\,{\rm K\,cm^{-3}} \left( \frac{\Sig}{\rm \Msun\,pc^{-2}} \right) \left( \frac{\sig}{\rm km\,s^{-1}} \right)^2 \left( \frac{\rbeam}{40\rm\,pc} \right)^{-1}.
\end{equation}

Figure~\ref{fig:Pturb} shows the distribution of \Pturb\ measured at $120$~pc scale in all 15 galaxies. In each panel, the distribution for the entire field of view is shown by the solid curve, while that for the disk and center regions are shown by the dashed and dotted curves. The measured median values and widths of the \Pturb\ distribution are reported in Table~\ref{tab:Pturb}.

Based on the functional form of Equation~\ref{eq:pressure} and the observed $\sig$-$\Sig$ correlation, one would expect the properties of the \Pturb\ distributions to resemble those of \Sig\ and \sig . Indeed, in Figure~\ref{fig:Pturb} we see a qualitatively similar behavior to that in Figures~\ref{fig:Sig-80pc}--\ref{fig:sig}, including the multimodal shape of the distribution functions and the contrast between disk and central populations.

Lines of fixed turbulent pressure \Pturb , a.k.a. isobars, appear as diagonal lines ($\sig \propto \Sig^{-0.5}$) running from top left to bottom right in Figures~\ref{fig:scaling}. These lines lie almost perpendicular to the actual distribution of data. In other words, \Pturb\ represents the reprojection of the observed $\sig$-$\Sig$ distribution along its principal axis. Sightlines located up and to the right in the plot (e.g., galaxy centers or the densest regions in disks) correspondingly have larger $\Sig\,\sig^2$, and thus higher turbulent pressure.

As expected, we measure the $16$-$84\%$ range for \Pturb\ to be wide even in individual galaxy disks (typical width are $\sim 1$-$2$\,dex). The distributions for the centers of strongly barred galaxies sometimes show $2$-$3$ orders of magnitude excess in median \Pturb , which makes them appear as a distinct peak in the \Pturb\ distribution plots (NGC~1672, NGC~3351, NGC~3627, NGC~4535, NGC~4303, NGC~4321). Such large dynamical range of turbulent pressure in the molecular ISM has been claimed for both the Milky Way and other nearby galaxies~\citep[see][for compilations]{Hughes_etal_2013a, Leroy_etal_2015,Schruba_etal_2018}.

The dynamical range in \Pturb\ across our sample is further boosted by strong inter-galaxy variations. We find $\Pturb \sim 10^3$-$10^5 \rm\;K\,cm^{-3}$ in the disks of low mass galaxies, while $\Pturb \sim 10^5$-$10^7 \rm\;K\,cm^{-3}$ for high mass ones. Such a correlation between \Pturb\ and galaxy stellar mass can be either a consequence of the $\Sig$-$M_\star$ correlation we mentioned in Section~\ref{sec:dist}, or a more fundamental relation that links the local gas properties to the ambient galactic environment. This topic will be investigated in a forthcoming paper. Here we emphasize that \Pturb\ varies dramatically within individual galaxies, and strongly and systematically among galaxies. The internal pressure of a patch of molecular gas is clearly a strong function of environment.

\section{Discussion}
\label{sec:discussion}

\begin{figure*}[htb]
\fig{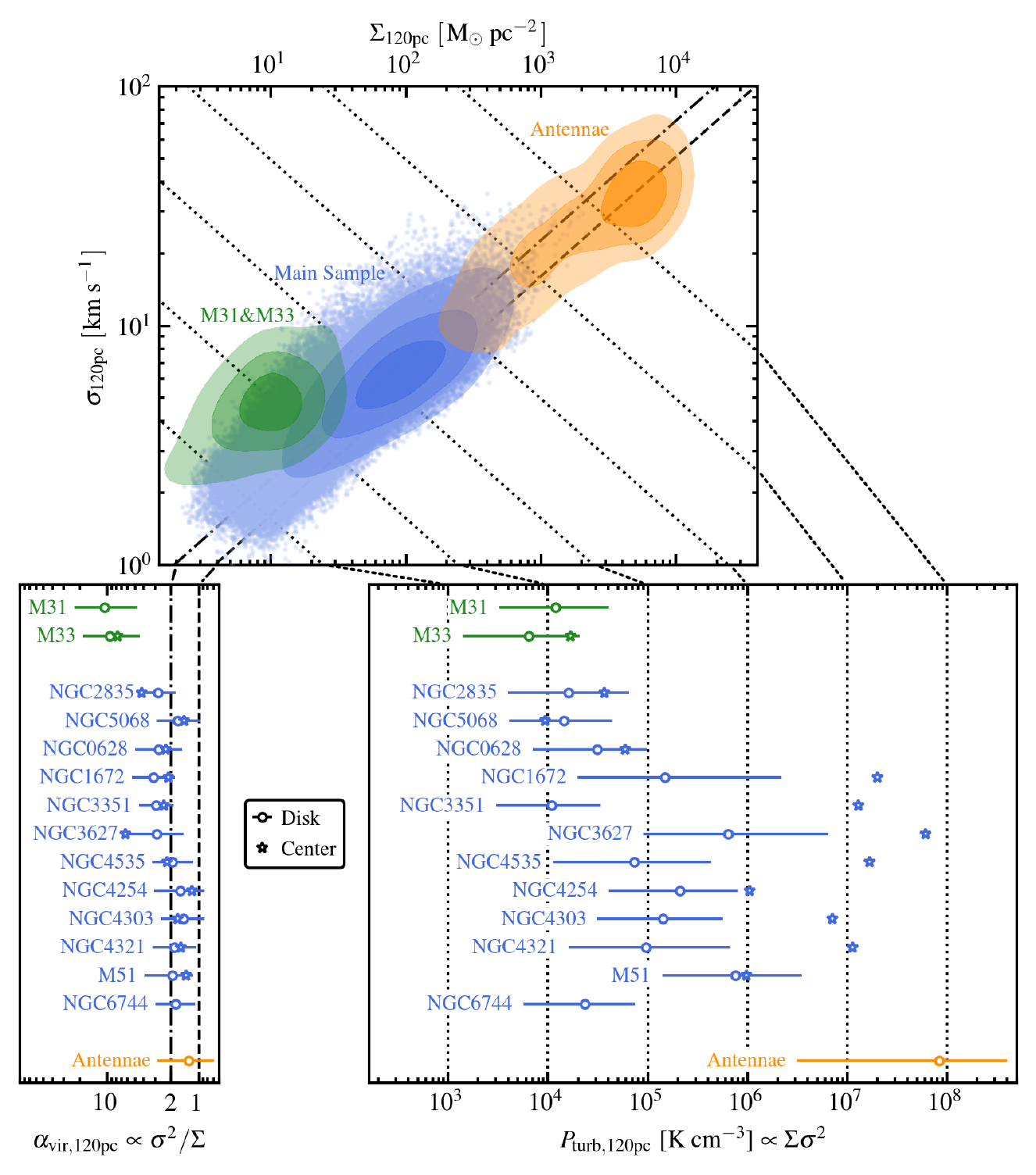}{0.7\textwidth}{}
\vspace{-2em}
\caption{
This figure summarizes the main results of this work. In the top panel, we show the measured \sig-\Sig\ scaling relation at 120~pc scale ($\rbeam = 60$~pc) for the disks of the 12 galaxies in the main sample (blue dots and contours), M31 and M33 disks (green contours), and the interacting region of the Antennae galaxies (orange contours) separately. Contours here show the mass weighted data density at levels including 20\%, 50\%, and 80\% of total gas mass. We show grid-lines indicating canonical values of \alphavir\ (dashed and dash-dot lines representing $\alphavir = 1$ and $2$ respectively) and fixed \Pturb\ values (dotted lines representing $\Pturb = 10^3$-$10^8\rm\;K\,cm^{-3}$ with 1~dex spacing) in the background, emphasizing the observed spread in molecular gas dynamical state and internal turbulent pressure. The galaxy-by-galaxy, mass weighted \alphavir\ distribution is presented in the bottom left panel (note the reversed abscissa), and the corresponding plot for \Pturb\ shown in the bottom right panel. In both panels, we group and color-code galaxies using the same scheme as for the top panel. We label galaxies by their names, and order the targets in the main sample by their total stellar mass. For each galaxy, we show the mass weighted median \alphavir\ or \Pturb\ in the galaxy center (star symbols), as well as the distribution observed in the disk (circles with error bars representing the mass weighted median and 16-84\% range). See the first paragraph in Section \ref{sec:discussion} for a brief summary.}
\label{fig:summary}
\end{figure*}

Figure \ref{fig:summary} summarizes many of our key findings. We show the line width as a function of surface density (top panel), as well as the mass weighted distribution of virial parameter (bottom left panel) and turbulent pressure (bottom right panel), for all targets at $120$~pc resolution. Molecular gas in these galaxies exhibits a wide range of surface density \Sig\ and velocity dispersion \sig\ at this fixed spatial scale. \Sig\ and \sig\ follow a universal scaling relation across the 12 galaxies in the main sample (blue dots and contours), while the two Local Group galaxies M31 and M33 (green contours) manifest a curvature towards shallower slope at low \Sig . Gas in the interacting region of the Antennae galaxies (orange contours) lies close to the extrapolated \sig-\Sig\ relation but shifted to higher apparent internal pressure. The observed molecular gas mass distribution in this \sig-\Sig\ parameter space implies a wide dynamical range of gas internal turbulent pressure \Pturb , and a narrow range of virial parameter \alphavir\ across our sample.

\subsection{Comparison with Previous Studies}

This work reveals that molecular gas surface density \Sig\ shows a wide distribution across the local star-forming galaxy population at fixed spatial scales. Recent molecular cloud surveys also find variations in the cloud average surface density across galactic and extragalactic systems \citep[][among many others]{Heyer_etal_2009,DonovanMeyer_etal_2013,Colombo_etal_2014a,Leroy_etal_2015,Egusa_etal_2018}. 

The \sig-\Sig\ scaling relation we find under the fixed-scale framework (see Figure \ref{fig:summary}) is the analog of the $\sig^2/R \propto \Sig$ relation frequently quoted in recent cloud studies. This relation is often discussed in the context of molecular clouds in virial equilibrium. However, subsequent works have suggested alternative dynamical states, including marginally bound, free-falling \citep{BallesterosParedes_etal_2011,IbanezMejia_etal_2016,Camacho_etal_2016}, or pressure-confined clouds \citep{Heyer_etal_2001,Oka_etal_2001,Field_etal_2011,Hughes_etal_2013a,Schruba_etal_2018}. Each of these scenarios predicts a particular value of \alphavir\ or some dependence of \alphavir\ on environment.

We found typical values of $1.5$-$3.0$ for the virial parameter $\alphavir \propto \sig^2/\Sig$, which implies that the molecular ISM often possesses slightly more kinetic energy than gravitational potential energy at the scales we study. However, uncertainties in the pre-factors and sub-beam distribution make the normalization of \alphavir\ a poor discriminant among these ideas. A main goal of this paper is to lay the groundwork for direct correlation of molecular gas structure with disk structure, kinematics, and host galaxy properties. We expect such analysis to help distinguish among these scenarios.

We do not see evidence in our sample for the strikingly low line widths recently observed in the inner regions of two early-type galaxies by \citet{Davis_etal_2017,Davis_etal_2018}. As they discuss, their measurements would fall far below our measured \sig-\Sig\ scaling relation, which implies $\alphavir < 1$. This region appears almost totally unpopulated by data in our sample. Their measurements carefully accounted for the rotation of the gas disk and had higher spatial resolution ($29$ and $13$~pc, respectively) than we achieve here. The easiest explanation to reconcile the two sets of observations seems to be invoking differences in sub-beam structure or line of sight depth in one or both data sets, but several other effects (e.g., missing short-spacing information, choice of CO excitation line) may also play a role. This remains a topic for future research.

We observe a wide range of turbulent pressure $\Pturb \propto \Sig\,\sig^2$ across our sample. This wide spread in \Pturb\ agrees with previous observations, which reveal vastly different internal pressures in different parts of the Milky Way \citep[e.g.,][]{Bertoldi_McKee_1992,Heyer_etal_2001,Oka_etal_2001,Field_etal_2011}, and in selected extragalactic systems across the cosmic distance scale \citep[e.g.,][]{Swinbank_etal_2012,Kruijssen_Longmore_2013,Livermore_etal_2015}. There appears to be no doubt that different galaxies drive their molecular gas to vastly different pressures at cloud scales.

Though not a unique explanation, a correlation of \Pturb\ with environment might be expected if molecular clouds represent over-densities in a self-regulated disk. In self-regulated disk models, gas maintains radial \citep[e.g.,][]{Silk_1997} or vertical \citep[e.g.,][]{Elmegreen_1989} force balance in the galaxy potential. The ISM regulates itself around this equilibrium state, with feedback from star formation coupling the large scale gas velocity dispersion and density. In recent years, this idea has been explored in a series of works using increasingly sophisticated simulations and analytic theories \citep{Koyama_Ostriker_2009,Ostriker_etal_2010,Ostriker_Shetty_2011,Kim_etal_2011,Kim_etal_2013,Kim_Ostriker_2015}.

Focusing on equilibrium in the vertical direction (i.e., perpendicular to the disk), these theories predict that over large spatial scales and long time scales, internal pressure in the ISM will balance its weight inside the disk gravitational potential. In this case, the large-scale dynamic equilibrium pressure, $\PDE$, may represent some local baseline value that is also relevant to the internal pressure of molecular clouds. This equilibrium pressure correlates positively with the stellar and gas surface densities, and thus is higher in massive galaxies and the inner parts of galaxies \citep[][]{Wong_Blitz_2002,Blitz_Rosolowsky_2004,Blitz_Rosolowsky_2006,Leroy_etal_2008}.

If $\PDE$ represents a long term and large scale average, we expect that the molecular gas that we observe represents an over-pressurization relative to this environment-dependent baseline value. This is qualitatively consistent with our observed trend of higher \Pturb\ in high stellar mass galaxies and in galaxy centers, but it also can be directly tested by combining our data with measurements of local disk structure at various spatial scales. We will present a quantitative investigation of this scenario in a forthcoming paper.

\subsection{Stepping Back to the Observable Quantities}

In the previous sections, we present our results in terms of physical quantities, such as molecular gas surface density, \Sig , and velocity dispersion, \sig . These quantities are inferred from two direct observables, the surface brightness and effective width of a low-$J$ CO line, mostly CO~(2-1).
The simple translations between CO line measurements and physical quantities also means that our results can be easily converted back to their observable form whenever necessary. This, together with our fixed-spatial-scale approach, allows for a straightforward comparison between our observations and predictions of the structure of CO emission from simulations of galaxies or analytic theory. Given the recent progress in modeling CO emission from simulations \citep[e.g.,][M. Gong et al. submitted]{Smith_etal_2014,Pan_etal_2015,Duarte-Cabral_Dobbs_2016}, we expect this to be a fruitful application of our measurements in the next few years. In this sense, our results represent the current best available measurements of the distribution of CO emission from star-forming disk galaxies at $45$-$120$~pc resolution, and should offer a valuable benchmark to identify the physics needed to produce a realistic molecular medium.

\begin{figure}[htb]
\fig{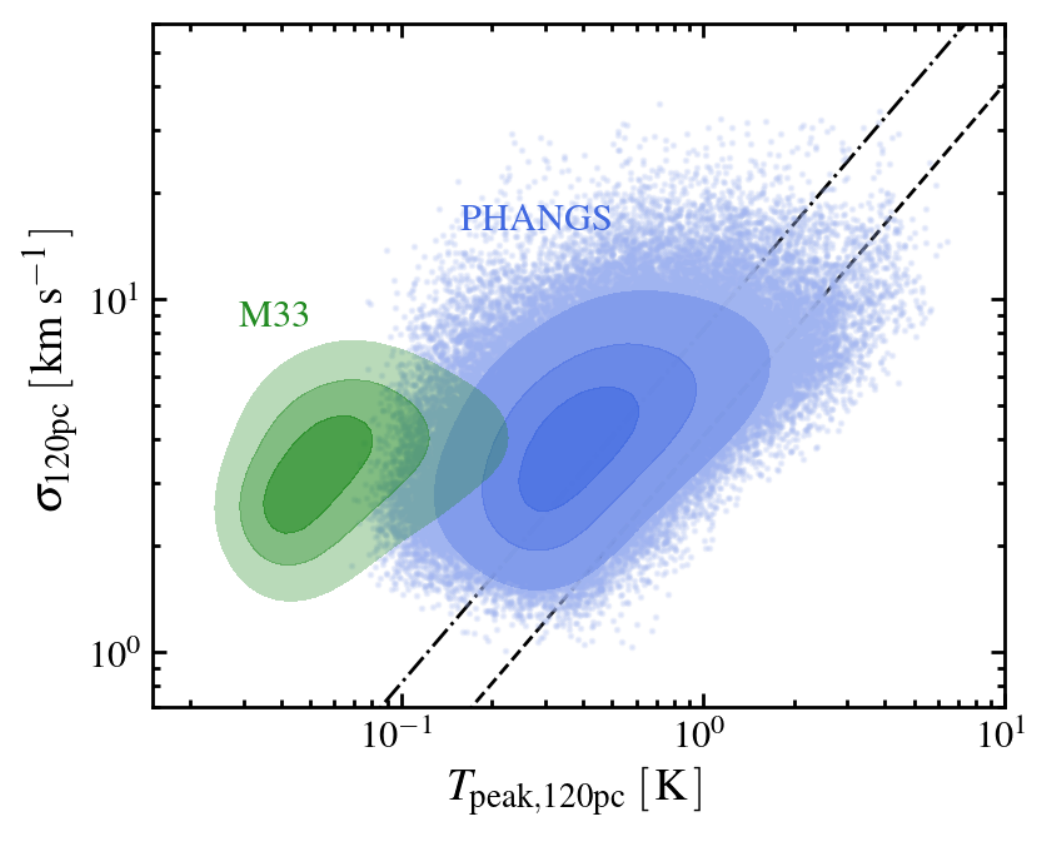}{0.46\textwidth}{}
\vspace{-2em}
\caption{
\sig-\Tpeak\ relation at 120~pc scale for all the 12 targets observed in CO~(2-1) line emission (11 PHANGS galaxy disks as blue dots and contours, M33 disks as green contours). Contours here show the data density levels including 20\%, 50\%, and 80\% of all measurements.
We show the expected $\sig \propto \Tpeak$ relation from $\alphavir = 1$ and $2$ as black dashed and dot-dash lines in the background.
Correlation between \sig\ and \Tpeak\ across the PHANGS sample is consistent with the expected $\sig \propto \Tpeak$ relation from an intrinsic $\sig \propto \Sig^{0.5}$ relation.
Measurements in M33 show very low \Tpeak , signifying the presence of severe beam dilution.}
\label{fig:sig-Tpeak-summary}
\end{figure}

In terms of observables, the correlation shown in Figure \ref{fig:scaling} is fundamentally a correlation between CO line-integrated intensity \Ico\ and effective width. Since the effective width is directly proportional to \Ico\ by construction (Equation~\ref{eq:ew}; ignoring the insignificant broadening correction for this discussion), one may wonder how much of the observed $\sig$-$\Sig$ correlation results from this by construction correlation. To address this question, we show in Figure~\ref{fig:sig-Tpeak-summary} the correlation between peak temperature \Tpeak\ and line width \sig , which are statistically independent quantities for resolved line profiles. Here we only include the 12 targets with CO~(2-1) data to make a fair comparison.

Given that $\Sig \propto \Ico \propto \sig \Tpeak$, an intrinsic $\sig \propto \Sig^{0.5}$ relation should manifest itself as a slope=1 correlation between effective width \sig\ and line peak intensity \Tpeak\ (a.k.a. ``peak temperature''). Conversely, if the $\sig$-$\Sig$ relation is purely artificial, we expect no correlation between \sig\ and \Tpeak . The observed correlation across the PHANGS sample is indeed consistent with the expectation of $\sig \propto \Tpeak$, although the relationship has larger scatter and, as shown in Appendix~\ref{apdx:sig-tpeak}, suffers from more severe selection effects than the $\sig$-$\Sig$ correlation. This result confirms that our observed line width-intensity relation is not artificial in nature, but rather conveys physical information about cloud-scale molecular gas kinematics. We show the galaxy-by-galaxy $\sig$-$\Tpeak$ relation in Appendix~\ref{apdx:sig-tpeak}.

As a side note, The low \Tpeak\ found in M33 (widely below $0.1$~K) further illustrates the existence of severe beam dilution effect in this galaxy. As we have mentioned in Section~\ref{sec:expectation} and Section~\ref{sec:scaling:low-end}, along with the influence of missing gas and external pressure \citep{Schruba_etal_2018}, this beam dilution likely contributes to producing the shallow \sig-\Sig\ relation at low \Sig .

\subsection{Key Caveats and Next Steps}

Much of our interpretation relies on the adopted CO-to-H$_2$ conversion factor, \alphaco , for the appropriate transition. We do expect \alphaco\ to vary between the central regions and disks of our targets \citep{Israel_2009,Sandstrom_etal_2013} and CO-dark gas may play an important role at low \Sig\ \citep[see, e.g.,][]{Smith_etal_2014}. We adopt fixed \alphaco\ for this study in order to link our measurements closely to observables. We expect improved measurements of CO excitation and comparison to high resolution dust maps to help refine the interpretation of our measurements in the coming years. As mentioned above, comparisons to the predicted CO emission from recent simulations \citep[e.g.,][M. Gong et al. submitted]{Smith_etal_2014,Pan_etal_2015,Duarte-Cabral_Dobbs_2016} should be increasingly feasible in the future, and our approach makes this comparison relatively straightforward.

This analysis uses the initial PHANGS-ALMA sample. This sample is currently being expanded to cover almost all southern, nearby, low-inclination, star-forming galaxies. With the final sample of $74$ galaxies, it will be possible to make definitive measurements of distribution functions and scaling relations across the whole local star-forming galaxy population.

In addition, many aspects of our analysis would benefit from better knowledge of the sub-beam gas distribution and the line of sight depth through the gas layer. This would improve our translation from surface to volume density (and so pressure), allow for better absolute estimates of the virial parameter, and help distinguish low beam-filling from variations in physical properties. We consider this to be a strong argument for even higher resolved case studies of the nearby galaxy population.

Last but not least, we expect to continue to improve our methodology. The ongoing parallel effort following a cloud identification scheme (\citeinprep{E. Rosolowsky et al. in prep.}) will provide a complementary description of the molecular gas structures, whereas more sophisticated characterization of CO intensity (e.g., forward-modeling at the cube level) will help improve both the statistical rigor and completeness of our measurements.

\section{Summary}\label{sec:summary}

We measure the velocity dispersion, \sig , and surface density, \Sig , of molecular gas at spatial scales of individual giant molecular clouds (GMCs) across a sample of $15$ nearby galaxies. We extract these measurements from sensitive, high resolution CO data cubes, including 11 CO~(2-1) cubes from the PHANGS-ALMA survey (\citeinprep{A. K. Leroy et al. 2018, in prep.}) and 4 other CO cubes from the literature. We convolve all data to a set of common spatial resolutions ($45$-$120$~pc), comparable to the size of GMCs with radii of $22.5$-$60$~pc. After convolving, we measure \sig\ and \Sig\ at these fixed spatial scales for all sightlines with robust CO detections. Our measurements represent a straightforward characterization of the statistics of all detected CO emission, without relying on any cloud segmentation approach. This facilitates inter-galaxy comparisons, is minimally-interpretive and straightforward to apply to simulations. 

We show the derived \Sig\ and \sig\ maps (Figure~\ref{fig:maps}), and report all our measurements in machine readable form (Table~\ref{tab:data}). In total, we obtain $\sim30,000$ independent \sig\ and \Sig\ measurements across our whole sample (see Table~\ref{tab:stats}). On average, these measurements characterize ${\sim} 70\%$ of the total CO flux at $80$~pc scales or ${\sim} 80\%$ of the CO flux at $120$~pc scales. This represents by far the largest compilation of cloud-scale measurements of \Sig\ and \sig\ in galaxies in the local universe. Our sample spans from low mass disk galaxies to massive spirals, significantly broadening the range of galactic environments in which the properties of molecular gas have been studied (Table~\ref{tab:galaxies}). Other than relatively high stellar mass, star-forming galaxies, we also include two Local Group galaxies M31 (\citeinprep{A. Schruba et al. in prep.}) and M33 \citep{Druard_etal_2014}, and the nearest major merger: the Antennae galaxies \citep{Whitmore_etal_2014}.

We use these measurements to characterize the physical state of molecular gas at cloud scales \citep[``Larson's Laws'';][]{Larson_1981}. We calculate the distribution of molecular mass, traced by CO flux, as functions of \Sig\ and \sig\ (Section~\ref{sec:dist}), virial parameter ($\alpha_{\rm vir} \propto \sig ^2/ \Sig$, Section~\ref{sec:alphavir}), and internal turbulent pressure ($\Pturb \propto \Sig\,\sig^2$, Section~\ref{sec:Pturb}). We find a strong positive correlation between \sig\ and \Sig\ (Section~\ref{sec:scaling}). Though we defer a detailed study of how the molecular gas properties vary with galactic environment to the next paper in this series, we highlight the striking difference between the central kpc and the disks of strongly barred star-forming galaxies.

Our main results are as follows: 

\begin{enumerate}

\item Across our whole sample, we observe a $2$-$3$ orders of magnitude variation in the molecular gas surface density \Sig . We observe strong galaxy-to-galaxy variations in the mass weighted median \Sig , from $30$ to $200\rm\;M_\odot\,pc^{-2}$ across our sample (not including our most extreme sample member -- the Antennae galaxies). Within the disk (excluding the central kpc) of an individual galaxy at fixed spatial resolution, 68\% of the molecular gas mass lies within $\pm0.3$-$0.6$~dex about this median value (Figures~\ref{fig:Sig-80pc} and \ref{fig:Sig-120pc}, Table~\ref{tab:Sig}).

\item The median velocity dispersion \sig\ (by mass) in each galaxy is typically $3$ to $10\;\kms$ (Figures~\ref{fig:sig} and Table~\ref{tab:sig}). The dynamical range in \sig\ is about a factor of $2$ narrower than the dynamical range in \Sig .

\item \sig\ and \Sig\ show strong correlation across our sample (Figures~\ref{fig:scaling}). We fit this \sig-\Sig\ relation with a power law model for each galaxy, and find the best-fit power law slopes $\beta$ to be $0.34$-$0.63$. Combining data across our main sample (i.e., all PHANGS-ALMA targets plus M51), we find the following best-fit relation at 80~pc resolution:
\begin{equation*}
\log_{10} \left( \frac{\sig}{\kms} \right) = 0.47\,\log_{10} \left( \frac{\Sig}{10^2\;\Msun\,\rm pc^{-2}} \right) + 0.85~,
\end{equation*}
\noindent with (systematic-dominated) uncertainties of $\sim 0.05$ in the slope and $\sim 0.05$~dex in the intercept. This scaling relation matches the expectation of a fixed virial parameter \alphavir , i.e., fixed ratio of kinetic to self-gravitational potential energy, with a small, $\lesssim 0.10$~dex intrinsic scatter around the average relation.

\item In many strongly barred galaxies, the central regions contribute a significant fraction of the overall gas mass (more than 30\% for several galaxies), and have higher apparent \Sig\ (Figures~\ref{fig:Sig-80pc} and \ref{fig:Sig-120pc}) as well as enhanced \sig\ at a given \Sig\ (Figures~\ref{fig:scaling}) relative to the disk population. This may reflect a higher density and more pervasive (less bound) medium in these regions, but we caution that part of this conclusion depends on our adopted assumptions on \alphaco\ , and beam smearing effects may contribute to the measured line width.

\item We extend our analysis towards both high and low \Sig\ by including the Antennae galaxies, M31 and M33. The Antennae have a $\sig$-$\Sig$ relationship consistent with an extrapolation of the mean relation found for our main sample, while both M31 and M33 show enhanced \sig\ at low \Sig . Artificially beam-diluted Milky Way data also show a similar trend.
The higher sensitivity of the M31 and M33 CO data reveals this behavior, but we show that we could not detect it in our other CO data sets.
This leads us to expect the flattening at low \Sig\ to be a genuine feature in low molecular gas density environments. 
The offset probably reflects a combination of low beam filling and line broadening due to the influence of the ambient medium and the galaxy potential.

\item Within our sample, the \alphavir\ distribution shows only small variations from galaxy to galaxy (excluding M31 and M33) and weak dependence on the physical resolution, suggesting a common dynamical state for molecular gas in most of our targets at $45$-$120$~pc spatial scales. The median $\alphavir$ value is typically ${\sim}1.5$-$3.0$, suggesting that most molecular gas structures are close to energy equipartition, though with slightly higher kinetic energy than self-gravitational potential energy (i.e. $K \gtrapprox U_g$). Note that we assume simple geometry and density profiles, and set the beam size to the cloud size, which implies a systematic uncertainty in the absolute values of \alphavir . Nevertheless, we expect the relative values of \alphavir\ in the disk regions of individual targets to be robust.

\item Conversely, our targets show a wide (mass weighted) distribution in their turbulent pressure, \Pturb , with $16$-$84\%$ range of ${\sim}1.0$-$2.0$~dex in each galaxy disk. The mass weighted median \Pturb\ varies by more than $4$ orders of magnitude across our sample. Even though the line of sight depth remains a major uncertainty, it is clear that the mean \Pturb\ of gas at $80$-$120$~pc scales depends strongly on environment.

\end{enumerate}

This paper represents a first characterization of the structural properties of molecular gas in nearby galaxies observed by the PHANGS-ALMA project. Besides improving our methodology of physical parameter estimation (e.g., accounting for CO-to-H$_2$ conversion factor variation, sub-beam and line of sight gas distribution; see Section \ref{sec:discussion}), the immediate next step in science will be to examine the correlation between cloud-scale molecular gas properties and the galactic-scale environment, including the stellar and gas surface densities, local kinematics, host galaxy properties, and estimates of the ``dynamical equilibrium'' pressure needed to sustain the gas disk in equilibrium \citep[e.g.,][]{Elmegreen_1989,Koyama_Ostriker_2009,Ostriker_etal_2010}. We expect to be able to quantitatively measure how the environment regulates molecular gas density, pressure and dynamical state, and to constrain the contributions of the galaxy potential to the observed line width (see also \citealt{Jeffreson_Kruijssen_2018}, \citealt{Meidt_etal_2018a},  and \citealt{Schruba_etal_2018}).

In addition to the mechanisms controlling molecular gas properties, we also aim to address the impact of the molecular gas dynamical state on local star formation activity in future papers \citep[][\citeinprep{D. Utomo et al., 2018, in prep.}]{Schruba_etal_2018}. In this domain, the logical next step will be to combine our molecular gas data with high resolution local star formation rate measurements, and examine the correlation between cloud-scale gas properties and observed star formation efficiency. As an extension to the work presented by \citet{Leroy_etal_2017}, our unprecedented sample size will enable systematic analysis across a range of galactic environments, and provide unique opportunities to test predictions from star formation theories.

\acknowledgments

We would like to thank the anonymous referee for valuable comments that helped us improve the manuscript. JS and AKL thank Eve Ostriker and David Weinberg for useful discussions.

This paper makes use of the following ALMA data: ADS/JAO.ALMA\# 2011.0.00876.S, 2012.1.00650.S, \hfill\break
2015.1.00925.S, and 2015.1.00956.S. These ALMA programs are described in \citet{Whitmore_etal_2014} and \citeinprep{A. K. Leroy et al. (2018, in prep.)}. ALMA is a partnership of ESO (representing its member states), NSF (USA) and NINS (Japan), together with NRC (Canada), NSC and ASIAA (Taiwan), and KASI (Republic of Korea), in cooperation with the Republic of Chile. The Joint ALMA Observatory is operated by ESO, AUI/NRAO and NAOJ. The National Radio Astronomy Observatory is a facility of the National Science Foundation operated under cooperative agreement by Associated Universities, Inc. 

This work also makes use of data from the CARMA M31 survey (\citeinprep{A. Schruba et al. in prep.}), the IRAM M33 large program \citep{Gratier_etal_2010,Druard_etal_2014} and the PdBI Arcsecond Whirlpool Survey \citep{Schinnerer_etal_2013,Pety_etal_2013}. Both the IRAM 30m telescope and PdBI are run by IRAM, which is supported by INSU/CNRS (France), MPG (Germany) and IGN (Spain).

We acknowledge the usage of the Extragalactic Distance Database\footnote{\url{http://edd.ifa.hawaii.edu/index.html}} \citep{Tully_etal_2009}, the HyperLeda database\footnote{\url{http://leda.univ-lyon1.fr}} \citep{Makarov_etal_2014}, the NASA/IPAC Extragalactic Database\footnote{\url{http://ned.ipac.caltech.edu}}, and the SAO/NASA Astrophysics Data System\footnote{\url{http://www.adsabs.harvard.edu}}.

JS, AKL and DU are supported by the National Science Foundation under Grants No. 1615105, 1615109, and 1653300. 
ER acknowledges the support of the Natural Sciences and Engineering Research Council of Canada (NSERC), funding reference number RGPIN-2017-03987.
AH acknowledges support from the Centre National d'Etudes Spatiales (CNES).
JMDK and MC gratefully acknowledge funding from the German Research Foundation (DFG) in the form of an Emmy Noether Research Group (grant number KR4801/1-1, PI Kruijssen). JMDK gratefully acknowledges funding from the European Research Council (ERC) under the European Union's Horizon 2020 research and innovation program via the ERC Starting Grant MUSTANG (grant agreement number 714907, PI Kruijssen). 
ES acknowledges funding from the European Research Council (ERC) under the European Union’s Horizon 2020 research and innovation program (grant agreement No. 694343). 
GB is supported by CONICYT/FONDECYT, Programa de Iniciaci\'on, Folio 11150220. 
APSH is a fellow of the International Max Planck Research School for Astronomy and Cosmic Physics at the University of Heidelberg (IMPRS-HD).
JP acknowledges support from the Programme National “Physique et Chimie du Milieu Interstellaire” (PCMI) of CNRS/INSU with INC/INP co-funded by CEA and CNES.
AU acknowledges support from Spanish MINECO grants ESP2015-68964 and AYA2016-79006. 

\facilities{ALMA, IRAM:30m, IRAM:Interferometer, CARMA}

\software{{\tt IPython} \citep{IPython}, {\tt NumPy} \citep{NumPy_SciPy}, {\tt SciPy} \citep{NumPy_SciPy}, {\tt Matplotlib} \citep{Matplotlib}, {\tt Astropy} \citep{Astropy}, {\tt emcee} \citep{Foreman-Mackey_etal_2013}, {\tt CASA} \citep{McMullin_etal_2007}.}

\bibliography{main.bib}

\newpage

\appendix

\section{Surface Density and Velocity Dispersion Maps at 120~pc Resolution}\label{apdx:maps}

\renewcommand\thefigure{\thesection\arabic{figure}}
\setcounter{figure}{0}

\begin{figure*}[!h]
\gridline{
\halfpanel{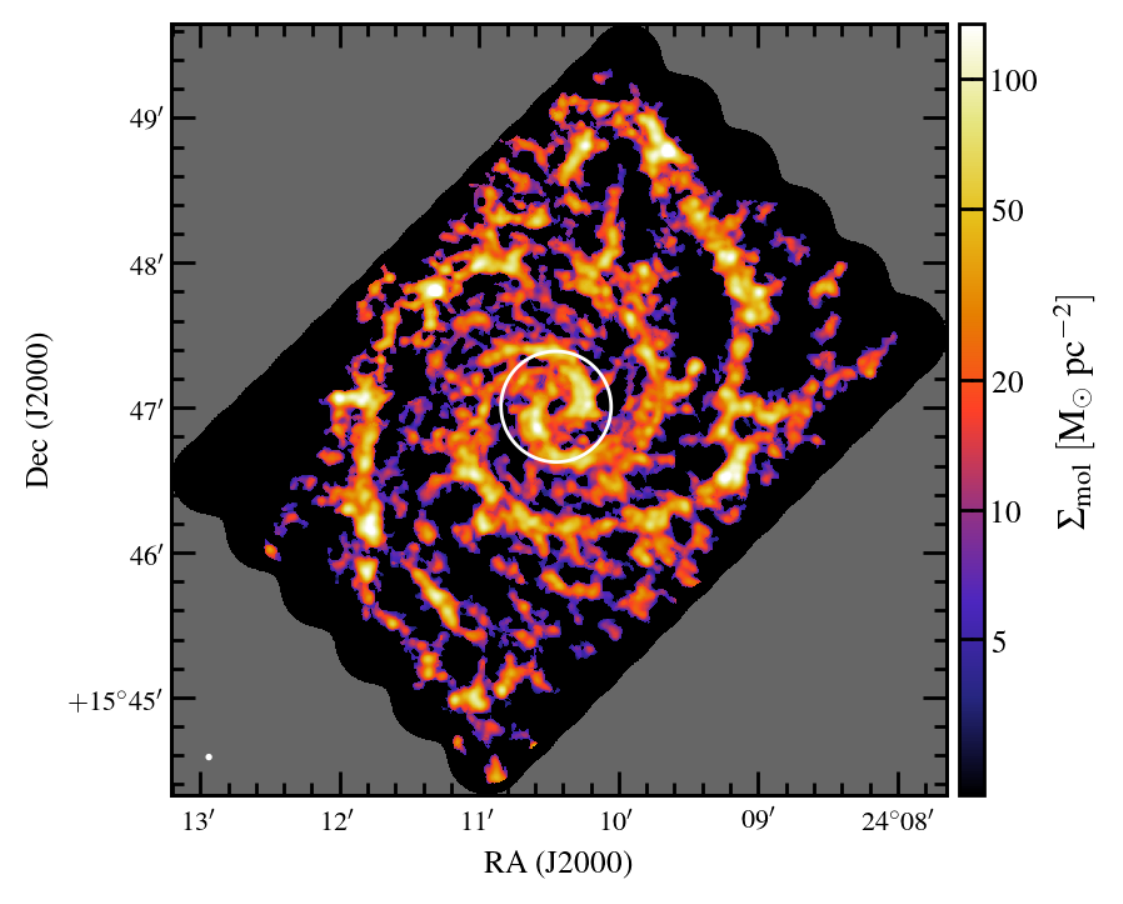}{0.32\textwidth}{(1a) NGC~628 at 120 pc resolution}
\halfpanel{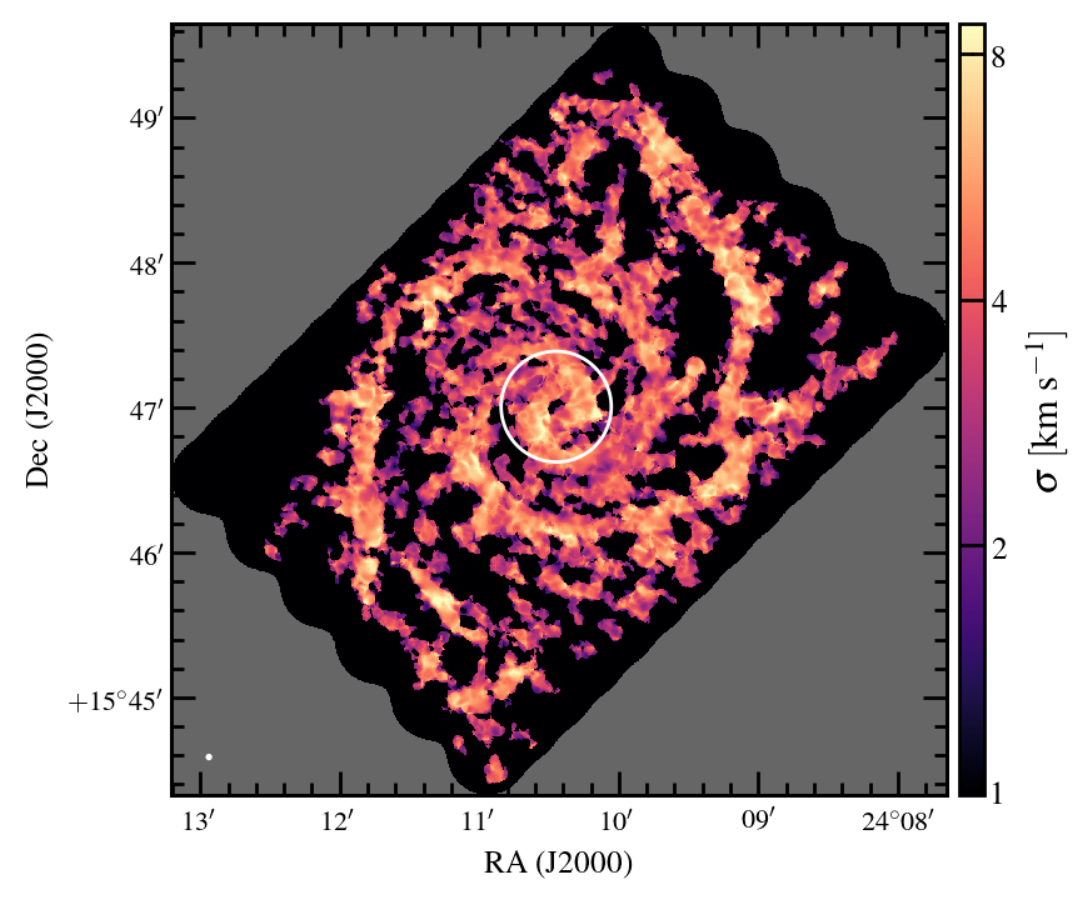}{0.32\textwidth}{(1b) NGC~628 at 120 pc resolution}
}
\gridline{
\halfpanel{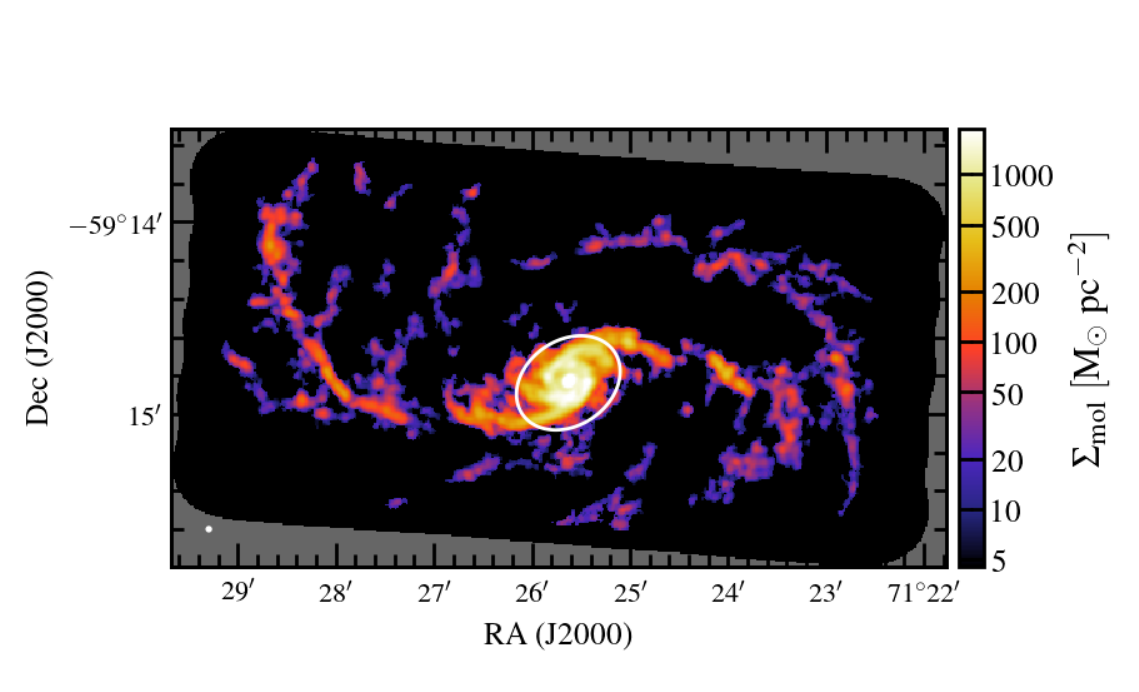}{0.25\textwidth}{(2a) NGC~1672 at 120 pc resolution}
\halfpanel{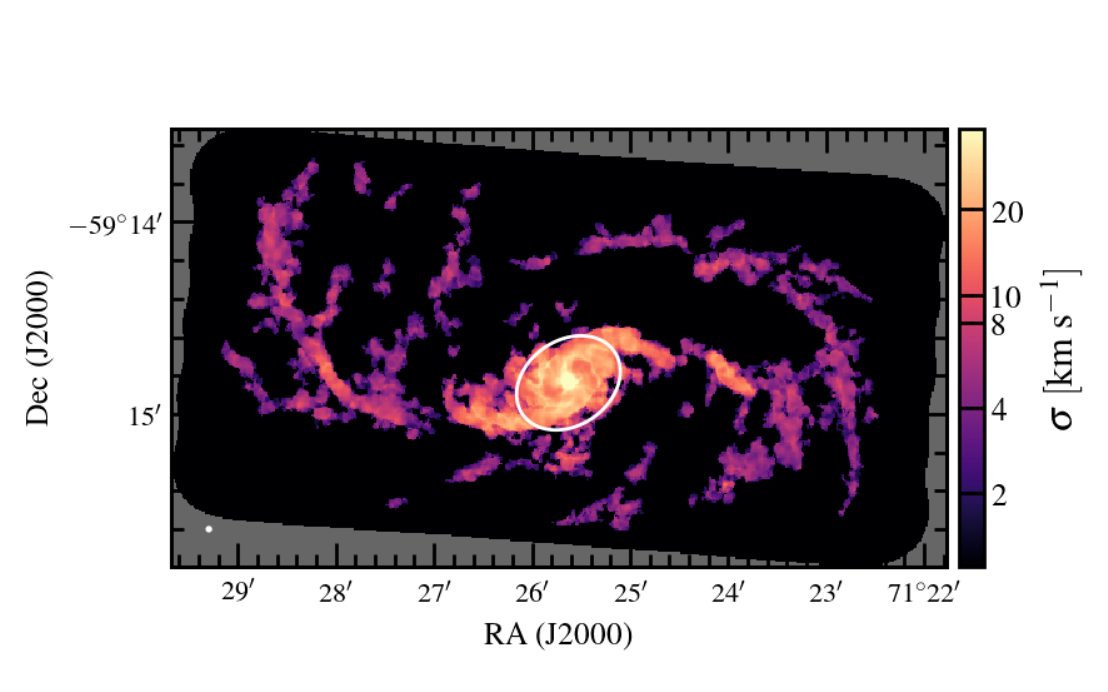}{0.25\textwidth}{(2b) NGC~1672 at 120 pc resolution}
}
\gridline{
\halfpanel{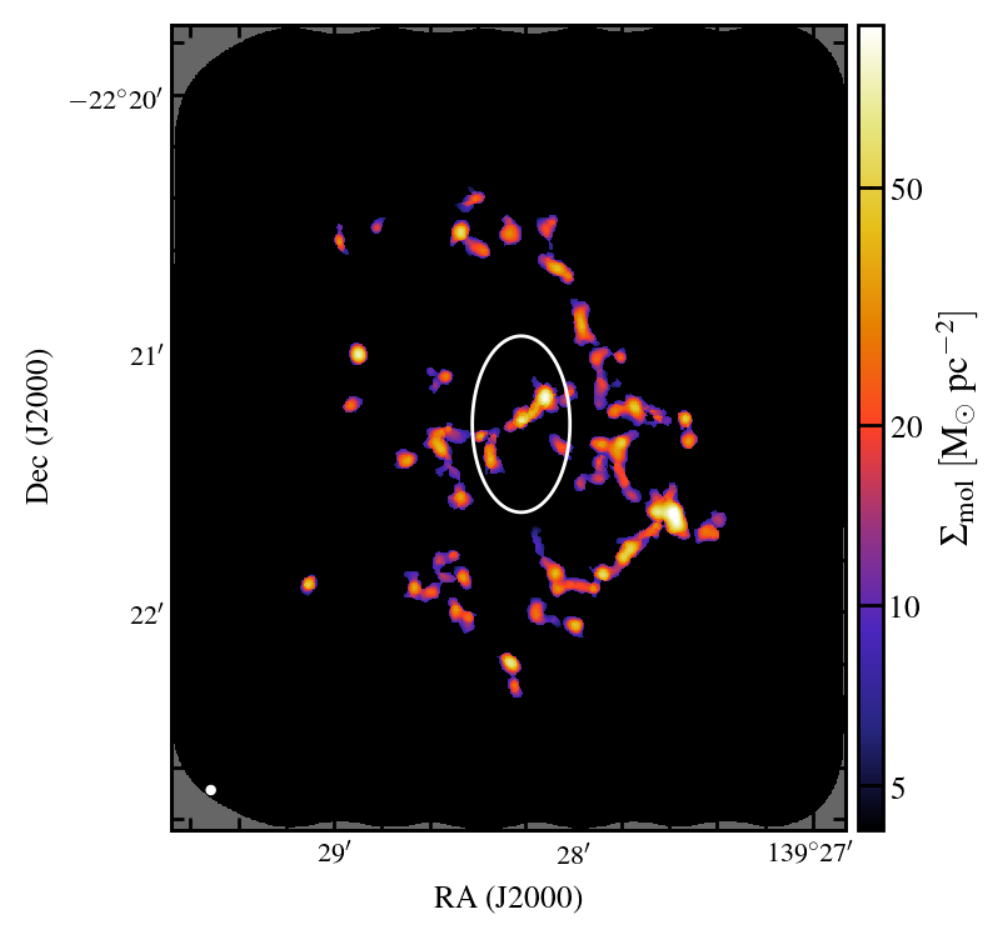}{0.31\textwidth}{(3a) NGC~2835 at 120 pc resolution}
\halfpanel{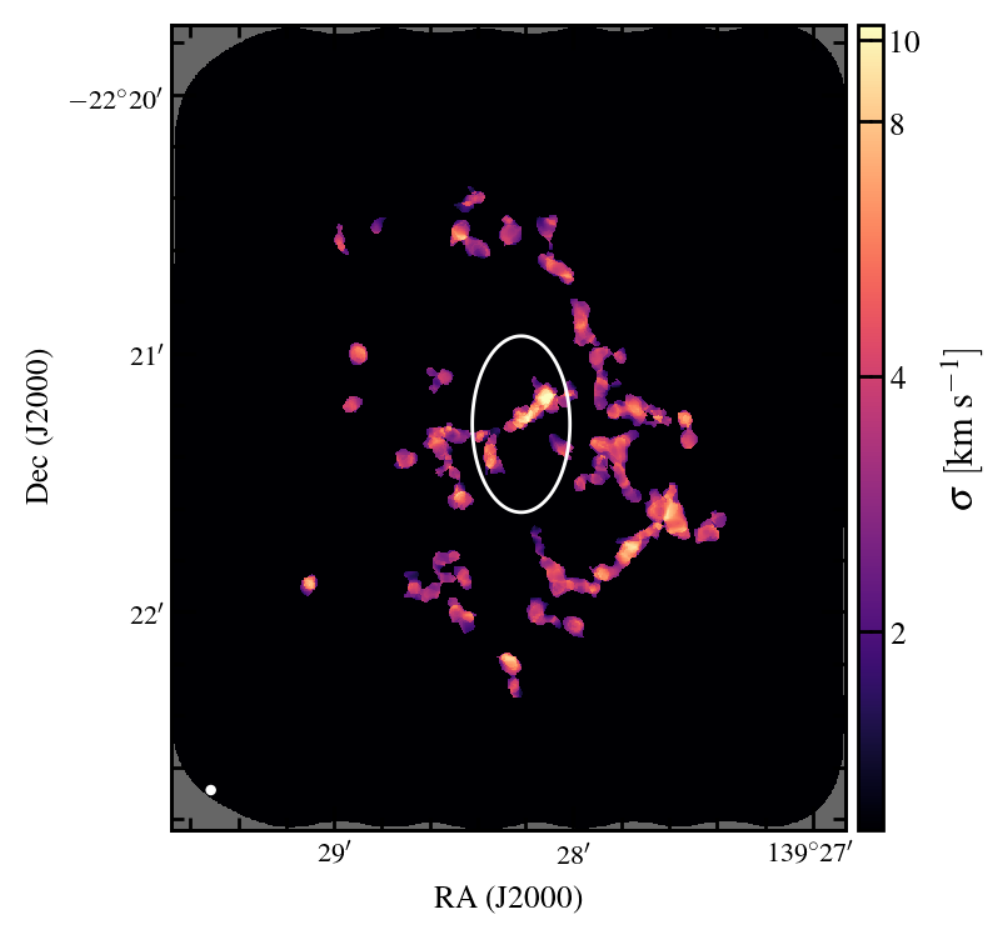}{0.31\textwidth}{(3b) NGC~2835 at 120 pc resolution}
}
\caption{Molecular gas surface density (left column) and velocity dispersion (right column) for all 15 galaxies at 120 pc resolution. The beam size appears as a white dot in the lower left corner of each panel. White elliptical contours mark the 1-kpc boundary between ``disk regions'' and ``central regions'' (see Section \ref{sec:dist:region}). Note that the gray regions lie outside the footprint of each CO survey, while the black regions show the sightlines that have no confident CO detection.}
\label{fig:maps}
\end{figure*}

\setcounter{figure}{0}
\begin{figure*}[p]
\gridline{
\halfpanel{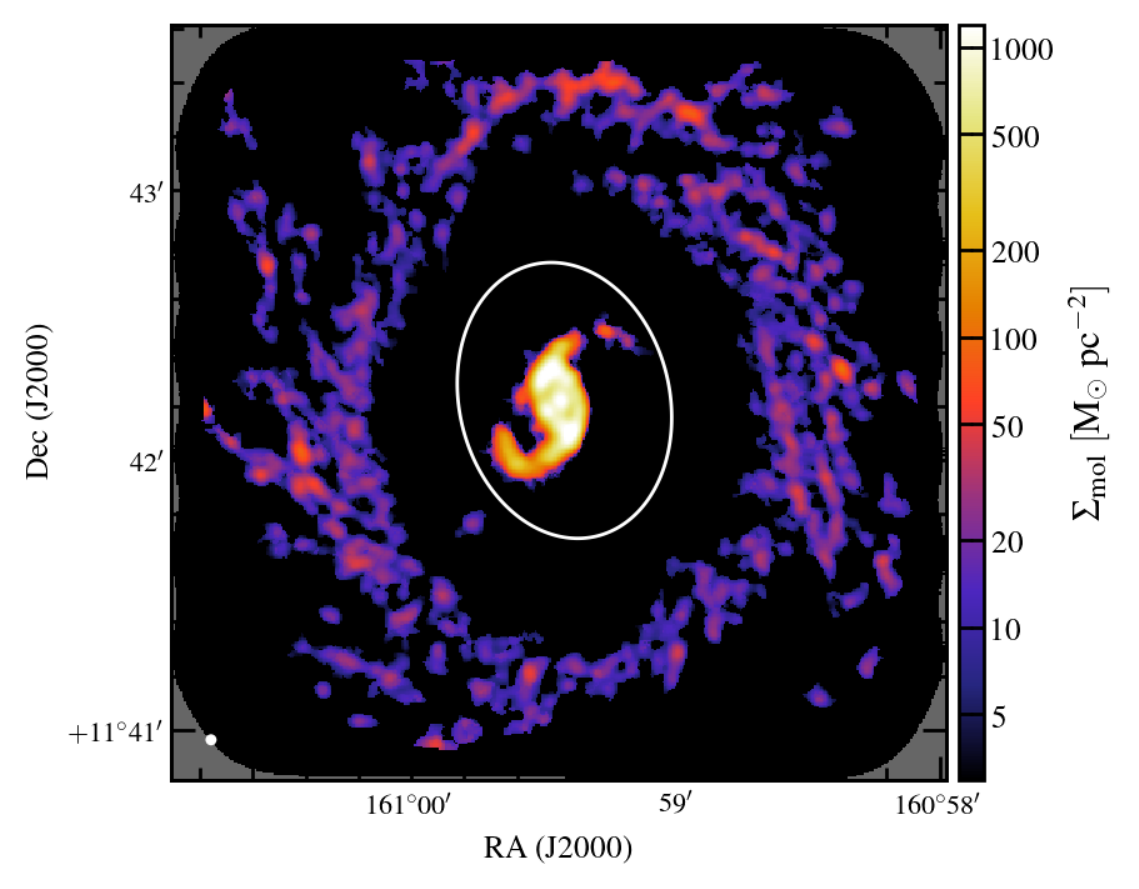}{0.37\textwidth}{(4a) NGC~3351 at 120 pc resolution}
\halfpanel{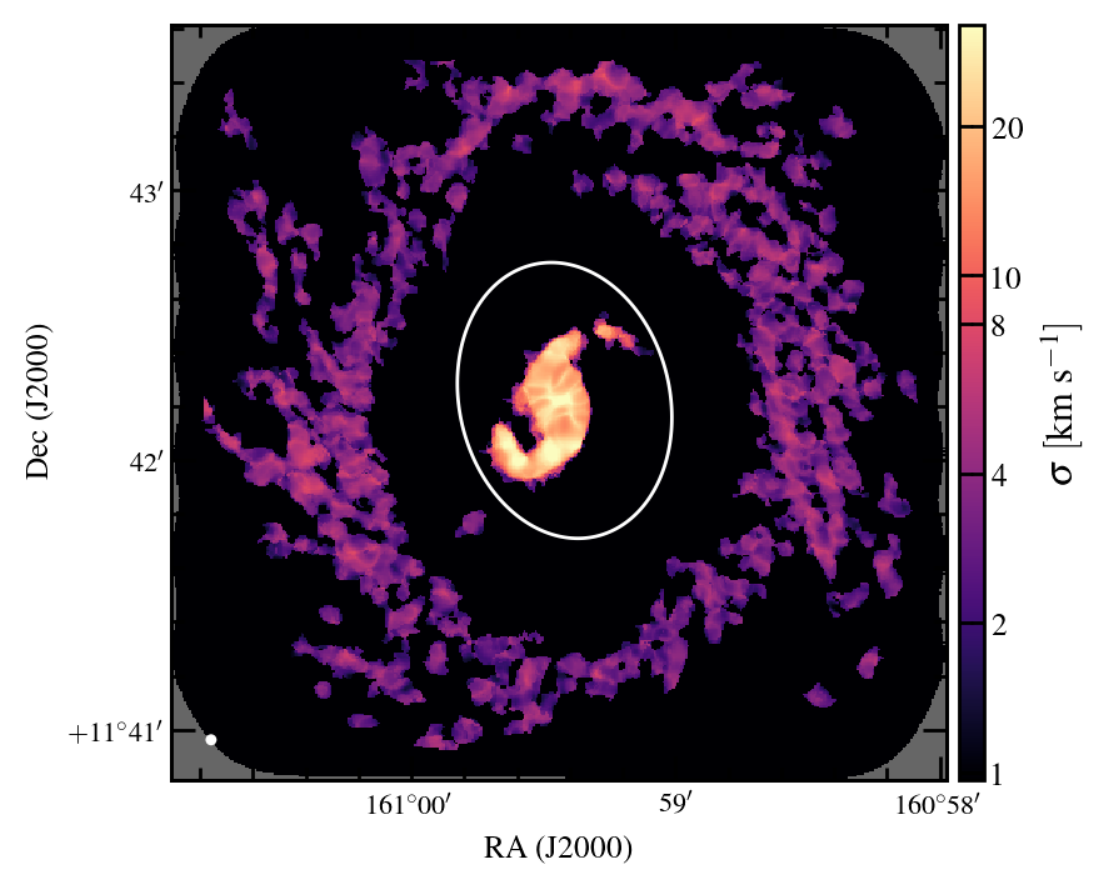}{0.37\textwidth}{(4b) NGC~3351 at 120 pc resolution}
}
\gridline{
\halfpanel{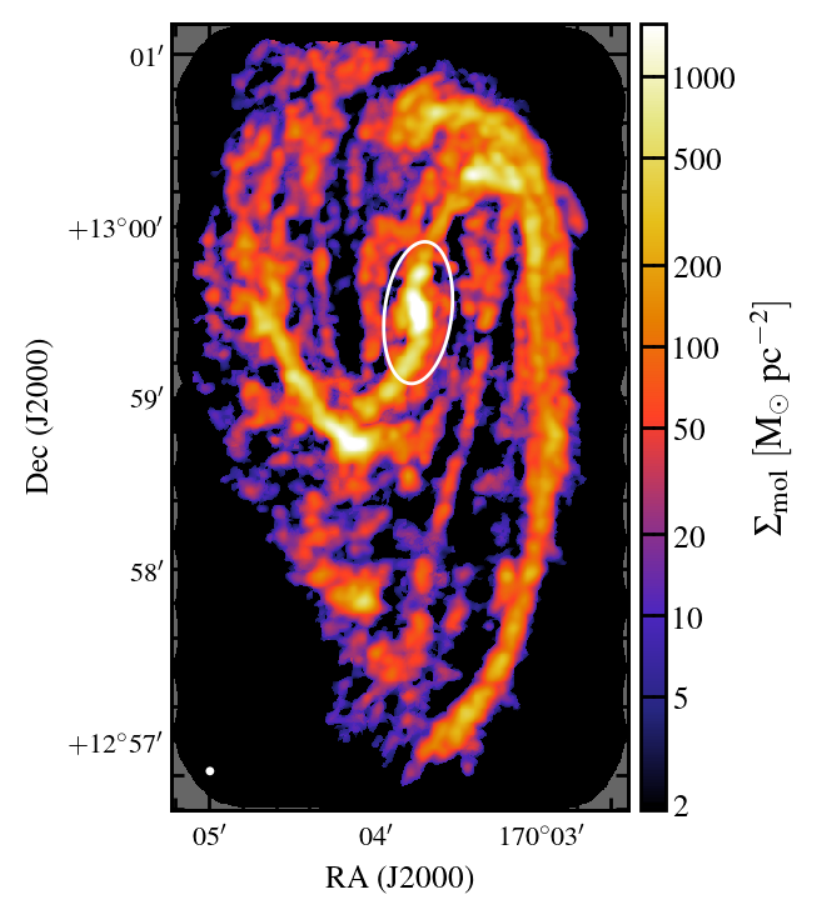}{0.40\textwidth}{(5a) NGC~3627 at 120 pc resolution}
\halfpanel{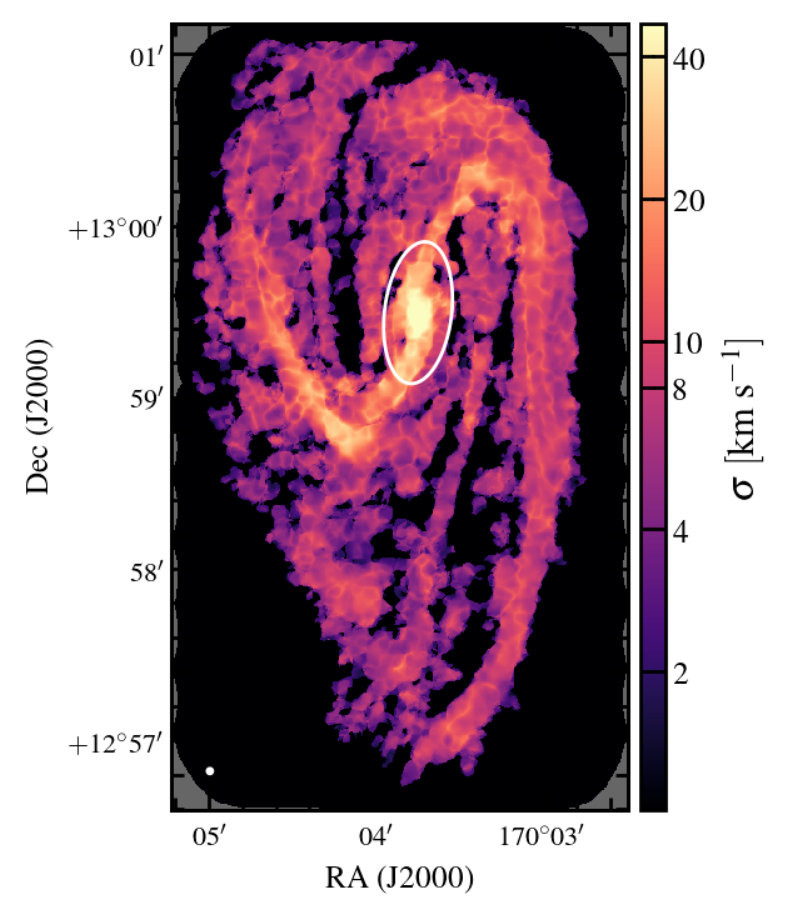}{0.40\textwidth}{(5b) NGC~3627 at 120 pc resolution}
}
\gridline{
\halfpanel{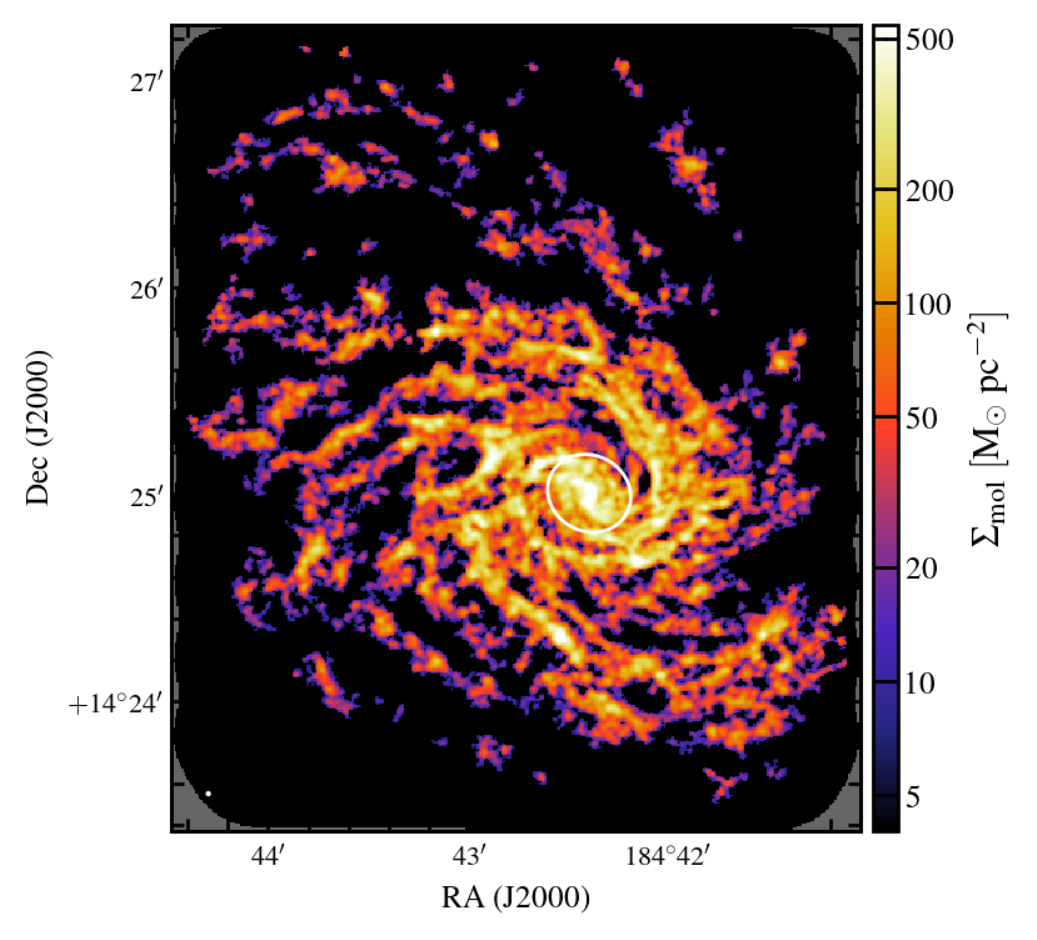}{0.40\textwidth}{(6a) NGC~4254 at 120 pc resolution}
\halfpanel{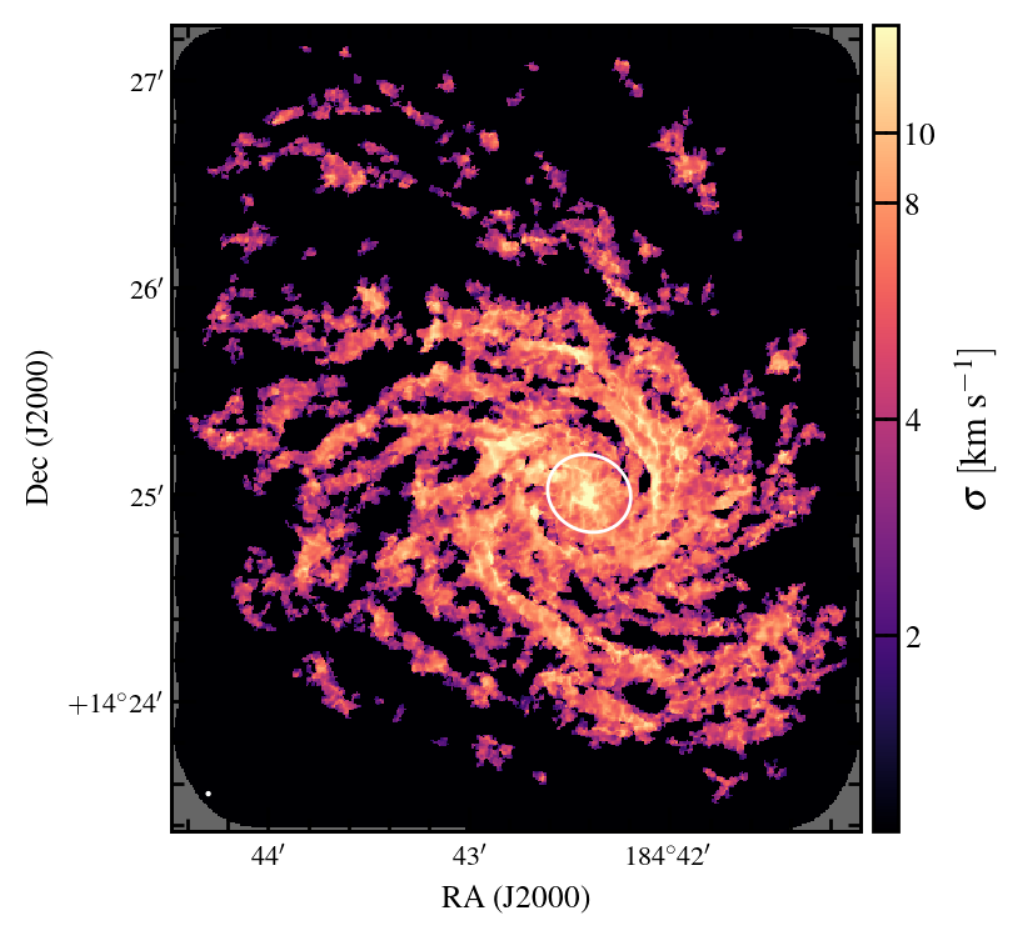}{0.40\textwidth}{(6b) NGC~4254 at 120 pc resolution}
}
\caption{(Continued)}
\end{figure*}

\setcounter{figure}{0}
\begin{figure*}[p]
\gridline{
\halfpanel{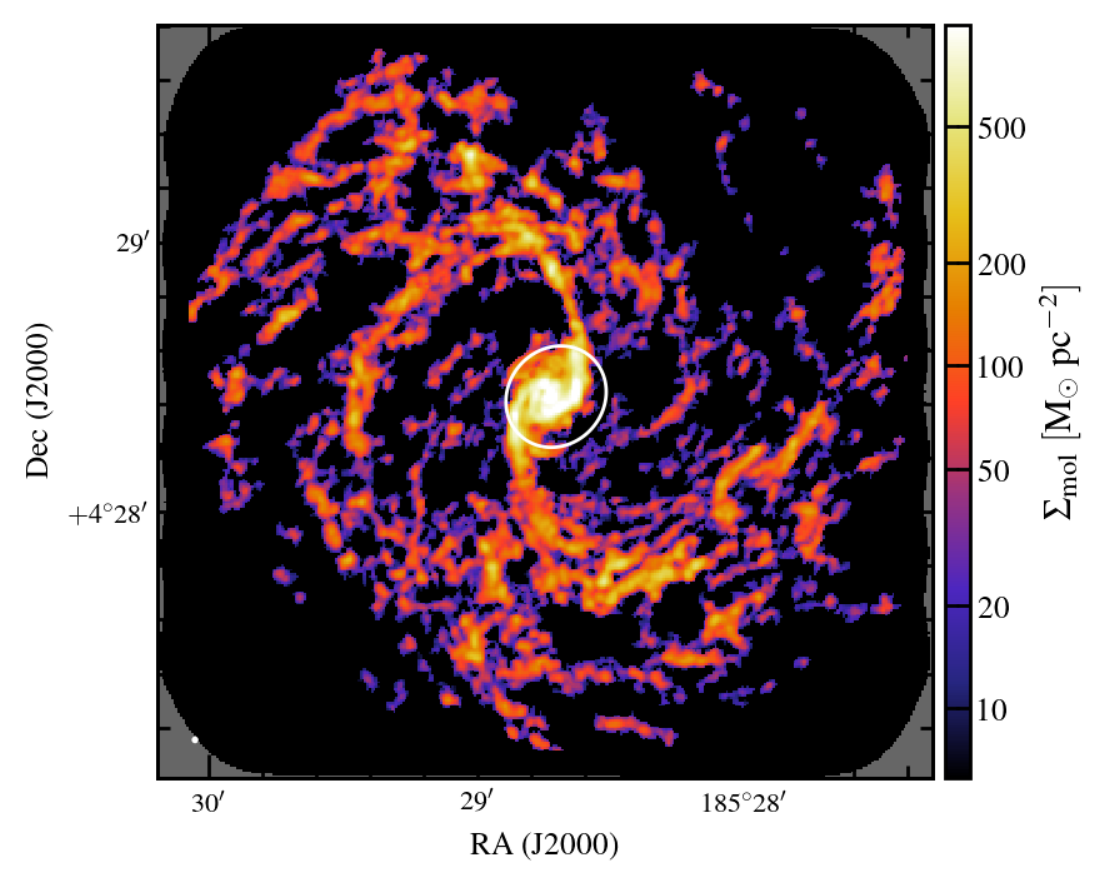}{0.35\textwidth}{(7a) NGC~4303 at 120 pc resolution}
\halfpanel{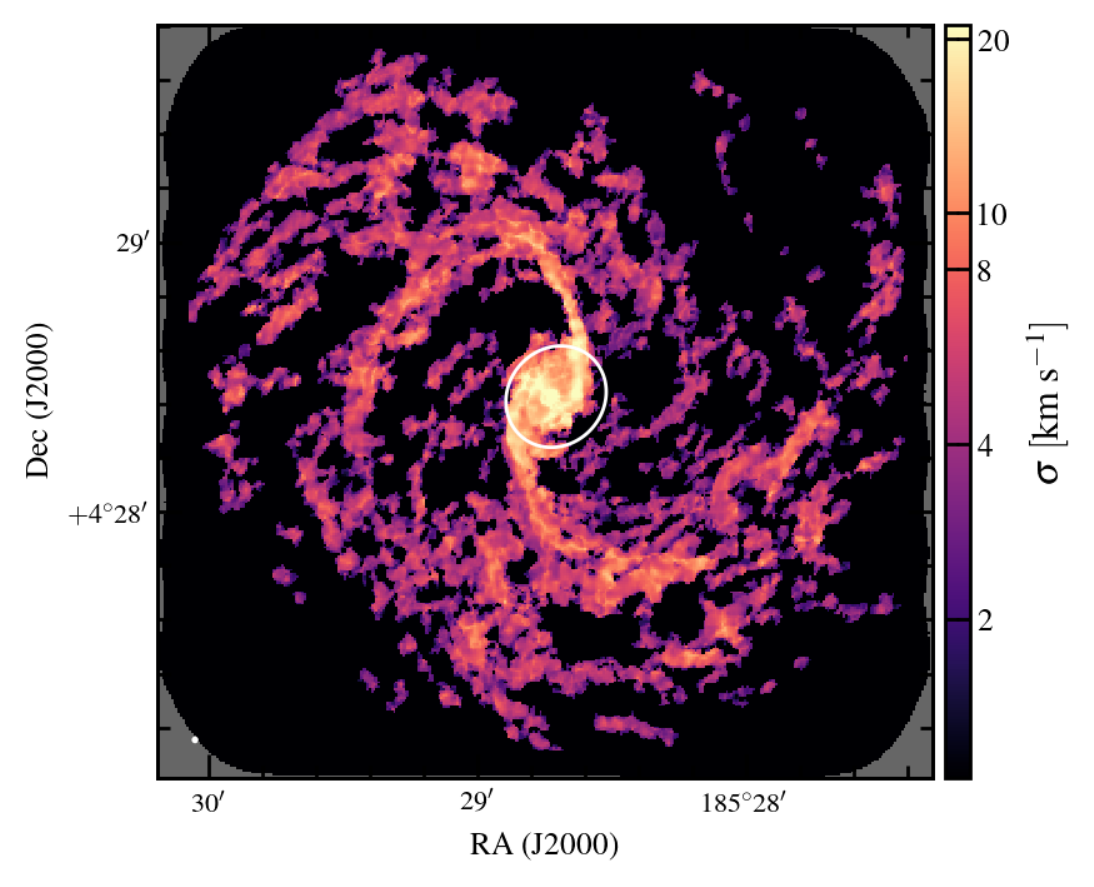}{0.35\textwidth}{(7b) NGC~4303 at 120 pc resolution}
}
\gridline{
\halfpanel{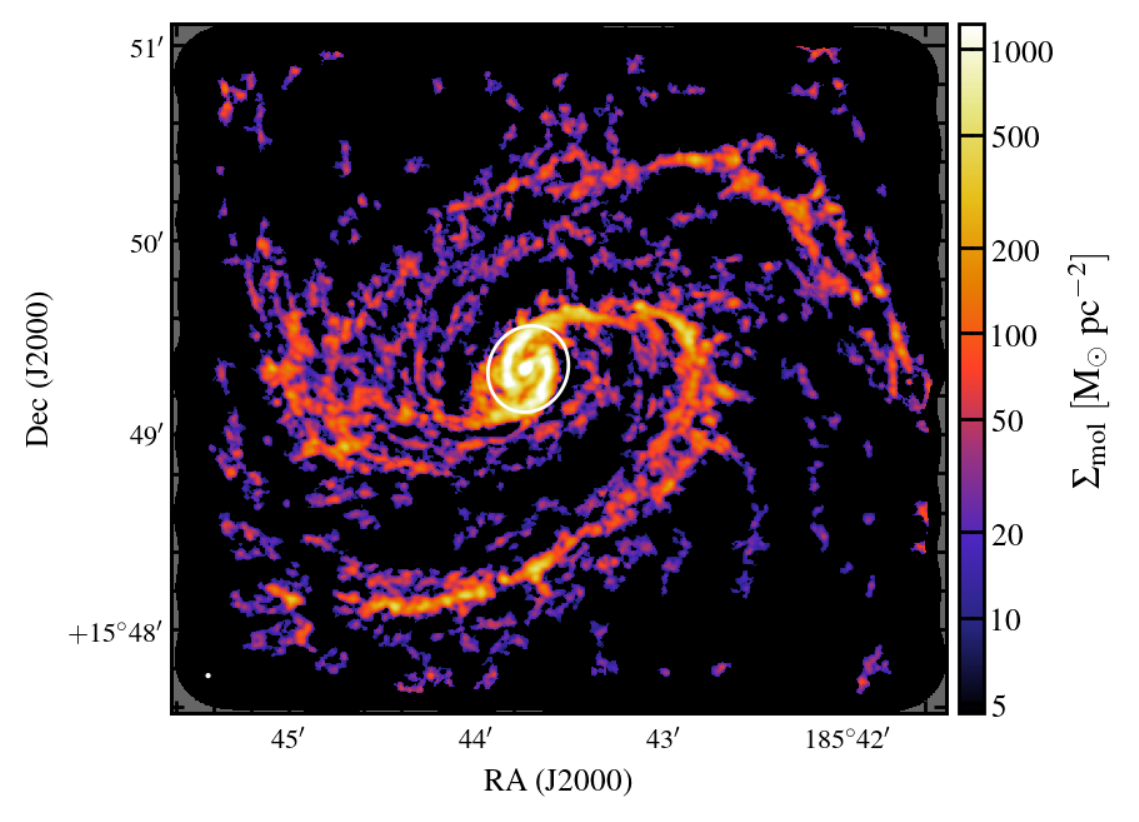}{0.35\textwidth}{(8a) NGC~4321 at 120 pc resolution}
\halfpanel{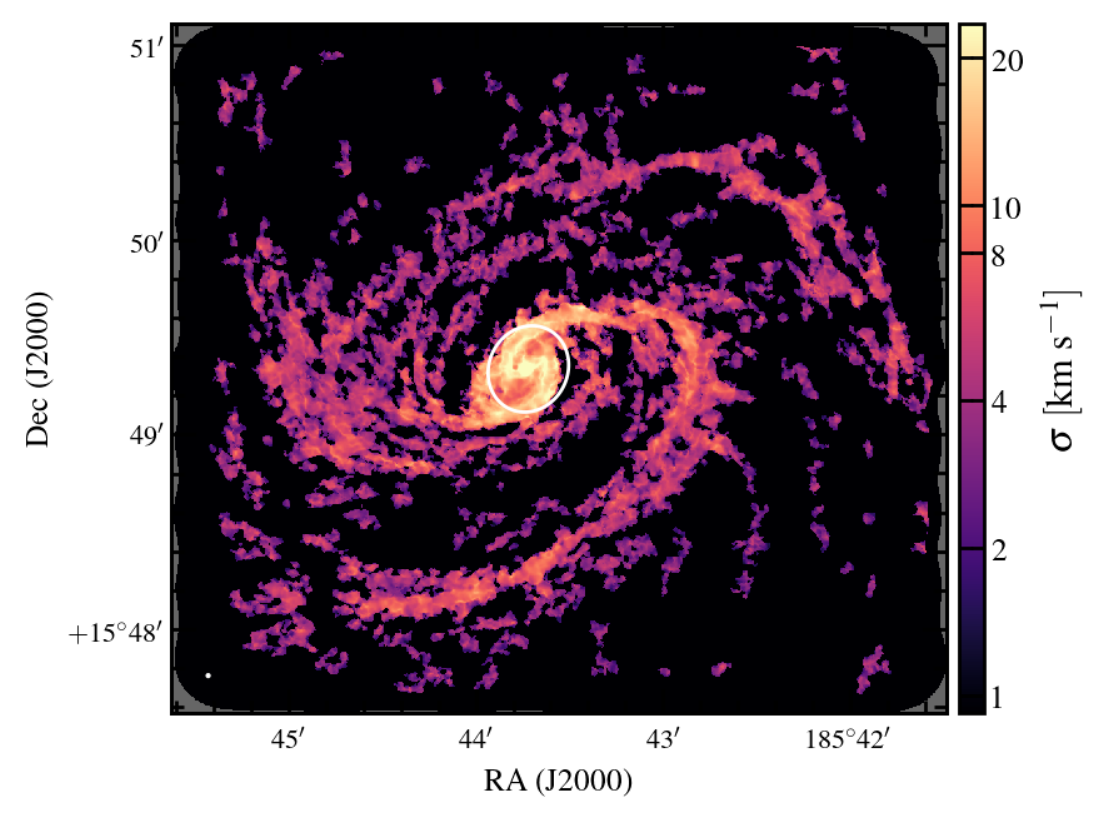}{0.35\textwidth}{(8b) NGC~4321 at 120 pc resolution}
}
\gridline{
\halfpanel{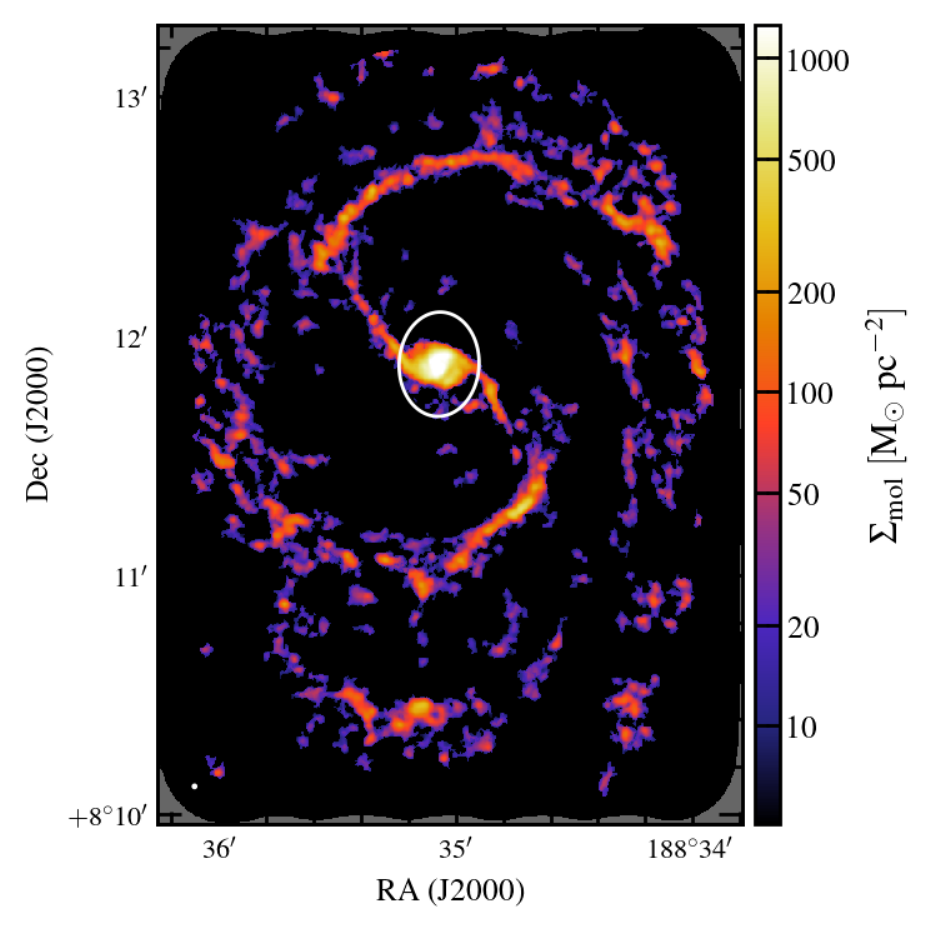}{0.45\textwidth}{(9a) NGC~4535 at 120 pc resolution}
\halfpanel{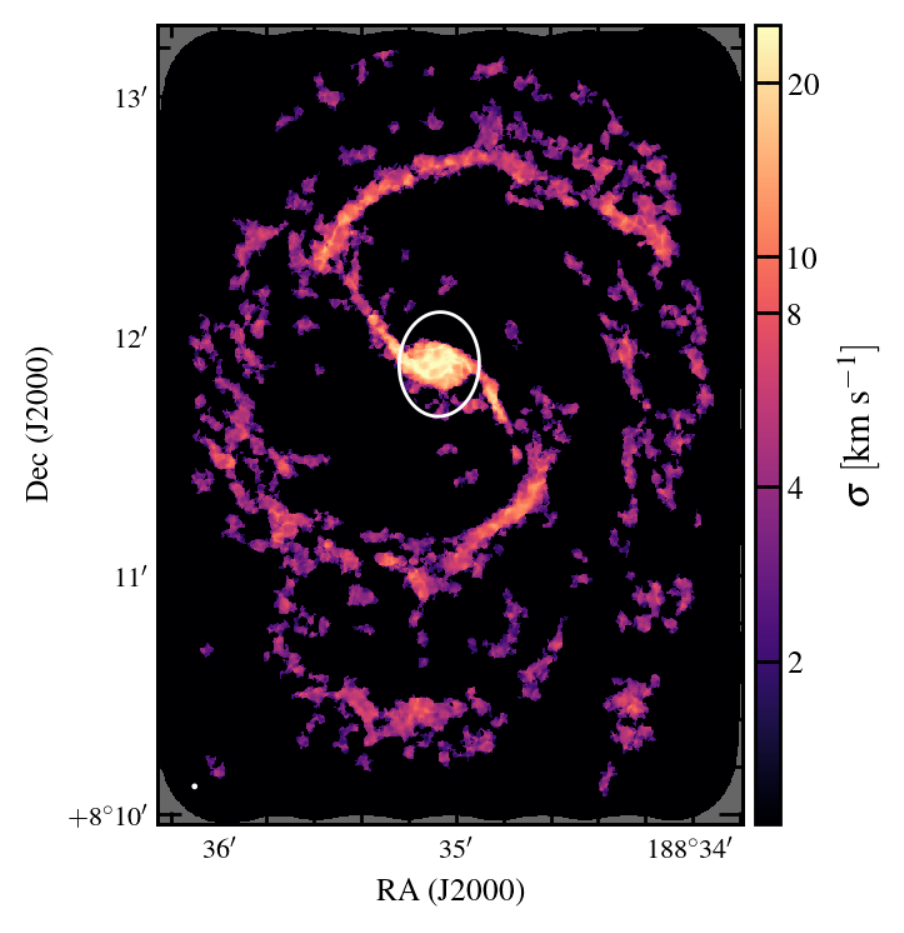}{0.45\textwidth}{(9b) NGC~4535 at 120 pc resolution}
}
\caption{(Continued)}
\end{figure*}

\setcounter{figure}{0}
\begin{figure*}[p]
\gridline{
\halfpanel{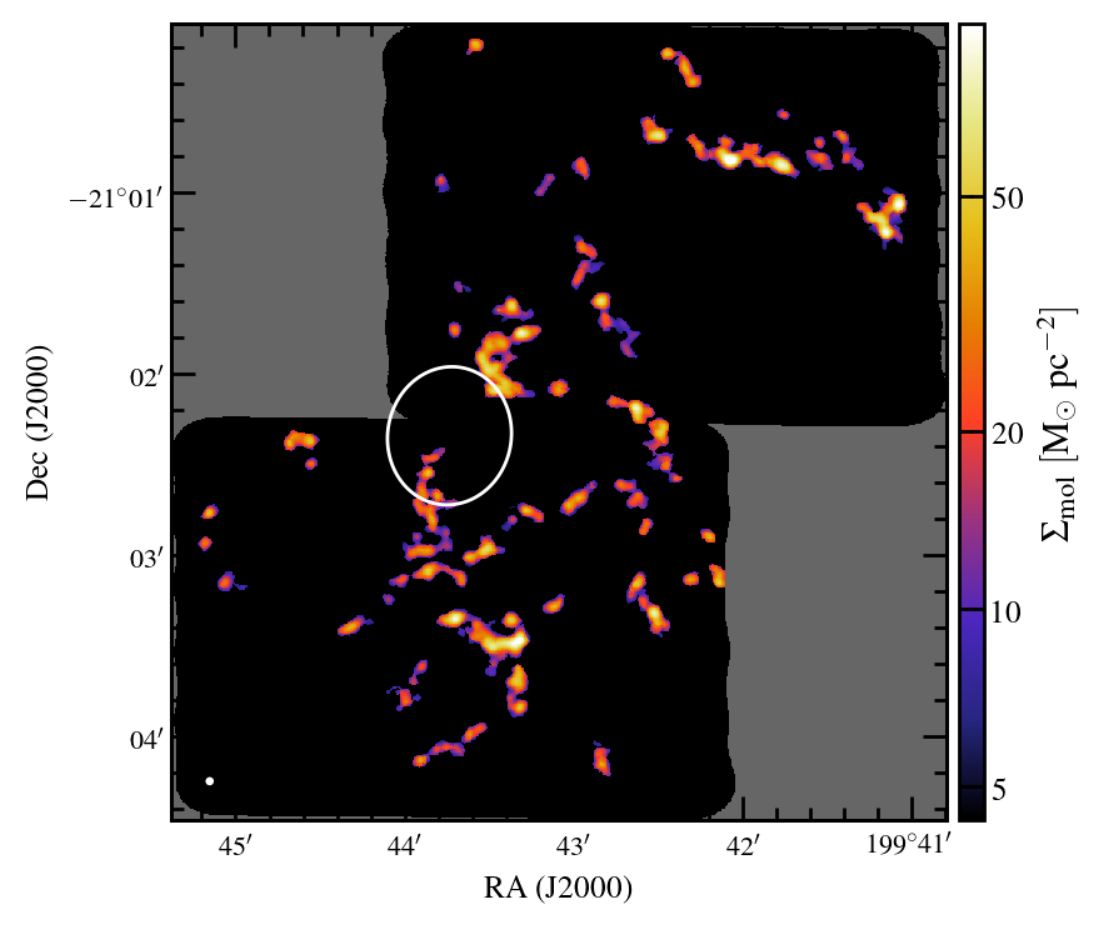}{0.40\textwidth}{(10a) NGC~5068 at 120 pc resolution}
\halfpanel{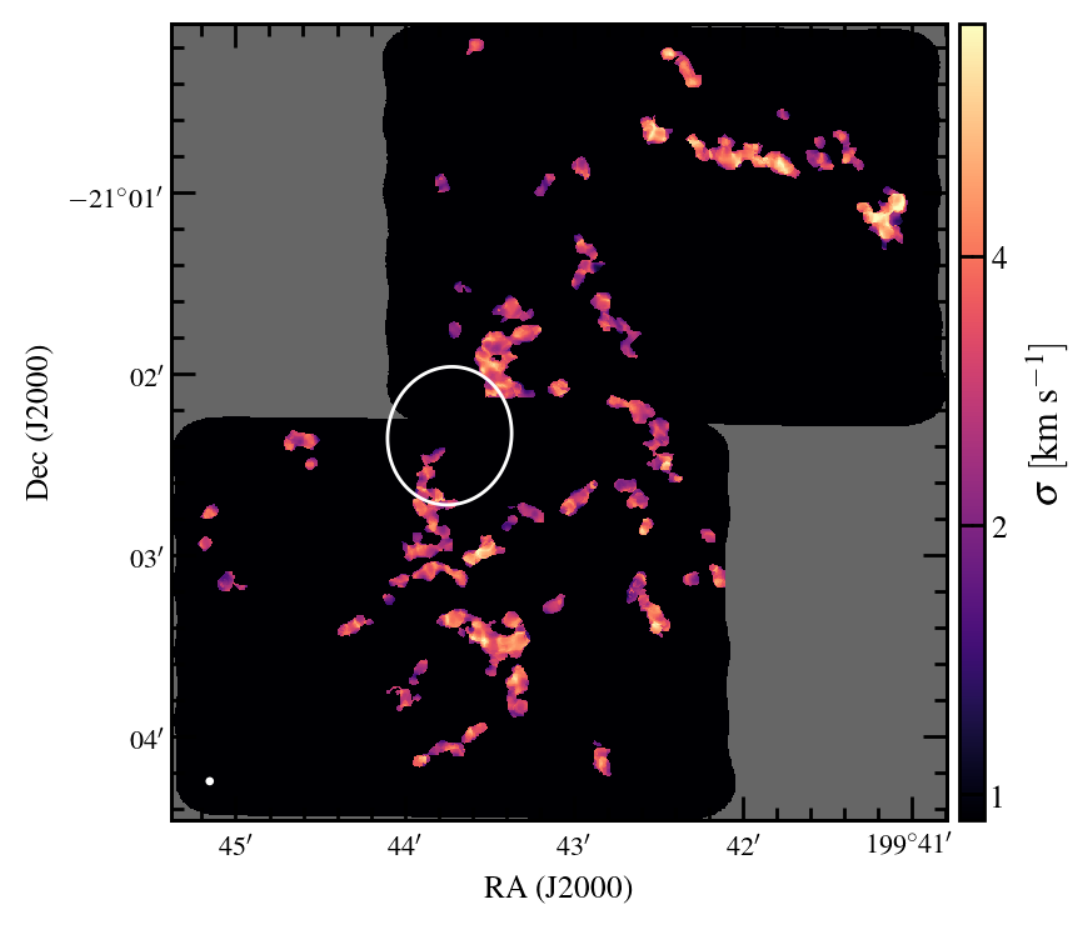}{0.40\textwidth}{(10b) NGC~5068 at 120 pc resolution}
}
\gridline{
\halfpanel{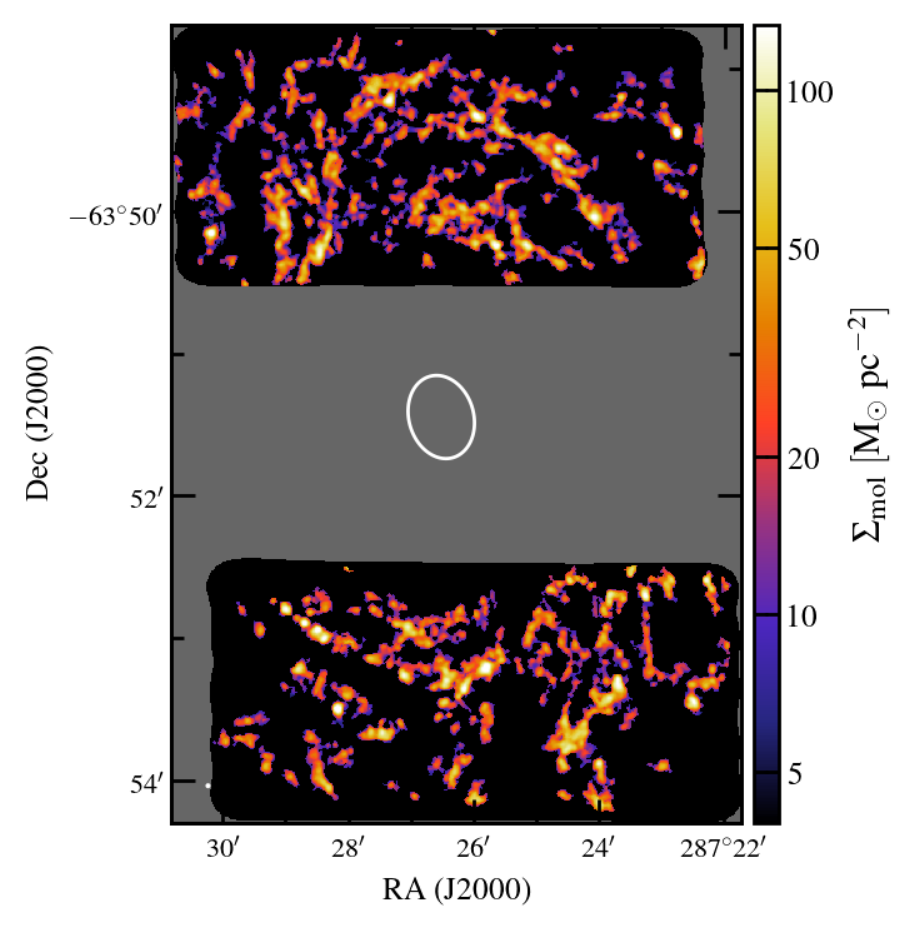}{0.48\textwidth}{(11a) NGC~6744 at 120 pc resolution}
\halfpanel{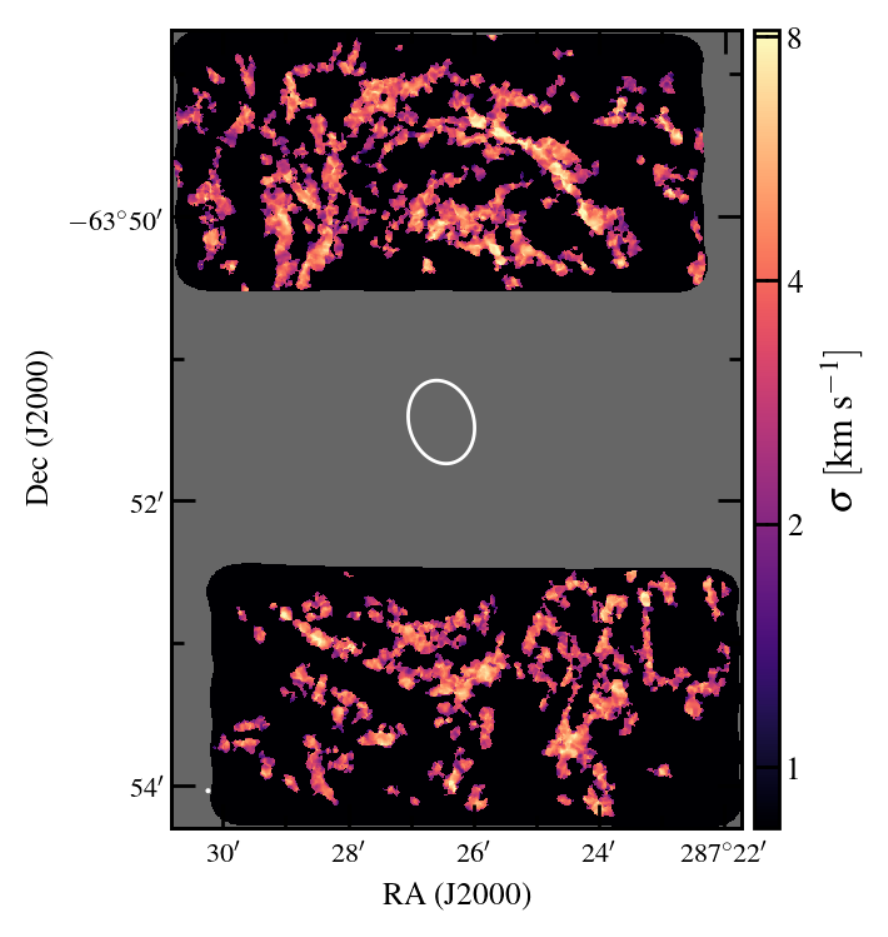}{0.48\textwidth}{(11b) NGC~6744 at 120 pc resolution}
}
\gridline{
\halfpanel{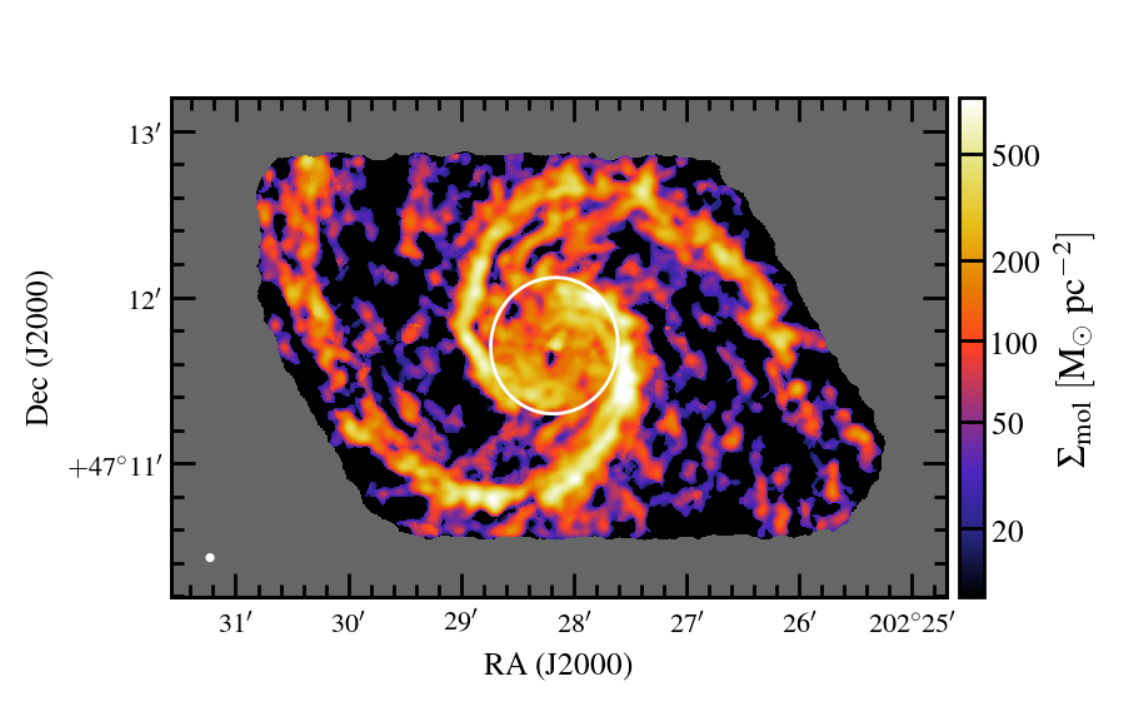}{0.29\textwidth}{(12a) M51 at 120 pc resolution}
\halfpanel{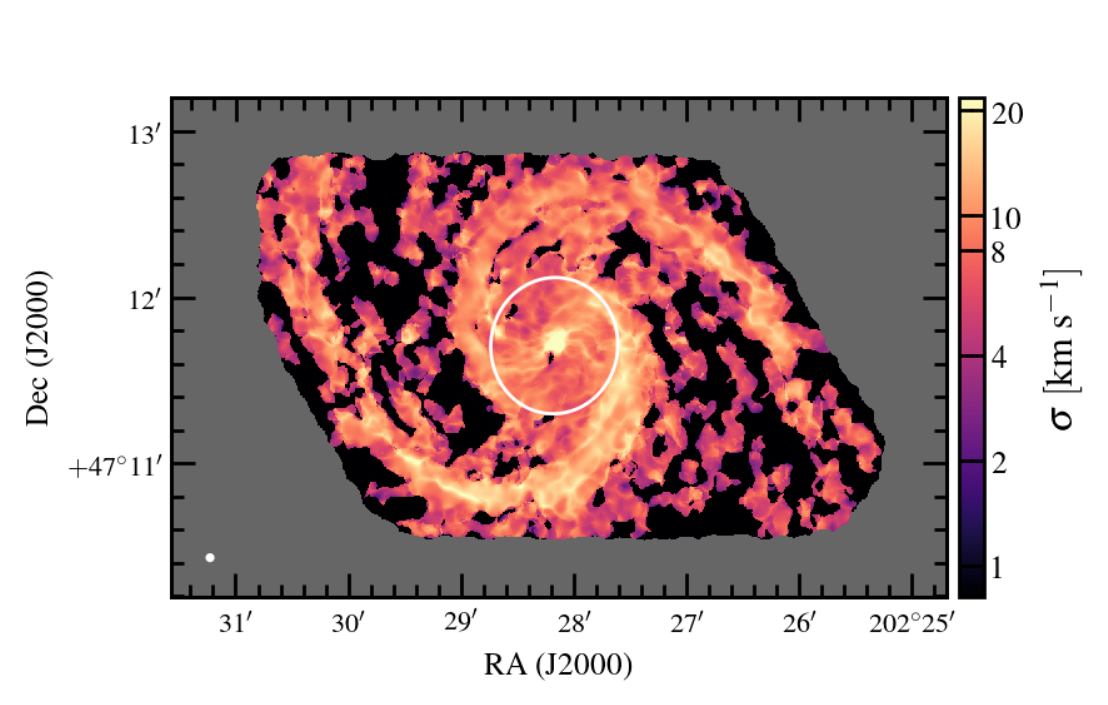}{0.29\textwidth}{(12b) M51 at 120 pc resolution}
}
\caption{(Continued)}
\end{figure*}

\setcounter{figure}{0}
\begin{figure*}[p]
\gridline{
\halfpanel{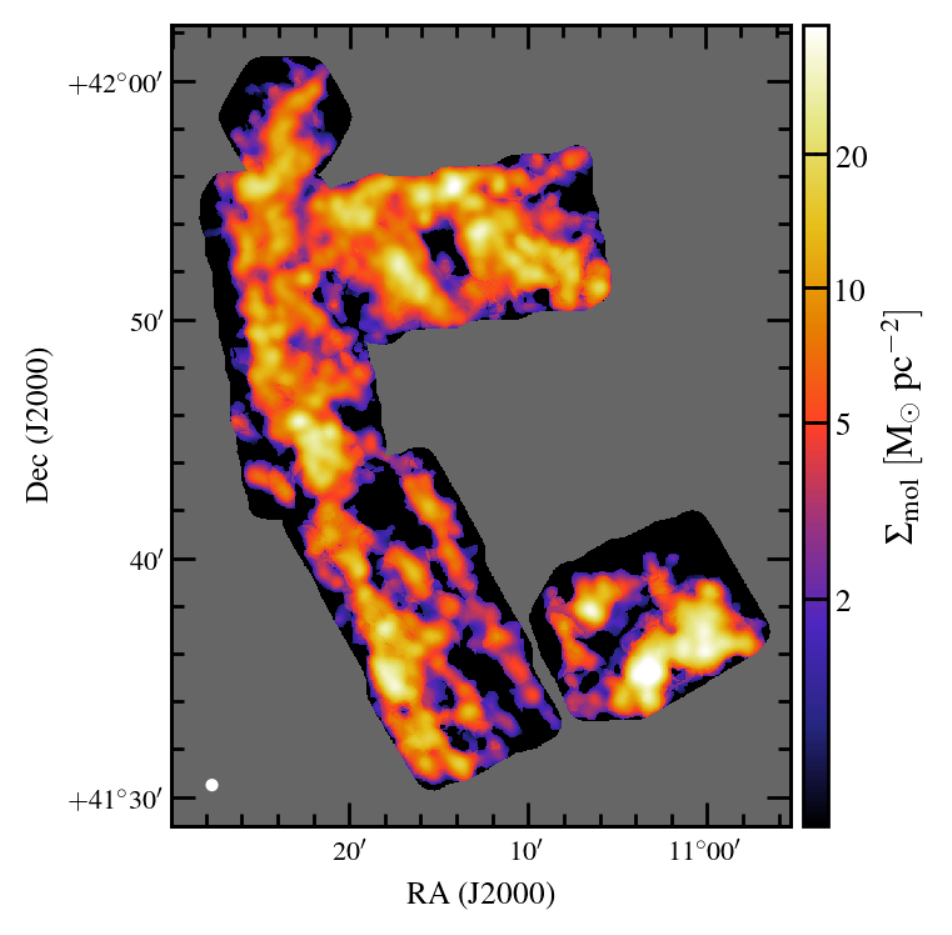}{0.35\textwidth}{(13a) M31 at 120 pc resolution}
\halfpanel{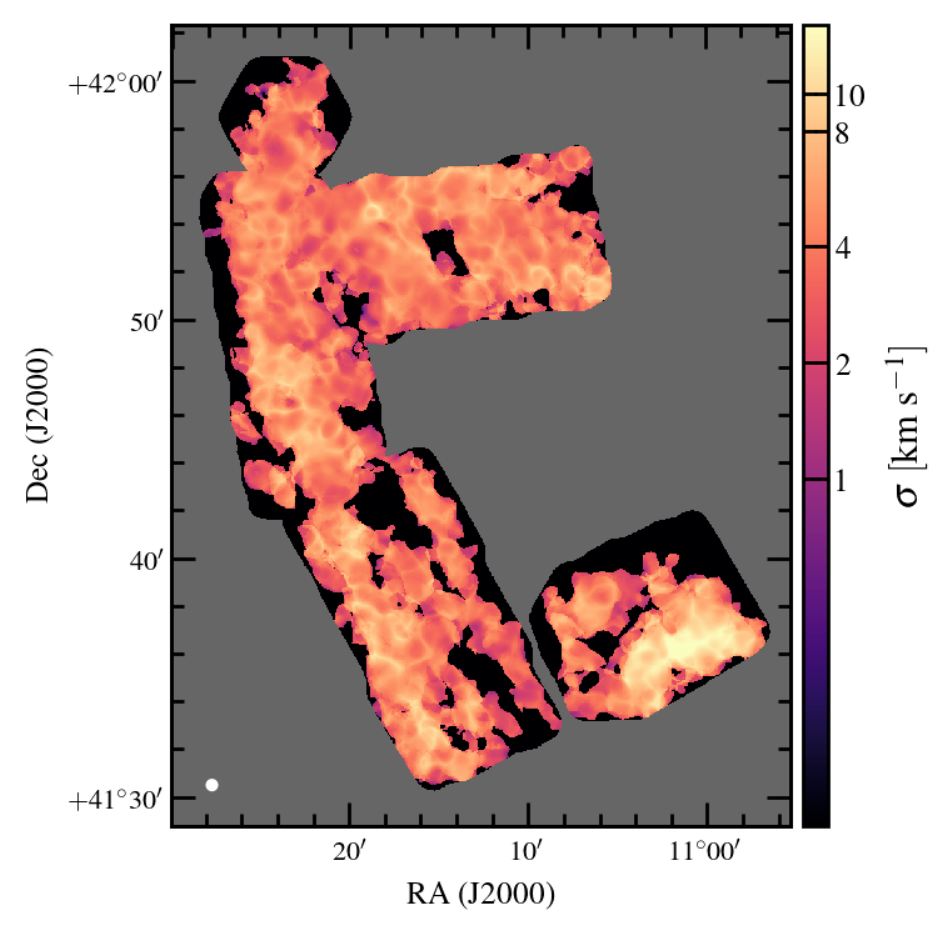}{0.35\textwidth}{(13b) M31 at 120 pc resolution}
}
\gridline{
\halfpanel{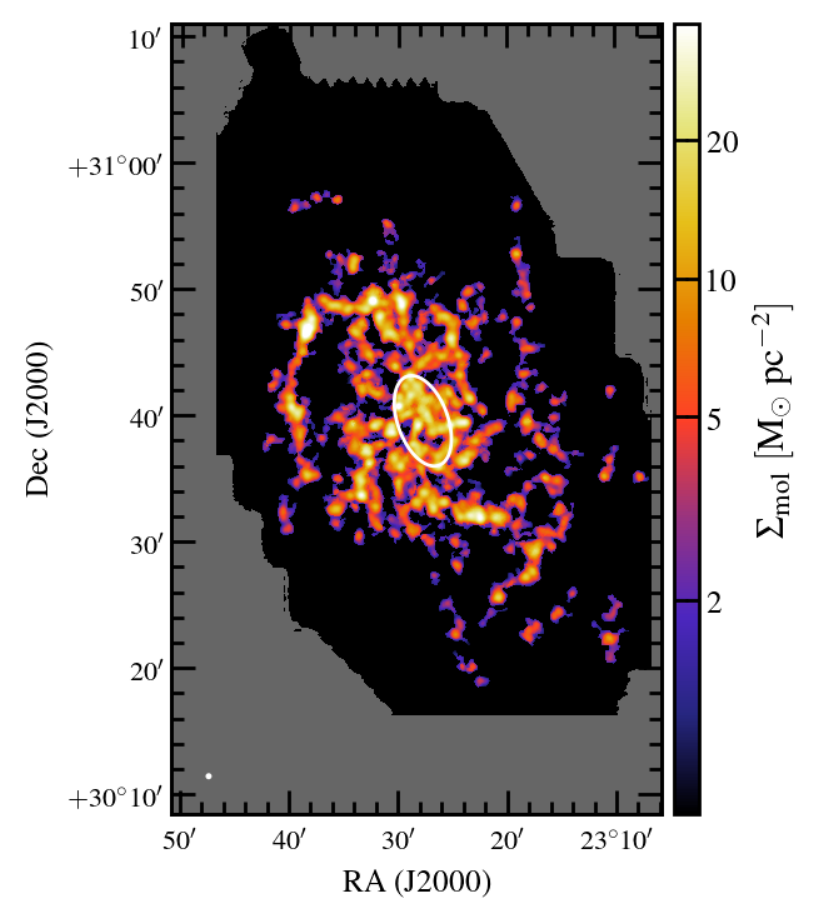}{0.45\textwidth}{(14a) M33 at 120 pc resolution}
\halfpanel{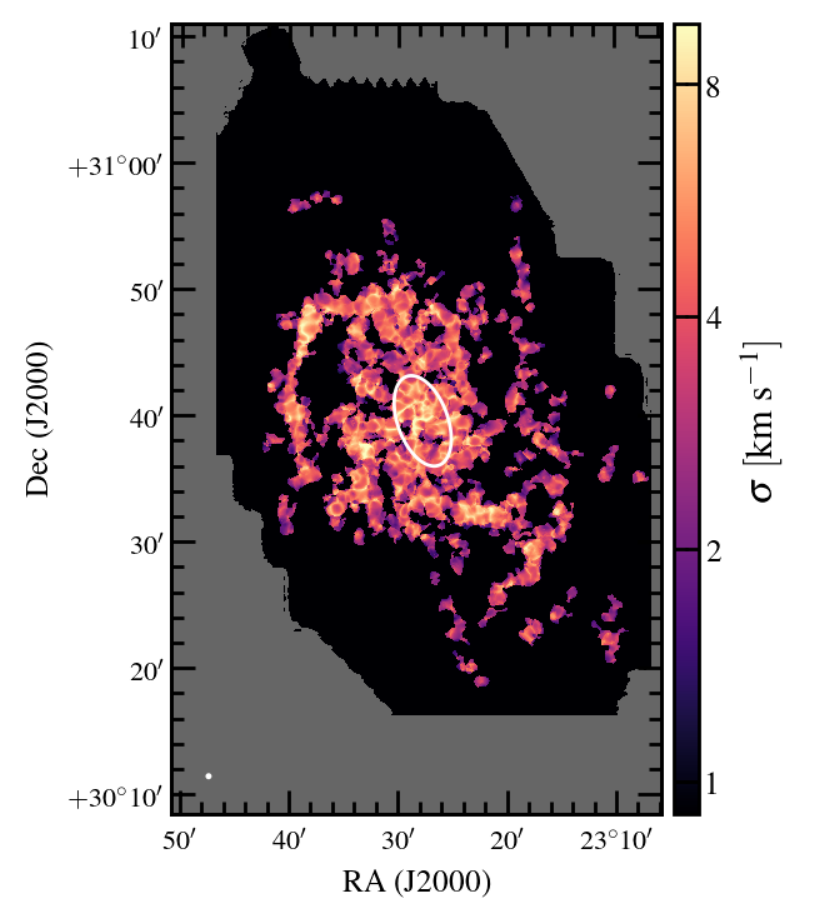}{0.45\textwidth}{(14b) M33 at 120 pc resolution}
}
\gridline{
\halfpanel{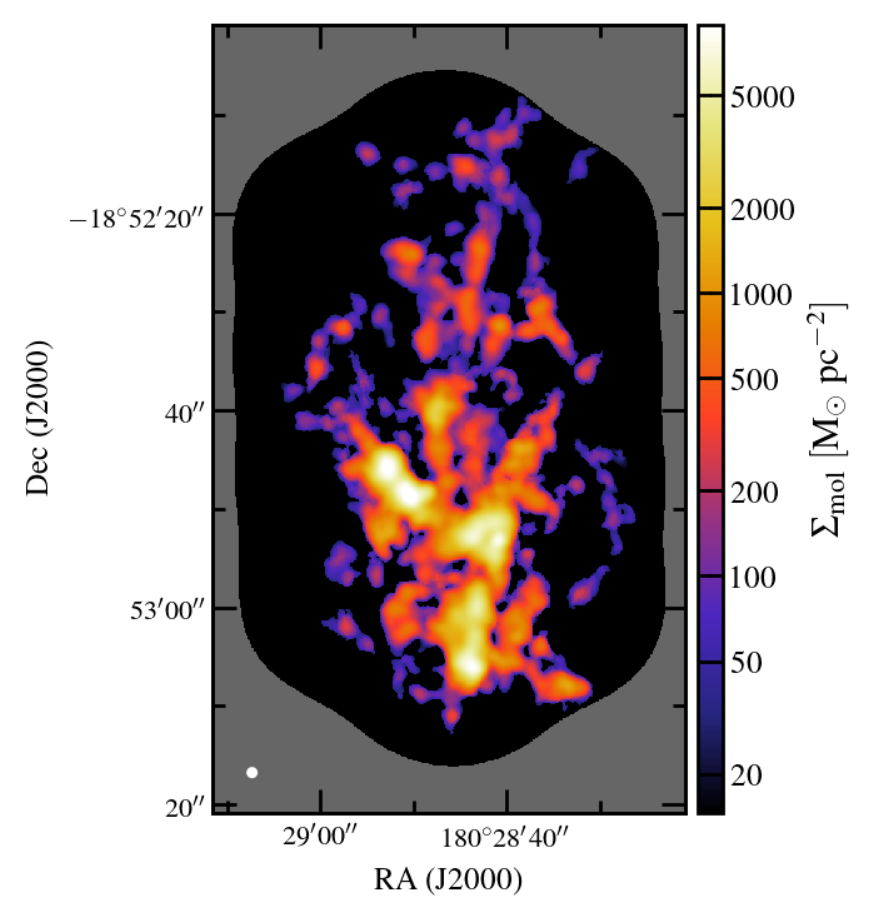}{0.35\textwidth}{(15a) The Antennae Galaxies at 120 pc resolution}
\halfpanel{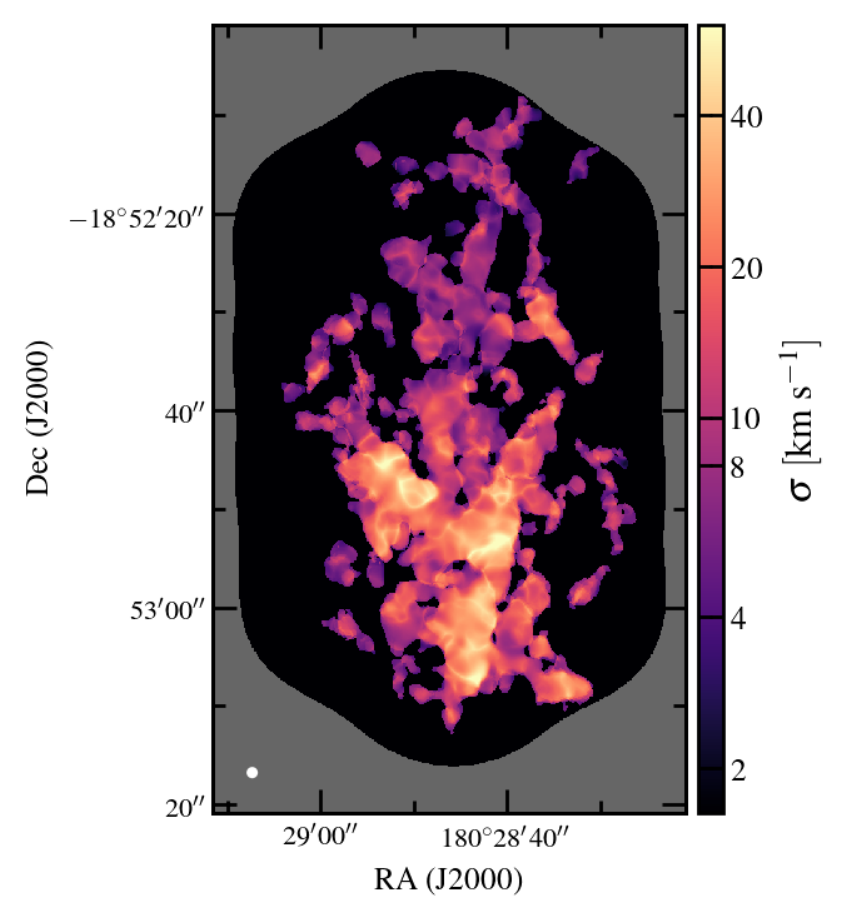}{0.35\textwidth}{(15b) The Antennae Galaxies at 120 pc resolution}
}
\caption{(Continued)}
\end{figure*}

\clearpage
\newpage

\section{Quantifying the Spectral Response Curve Width of the CO Maps}\label{apdx:response}

\renewcommand\thetable{\thesection\arabic{table}}
\setcounter{table}{0}

As mentioned in Section \ref{sec:method:sig}, when measuring the CO line width, we correct for the broadening caused by the instrumental spectral response curve using Equation \ref{eq:ew}. To estimate the width of this response curve, $\sig_{\rm response}$, we adopt the empirical approach suggested by \citet{Leroy_etal_2016}, shown as following:

\begin{eqnarray} \label{eq:response}
\sig_{\rm response} \approx \frac{\Delta v_{\rm channel}}{\sqrt{2\pi}} \times (1.0\,+\,1.18k\,+\,10.4k^2), \\
k \approx 0.0\,+\,0.47r\,-\,0.23r^2\,-\,0.16r^3\,+\,0.43r^4.
\end{eqnarray}

\noindent where $\Delta v_{\rm channel}$ denotes the channel width, $k$ quantifies the coupling between adjacent channels, and $r$ is the channel-to-channel correlation coefficient. This approach assumes that the spectral response outside each individual channel could be approximated by a Hanning-like kernel with shape $[k, 1-2k, k]$. In practice, we measure $r$ from the correlation between noise in successive empty spectral channels in the data cube, and then convert it to $k$ and $\sigma_{\rm response}$.

This empirically calibrated correction is shown to be effective when the channel-to-channel correlation is moderate \citep[or quantitatively, $r \lesssim 0.65$; see][]{Leroy_etal_2016}. To show that this condition holds for all the CO data analyzed here, in Table \ref{tab:corr} we report the measured channel-to-channel correlation coefficients for all CO maps at all available spatial resolutions. As a sanity check, for the four ancillary CO maps (M51, M31, M33, and Antennae), our measured correlation coefficients $r$ at $60$~pc resolution correspond to $k$ values of 0.14, 0.09, 0.11 and 0.07, which are consistent with those reported by \citet{Leroy_etal_2016} (see Table 3 therein).

\begin{deluxetable}{lccccc}[h]
\tabletypesize{\footnotesize}
\tablecaption{Channel-to-channel Correlation\label{tab:corr}}
\tablewidth{100pt}
\tablehead{
\colhead{Galaxy} &
\colhead{$r_{\rm 45pc}$} &
\colhead{$r_{\rm 60pc}$} &
\colhead{$r_{\rm 80pc}$} &
\colhead{$r_{\rm 100pc}$} &
\colhead{$r_{\rm 120pc}$}
}
\startdata
NGC~628 & 0.08 & 0.07 & 0.07 & 0.07 & 0.07 \\
NGC~1672 & -- & -- & -- & 0.11 & 0.11 \\
NGC~2835 & 0.15 & 0.16 & 0.16 & 0.16 & 0.17 \\
NGC~3351 & -- & 0.11 & 0.11 & 0.11 & 0.12 \\
NGC~3627 & -- & 0.10 & 0.10 & 0.10 & 0.10 \\
NGC~4254 & -- & -- & -- & -- & 0.11 \\
NGC~4303 & -- & -- & -- & -- & 0.10 \\
NGC~4321 & -- & -- & -- & 0.11 & 0.11 \\
NGC~4535 & -- & -- & -- & -- & 0.10 \\
NGC~5068 & 0.11 & 0.12 & 0.13 & 0.14 & 0.14 \\
NGC~6744 & -- & 0.08 & 0.09 & 0.10 & 0.10 \\
M51 & 0.35 & 0.35 & 0.35 & 0.36 & 0.37 \\
M31 & 0.13 & 0.21 & 0.33 & 0.44 & 0.50 \\
M33 & -- & 0.28 & 0.26 & 0.26 & 0.26 \\
Antennae & -- & 0.17 & 0.17 & 0.17 & 0.17 \\
\enddata
\end{deluxetable}

\clearpage
\newpage

\section{Power-law Fitting Strategy and Results}\label{apdx:mcmc}

\renewcommand\thefigure{\thesection\arabic{figure}}
\setcounter{figure}{0}
\setcounter{equation}{0}

Here we describe the methodology used to derive the best fit power-law parameters for the \sig-\Sig\ scaling relation (Section~\ref{sec:scaling:disk}). In broad terms, we take into account the intrinsic scatter around the best fit relation, the statistical error on the data, and the truncation of the data distribution due to the selection effect (see Section~\ref{sec:scaling:disk}). We adopt a Bayesian approach with an uninformative prior, and find the best fit model parameters, as well as their estimated uncertainties, by sampling the posterior distribution using a Markov chain Monte-Carlo (MCMC) method. Our implementation is based on the Python package {\tt emcee} \citep{Foreman-Mackey_etal_2013}, which implements the affine-invariant ensemble sampler \citep{Goodman_Weare_2010}.

We model the scaling relation between molecular gas velocity dispersion \sig\ and surface density \Sig\ as a power-law, or equivalently a straight line in logarithmic space, as parameterized by Equation~\ref{eq:scaling}. As we set up our Bayesian model in logarithmic space, we introduce the following notations for simplicity:

\begin{align}
s_0 &\equiv \log_{10}\left(\frac{\sig}{1\;\kms}\right)~, \\
S_2 &\equiv \log_{10}\left(\frac{\Sig}{10^2\;\rm\Msun\,pc^{-2}}\right)~.
\end{align}

\noindent The corresponding likelihood for a given $s_0$-$S_2$ pair can be expressed as

\begin{equation}
\mathrm{P}(s_0|\beta,A,\Delta_\mathrm{intr};S_2,\Delta_\mathrm{stat}) = (2\pi)^{-1/2} (\Delta_\mathrm{stat}^2+\Delta_\mathrm{intr}^2)^{-1/2}
\exp\left(
-\frac{(s_0 - \beta S_2-A)^2}
{2\,(\Delta_\mathrm{stat}^2+\Delta_\mathrm{intr}^2)}
\right)~.
\label{eq:like}
\end{equation}

\noindent The two terms in the denominator within the exponential term are two independent contributors to the overall scatter around the best fit model: $\Delta_\mathrm{intr}$ denotes the intrinsic dispersion in $s_0$, whereas $\Delta_\mathrm{stat}$ quantifies the contribution from the (statistical) measurement uncertainties in {\it both} $s_0$ {\it and} $S_2$. The statistical uncertainty term can be expressed as $\Delta_\mathrm{stat}^2 = \mathbf{x}\,\mathbf{C}\,\mathbf{x^T}$, where $\mathbf{x} = (1, -\beta)$, and $\mathbf{C}$ is the covariance matrix of the statistical error in ($s_0$, $S_2$). We refer the reader to a nice presentation of this concept in Section~7 in \citet{Hogg_etal_2010}.

We adopt uninformative priors for the fitting parameters $\beta$, $A$, and $\Delta_\mathrm{intr}$. The posterior probability $\mathrm{P}(\beta,A,\Delta_\mathrm{intr} | s_0,S_2,\Delta_\mathrm{stat})$ is thus proportional to the likelihood $\mathrm{P}(s_0 | \beta,A,\Delta_\mathrm{intr};S_2,\Delta_\mathrm{stat})$. To further simplify the notation, we use $I$ to represent the collection of all input information $s_0$, $S_2$, and $\Delta_\mathrm{stat}$, so the posterior distribution function can be expressed as $\mathrm{P}(\beta,A,\Delta_\mathrm{intr} | I)$.

The noise in the CO data cube prevents us from detecting faint-and-wide CO lines, meaning that there is a truncation in our data sample on the low \Sig , high \sig\ side, as illustrated in Figures~\ref{fig:scaling} and \ref{fig:scaling-ext}. We account for this selection effect by modifying the posterior distribution function as $\mathrm{P}^\mathrm{trunc}(\beta,A,\Delta_\mathrm{intr} | I) = f\;\mathrm{P}(\beta,A,\Delta_\mathrm{intr} | I)$. The normalization factor $f$ is itself {\it a function of $\beta$, $A$, $\Delta_\mathrm{intr}$, and $\Delta_\mathrm{stat}$}:

\begin{equation}
f =
\begin{cases}
\left[\iint_R\,\mathrm{d}s_0\,\mathrm{d}S_2\; \mathrm{P}(\beta,A,\Delta_\mathrm{intr} | s_0, S_2,\Delta_\mathrm{stat})\right]^{-1}~, & \text{$(s_0, S_2) \in R$~;}\\
0~, & \text{$(s_0, S_2) \notin R$~.}
\end{cases}
\label{eq:trunclike}
\end{equation}

\noindent Here $R$ denotes the entire region above the detection limit in the $s_0$-$S_2$ parameter space (i.e., the unshaded regions in Figure~\ref{fig:scaling} and \ref{fig:scaling-ext}). The detection limit is defined by the noise level in the CO data cube and the $\mathrm{S/N}>5$ criteria for CO signal identification (see Section~\ref{sec:method:procedure}).

For each galaxy at each resolution level, we express the overall posterior distribution function as 

\begin{equation}
\prod_i \mathrm{P}^\mathrm{trunc}(\beta,A,\Delta_\mathrm{intr} | I_i)~,
\end{equation}

\noindent where $I_i$ correspond to all the measurements and associated uncertainties along the $i$-th sightline, and we only include sightlines above the $\mathrm{S/N}>5$ threshold (i.e., those located in region $R$). We then sample this posterior distribution using an MCMC method, and find the best fit $\beta$, $A$, and $\Delta_\mathrm{intr}$, as well as their uncertainties.

As a sanity check on our fitting routine, we generate mock distributions that follow a given power law relation with given intrinsic scatter and truncation, and input these mock distributions to the fitting routine. We find that the input parameters can be accurately recovered as long as the truncation does not severely affect the whole distribution.

In addition to the galaxy-by-galaxy analysis, we also try to combine data at the same resolution across the main sample, and derive a set of best-fit parameters from the combined sample. For this purpose, we construct the posterior distribution function by calculating the product of the posterior functions for each individual galaxy. This naturally takes into account the galaxy-dependent truncation in the data. We then replicate the sampling and parameter estimation process as we do for individual galaxies.

The best-fit model parameters and their uncertainties are reported in Table~\ref{tab:fit}. The galaxy-by-galaxy best-fit results are shown at the top, and the best-fit for the combined sample (PHANGS-ALMA targets plus M51) are shown at the bottom. Here we also provide the distribution-correlation plots for each individual galaxy (at 120~pc resolution) in Figure Set \ref{figset:mcmc}.

\figsetstart
\figsetnum{C}
\figsettitle{Distribution-correlation plots for $\beta$, $A$, and $\Delta_\mathrm{intr}$, measured for all individual galaxies at 120~pc resolution.}
\figsetgrpstart
\figsetgrpnum{C.1}
\figsetgrptitle{NGC~628 at 120~pc resolution}
\figsetplot{NGC0628_120pc_5sigma.pdf}
\figsetgrpnote{}
\figsetgrpend
\figsetgrpstart
\figsetgrpnum{C.2}
\figsetgrptitle{NGC~1672 at 120~pc resolution}
\figsetplot{NGC1672_120pc_5sigma.pdf}
\figsetgrpnote{}
\figsetgrpend
\figsetgrpstart
\figsetgrpnum{C.3}
\figsetgrptitle{NGC~2835 at 120~pc resolution}
\figsetplot{NGC2835_120pc_5sigma.pdf}
\figsetgrpnote{}
\figsetgrpend
\figsetgrpstart
\figsetgrpnum{C.4}
\figsetgrptitle{NGC~3351 at 120~pc resolution}
\figsetplot{NGC3351_120pc_5sigma.pdf}
\figsetgrpnote{}
\figsetgrpend
\figsetgrpstart
\figsetgrpnum{C.5}
\figsetgrptitle{NGC~3627 at 120~pc resolution}
\figsetplot{NGC3627_120pc_5sigma.pdf}
\figsetgrpnote{}
\figsetgrpend
\figsetgrpstart
\figsetgrpnum{C.6}
\figsetgrptitle{NGC~4254 at 120~pc resolution}
\figsetplot{NGC4254_120pc_5sigma.pdf}
\figsetgrpnote{}
\figsetgrpend
\figsetgrpstart
\figsetgrpnum{C.7}
\figsetgrptitle{NGC~4303 at 120~pc resolution}
\figsetplot{NGC4303_120pc_5sigma.pdf}
\figsetgrpnote{}
\figsetgrpend
\figsetgrpstart
\figsetgrpnum{C.8}
\figsetgrptitle{NGC~4321 at 120~pc resolution}
\figsetplot{NGC4321_120pc_5sigma.pdf}
\figsetgrpnote{}
\figsetgrpend
\figsetgrpstart
\figsetgrpnum{C.9}
\figsetgrptitle{NGC~4535 at 120~pc resolution}
\figsetplot{NGC4535_120pc_5sigma.pdf}
\figsetgrpnote{}
\figsetgrpend
\figsetgrpstart
\figsetgrpnum{C.10}
\figsetgrptitle{NGC~5068 at 120~pc resolution}
\figsetplot{NGC5068_120pc_5sigma.pdf}
\figsetgrpnote{}
\figsetgrpend
\figsetgrpstart
\figsetgrpnum{C.11}
\figsetgrptitle{NGC~6744 at 120~pc resolution}
\figsetplot{NGC6744_120pc_5sigma.pdf}
\figsetgrpnote{}
\figsetgrpend
\figsetgrpstart
\figsetgrpnum{C.12}
\figsetgrptitle{M51 at 120~pc resolution}
\figsetplot{NGC5194_120pc_5sigma.pdf}
\figsetgrpnote{}
\figsetgrpend
\figsetgrpstart
\figsetgrpnum{C.13}
\figsetgrptitle{M31 at 120~pc resolution}
\figsetplot{NGC0224_120pc_5sigma.pdf}
\figsetgrpnote{}
\figsetgrpend
\figsetgrpstart
\figsetgrpnum{C.14}
\figsetgrptitle{M33 at 120~pc resolution}
\figsetplot{NGC0598_120pc_5sigma.pdf}
\figsetgrpnote{}
\figsetgrpend
\figsetgrpstart
\figsetgrpnum{C.15}
\figsetgrptitle{The Antennae at 120~pc resolution}
\figsetplot{NGC4038_120pc_5sigma.pdf}
\figsetgrpnote{}
\figsetgrpend
\label{figset:mcmc}
\figsetend

\begin{figure*}[!htb]
\fig{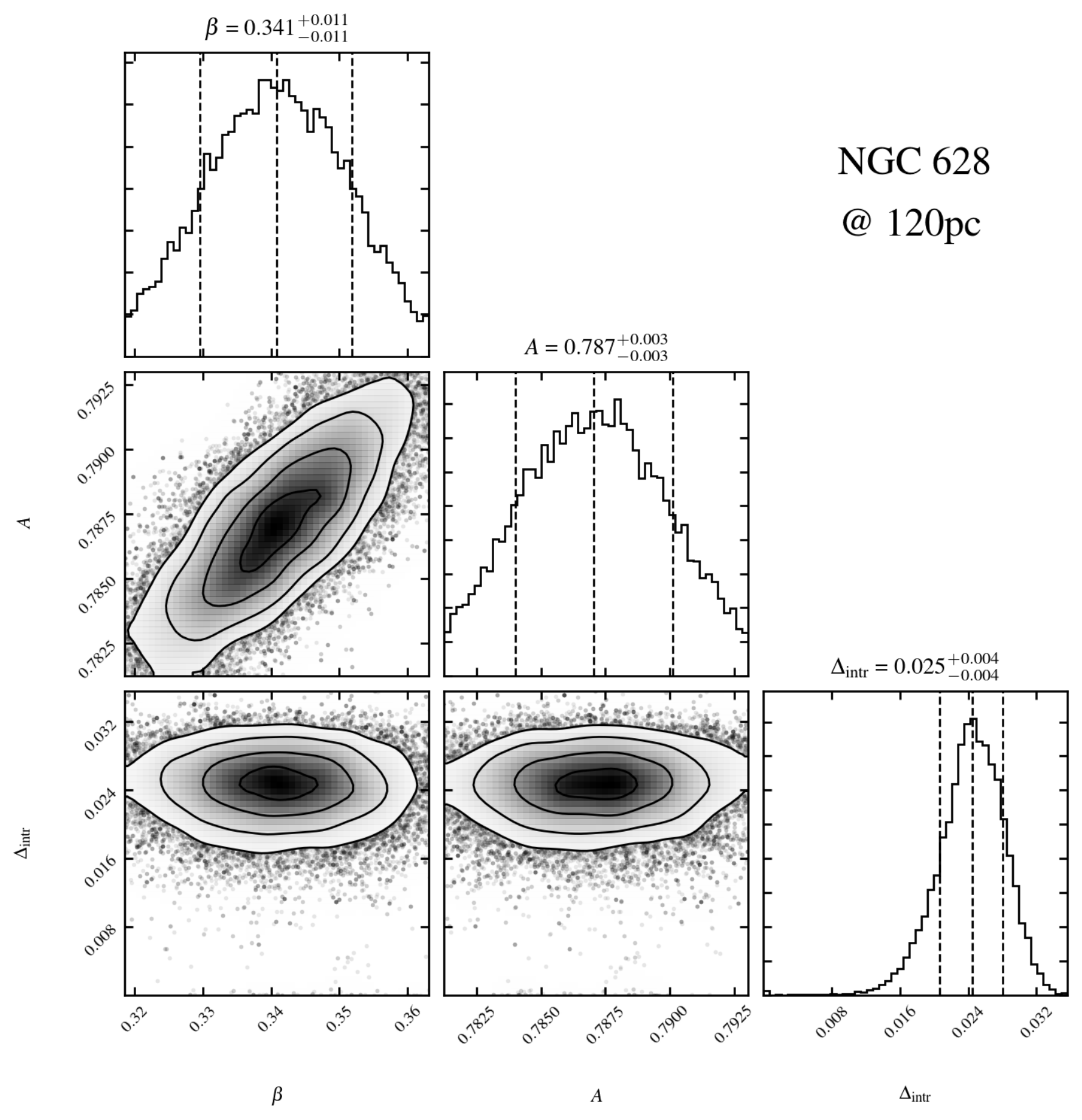}{0.8\textwidth}{}
\vspace{-2em}
\caption{An example figure from Figure Set \ref{figset:mcmc}, showing the distribution-correlation plot for NGC~628 at 120~pc resolution.
The complete figure set (including 15 figures) is available in the online journal.}
\label{fig:mcmc}
\end{figure*}

\clearpage
\newpage

\section{Correlation between CO Line Width and Peak Intensity in Individual Galaxies}\label{apdx:sig-tpeak}

\renewcommand\thefigure{\thesection\arabic{figure}}
\setcounter{figure}{0}
\setcounter{equation}{0}

\begin{figure*}[!htb]
\fig{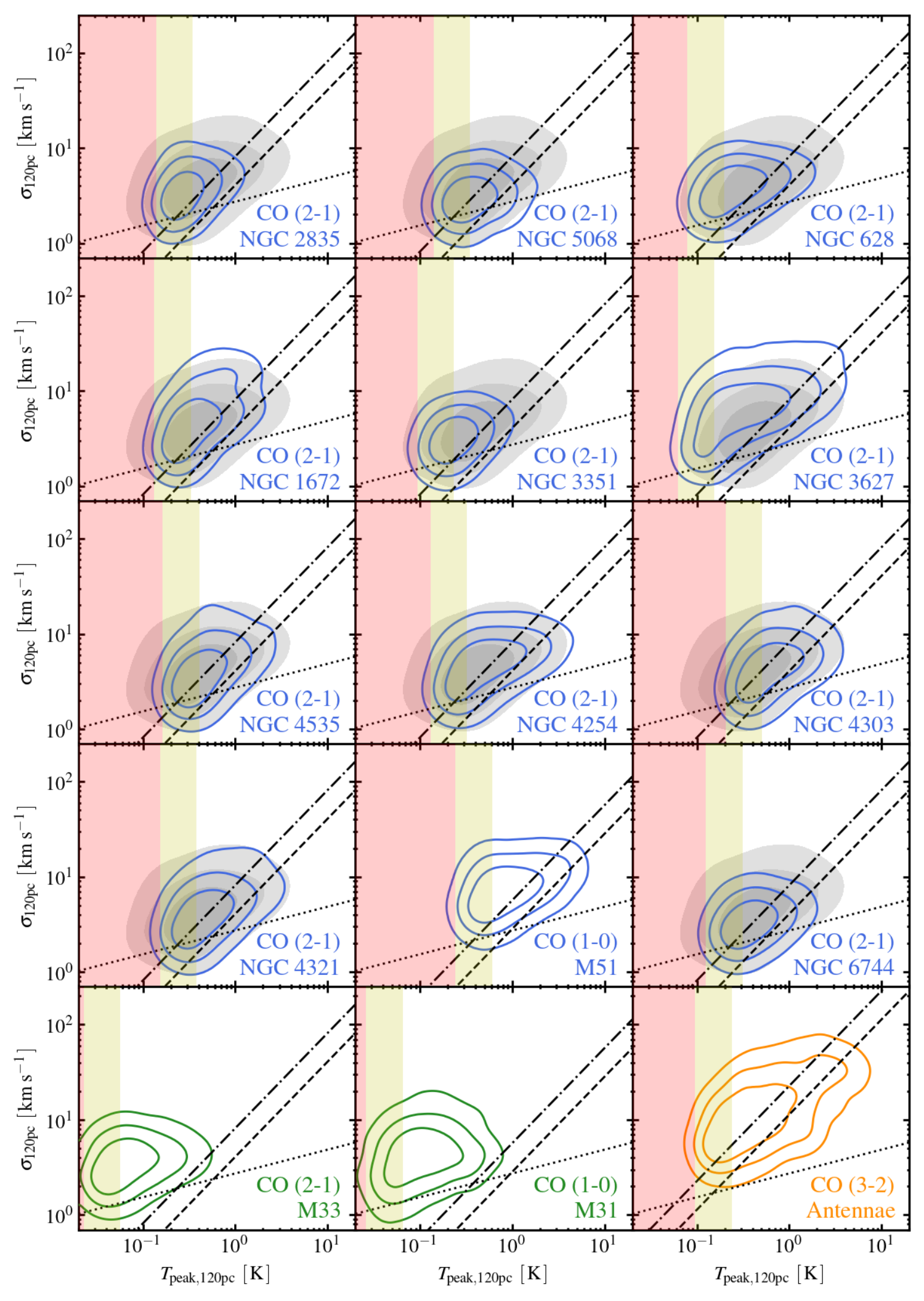}{0.78\textwidth}{}
\vspace{-2em}
\caption{Correlations between CO line width \sig\ and peak specific intensity \Tpeak\ measured for all 15 galaxies at 120~pc resolution.
The colored contours in each panel show the measurements for each galaxy disk, whereas the gray filled contours in the background show the data density of all measurements in the disk region of all PHANGS-ALMA targets.
The black dashed (resp. dash-dot) line represents the expected $\sig \propto \Tpeak$ relation from $\alphavir = 1$ (resp. $2$), and the black dotted line is the $\sig \propto \Tpeak^{0.25}$ relation as expected for clouds with fixed brightness temperature and being completely unresolved (see discussion above).
The yellow and red hatched region represent the sensitivity limit for each galaxy (5-$\sigma$ and 2-$\sigma$ respectively).}
\label{fig:sig-Tpeak}
\end{figure*}

In Section~\ref{sec:discussion}, we present the \sig-\Tpeak\ correlation for the PHANGS-ALMA sample and M33, which supports the point that the observed \sig-\Sig\ scaling relation does not originate from the by construction correlation between $\sig$ and $\Sig \propto \sig T_{\rm peak}$. Here we show the galaxy-by-galaxy \sig-\Tpeak\ relation for all 15 targets in Figure~\ref{fig:sig-Tpeak}.

We further expand the discussion in Section~\ref{sec:discussion} by considering another scenario, that we capture an isolated, unresolved molecular gas cloud inside each beam. If the true brightness temperature of clouds remains fixed, the observed $\Tpeak$ variation mainly reflects the variation in the beam filling factor. In this case, we have

\begin{equation}
\Tpeak = \frac{A_{\rm cloud}}{A_{\rm beam}}\,T_{\rm intrinsic}~,
\label{eq:dilution}
\end{equation}

\noindent where $A_{\rm beam}$ is the beam area, $A_{\rm cloud}$ the projected area, and $T_{\rm intrinsic}$ the intrinsic CO brightness temperature of the cloud. If we assume that the cloud follows a fiducial size-line width relation of $\sig = 1.0\;\kms\,(S/1\rm\;pc)^{0.5}$ \citep[following the notation in][]{Solomon_etal_1987}, we can infer the cloud area $A_{\rm cloud}$ from $\sig$ as

\begin{equation}
A_{\rm cloud} = 11.6\rm\;pc^2\left(\frac{\sig}{1\;\kms}\right)^4~.
\label{eq:area-lw}
\end{equation}

\noindent Note that $S$ is just a parametrization of the cloud linear size defined in \citep{Solomon_etal_1987}. Equation~\ref{eq:dilution} and \ref{eq:area-lw} together imply a \sig-\Tpeak\ relation

\begin{equation}
\sig = 6.1\;\kms\;\left(\frac{\rbeam}{60\rm\;pc}\right)^{0.5}\;\left(\frac{\Tpeak}{T_{\rm intrinsic}}\right)^{0.25}~.
\end{equation}

\noindent Here \rbeam\ is the half-width-at-half-maximum of the beam, so that $\rbeam = 60\rm\;pc$ corresponds to 120~pc resolution.

In Figure~\ref{fig:sig-Tpeak} we show this $\sig \propto \Tpeak^{0.25}$ relation under the assumption of $T_{\rm intrinsic} = 25$~K (black dotted line), together with the $\sig \propto \Tpeak$ relations as expected from fixed $\alphavir$ values (black dashed and dash-dot lines). About half of the PHANGS-ALMA targets show \sig-\Tpeak\ joint distributions that are consistent with $\sig \propto \Tpeak$, with the caveat that the measurement error on $\Tpeak$ is usually fractionally larger than that on $\Sig$, and the sensitivity cut is limiting our ability in probing low \Tpeak\ regimes. Nevertheless, several targets do have apparently shallower $\sig \propto \Tpeak$ relations (NGC~5068, NGC~628, NGC~3627, NGC~4254, M51, M33, M31). Four of them are low mass galaxies or galaxies with low molecular gas content (NGC~5068, NGC~628, M33, M31), in which we indeed expect severe beam dilution.

Our peak intensity results may be a useful alternative to the \sig-\Sig\ results for theories focused on CO radiative transfer or sub-beam clumping.

\end{document}